\documentclass[11pt,letterpaper]{article}
\pdfoutput=1

\usepackage{jcappub}

\usepackage{dcolumn}
\usepackage{bm}
\usepackage{natbib}
\usepackage{amssymb}
\usepackage{epstopdf}
\usepackage{hhline}
\usepackage[T1]{fontenc}
\usepackage{wasysym}
\usepackage{url}
\usepackage[flushleft]{threeparttable}
\usepackage{tabularx}
\usepackage{amsmath}
\usepackage{float}
\usepackage{datetime}
\usepackage{textpos}
\usepackage{booktabs}
\usepackage{graphicx}
\usepackage{mathrsfs}
\usepackage{adjustbox}
\usepackage{epsfig}
\usepackage{color}
\usepackage{bm}
\usepackage{graphicx}

\newcommand{\be}{\begin{equation}}
\newcommand{\ee}{\end{equation}}
\newcommand{\bea}{\begin{eqnarray}}
\newcommand{\eea}{\end{eqnarray}}

\newcommand{\sech}{{\rm ~sech~}}

\begin{document}

\title{Strong Evidence that the Galactic Bulge is Shining in Gamma Rays}

\author[a,b,c]{Oscar Macias,}
\affiliation[a]{Center for Neutrino Physics, Department of Physics, Virginia Tech, Blacksburg, VA 24061, USA}
\affiliation[b]{Kavli Institute for the Physics and Mathematics of the
Universe (WPI), University of Tokyo, Kashiwa, Chiba 277-8583, Japan}
\affiliation[c]{GRAPPA Institute, University of Amsterdam, 1098 XH Amsterdam, The Netherlands}

\author[a]{Shunsaku Horiuchi,}

\author[d]{Manoj Kaplinghat,}
\affiliation[d]{Center for Cosmology, Department of Physics and Astronomy, University of California, Irvine, Irvine, California 92697 USA}

\author[e]{Chris Gordon,}
\affiliation[e]{School of Physical and Chemical Sciences, University of Canterbury, Christchurch, New Zealand}

\author[f]{Roland~M.~Crocker,}
\affiliation[f]{Research School of Astronomy and Astrophysics, Australian National University, Canberra, Australia}

\author[g]{David M. Nataf}
\affiliation[g]{Center for Astrophysical Sciences and Department of Physics and Astronomy, 
The Johns Hopkins University, 
Baltimore, MD 21218}

\emailAdd{oscar.macias@ipmu.jp}

\date{today}

\abstract{There is growing evidence that the Galactic Center Excess identified in the \textit{Fermi}-LAT gamma-ray data arises from a population of faint astrophysical sources. We provide compelling supporting evidence by showing that the morphology of the excess traces the stellar over-density of the Galactic bulge. By adopting a template of the bulge stars obtained from a triaxial 3D fit to the diffuse near-infrared emission, we show that it is detected at high significance. The significance deteriorates when either the position or the orientation of the template is artificially shifted, supporting the correlation of the gamma-ray data with the Galactic bulge. In deriving these results, we have used more  sophisticated templates at low-latitudes for the \textit{Fermi} bubbles compared to previous work and the  three-dimensional Inverse Compton (IC) maps recently released by the {\tt GALPROP} team.  Our results provide strong constraints on Millisecond Pulsar (MSP) formation scenarios proposed to explain the excess. We find that an \textit{admixture formation} scenario, in which some of the relevant binaries are \textit{primordial} and the rest are formed \textit{dynamically}, is preferred over a primordial-only formation scenario at $7.6\sigma$ confidence level. Our detailed morphological analysis also disfavors models of the disrupted globular clusters scenario that predict a spherically symmetric distribution of MSPs in the Galactic bulge. For the first time, we report evidence of a high energy tail in the nuclear bulge spectrum that could be the result of IC emission from electrons and positrons injected by a population of MSPs and star formation activity from the same site.}

\maketitle

\section{Introduction}\label{sec:intro}

The center of the Milky Way Galaxy is a complex environment displaying an extremely rich phenomenology of energetic cosmic-ray production and interactions. Many types of transient and persistent cosmic-ray source candidates populate the region, including a supermassive black hole \citep{Genzel:2010zy}, active star formation \citep{Krumholz:2015aa}, supernova remnants \citep{Ponti:2015tya}, pulsars and black holes \citep{Muno:2003ne}, as well as new physics involving particle dark matter annihilations or decays (see, e.g., \citep{Bertone:2004pz} for a review). The bulk of the gamma-ray signal from the Galactic Center arises from the interactions of energetic cosmic rays with the interstellar gas and radiation fields. However, there is growing evidence that there exists an excess of extended gamma rays observed on top of models of cosmic-ray interactions in the Galactic Center. Dubbed the Galactic Center Excess (GCE), this signal has been detected by multiple analyses of the \textit{Fermi}-LAT data \citep{Goodenough:2009gk,Hooper:2010mq,Boyarsky:2010dr,Abazajian:2012pn,Gordon:2013vta,Macias:2013vya,Huang:2013pda,Abazajian:2014fta,Abazajian:2014hsa,Calore:2014xka,Daylan:2014rsa,TheFermi-LAT:2015kwa}.

The origin(s) of the GCE remain elusive. The main properties of the GCE---being centrally peaked with a roughly spherically symmetric spatial morphology, having an energy spectrum peaking at a few GeV, and total luminosity of $\sim 10^{37}$ erg/s---can be accommodated within a scenario where weakly interacting massive particle (WIMP) dark matter of $\mathcal{O}(10 \, {\rm GeV})$ mass distributed in a slightly steep Navarro-Frenk-White (NFW)-like profile self-annihilate. However, the GCE might also arise from a population of gamma-ray emitting pulsars, in the form of either old ``recycled'' millisecond pulsars \citep{Abazajian:2010zy,Abazajian:2014fta}, or young pulsar remnants from Galactic Center supernovae \citep{OLeary:2015qpx,OLeary:2016cwz}. The gamma-ray spectra of known pulsars are similar to the GCE, and the capture of pulsars onto the Galactic Center region by Globular Cluster disruptions could explain the GCE's quasi-spherical spatial distribution \citep{Gnedin:2013cda,Brandt:2015ula}. Other explanations for the GCE discussed in the literature include outbursts of cosmic-ray production by the central supermassive black hole (e.g., \citep{Cholis:2015dea}). 

While pulsars are a natural astrophysical candidate, there is ongoing debate regarding the consistency of the luminosity function required to explain the GCE with that measured for the pulsar population observed elsewhere as point sources~\citep{Cholis:2014lta,Hooper:2015jlu,Ploeg:2017vai,Bartels:2018xom}. Nevertheless, there are strengthening indications that the GCE may have an astrophysical origin. Cosmic-ray interactions in the Galactic Center region are notoriously complex to model. New analyses have cast doubt on some of the main properties claimed previously for the GCE. First, the energy spectrum of the GCE is subject to large systematic uncertainties arising from the incomplete understanding of cosmic-ray interactions in the Galactic Center region (e.g., \citep{Abazajian:2014fta,Abazajian:2014hsa,Calore:2014xka,TheFermi-LAT:2015kwa}). A variety of scenarios might, therefore, describe the GCE based on the energy spectrum alone. Second, it has been argued that the photon count distribution of the GCE is more consistent with arising from a population of faint point sources rather than being truly diffuse \citep{Lee:2015fea,Bartels:2015aea}. This observation supports an astrophysical origin in the form of faint point sources over a microphysics origin in the form of dark matter particle annihilation. Third, there are growing reports of a spatial variation of the GCE energy spectrum with latitude \citep{Horiuchi:2016zwu} showing a high-energy tail away from the plane, similar to that recently reported by Ref.~\cite{Linden:2016rcf}. Again, this disfavors a dark matter origin because particle properties would be expected to be spatially invariant (although secondary gamma-ray mechanisms might remain spatially-dependent). 

Finally, it has recently been argued that the spatial morphology of the GCE is better matched to the asymmetric (stellar) bulge of the Milky Way  than the spherically symmetric distribution expected for dark matter \citep{Macias:2016nev,Bartels2017}. The Galactic bulge is a triaxial bar-like structure in the central region of our Galaxy extending in size to a few kpc \citep{Dwek:1995xu,Freudenreich:1997bx,LopezCorredoira:1999dg}. While the exact morphological properties of the bulge still remain under investigation, studies agree that it is asymmetric on our sky plane given that its major axis approaches us at positive Galactic longitudes (see, e.g., \citep{Babusiaux:2005zi}). Reference~\cite{Macias:2016nev} adopted two bulge maps to analyze the GCE: (i) The X-shaped Galactic bulge template of Ness \& Lang \citep{Ness:2016aaa}, and (ii) the boxy shaped bulge of Freudenreich~\cite{Freudenreich:1997bx}. They found that the data overwhelmingly preferred either of the bulge maps over the spherically symmetric distribution expected by a dark matter origin. The bulge houses a broad mix of stellar populations from old to star forming \citep{Garzon:1997cs,Hammersley:2000mx}, and should contain ample candidates of astrophysical gamma-ray emitters. 

In this paper, we put the reported bulge-correlation of the Galactic Center Excess to further tests. First, we include the recent {\tt GALPROP} three-dimensional (3D) Inverse Compton (IC) maps constructed in Ref.~\cite{Porter:2017vaa} in our pipeline. The new IC maps do not assume the Galactocentric cylindrical symmetry adopted in all two-dimensional (2D) IC maps constructed with previous {\tt GALPROP} versions (v54 or older) and contain sophisticated model templates for the bulge/bar, spiral arms and stellar disk.  Second, compared to previous work in Ref.~\cite{Macias:2016nev}, our region of interest (RoI) is about a factor of $7$ larger. Third, we translate and rotate the bulge template to investigate how robustly it is  detected in the Fermi-LAT data. We find that  the Galactic bulge is detected strongly in gamma rays only when consistently positioned and aligned within one or two degrees of the known stellar bulge position~\cite{Freudenreich:1997bx}. 

Fourth, we perform a morphological analysis of the stellar bulge distribution in order to test the different MSP formation scenarios that have been discussed in the literature.
We consider 
 an \textit{
admixture formation} scenario in which some of relevant binaries are primordial and the remaining ones are the result of stellar interactions.
We find that the admixture formation scenario
is preferred to primordial-only formation with a confidence of $7.6\sigma$. Finally, we present the gamma-ray spectra of the boxy bulge and the nuclear bulge. For the first time, we report evidence of a high-energy tail in the nuclear bulge spectrum that could be the result of IC emission from electrons and positrons injected by a population of MSPs and/or star formation activity from the same site. The dark matter implications of our new results will be presented in a separate article.

The structure of this article is as follows. In Sec.~\ref{sec:setup} we provide details about the data, model templates, and pipeline used in this study. In Sec.~\ref{sec:tests} we show our results for the empirical tests of rotation and translation of the stellar templates as well as our morphological analysis. In Sec.~\ref{sec:bulge} we present the gamma-ray properties of the Galactic bulge. Finally, a discussion and our conclusions are given in Sec.~\ref{sec:discussion}.  

\begin{table*}[!htbp]\caption{{\bf Summary of the model maps considered in this study. \label{Tab:definitions}}}
\begin{adjustbox}{width=1.0\textwidth, center}
\centering\begin{threeparttable}
 \scriptsize 
\begin{tabular}{llr}
\hline\hline
Component  & Brief description & Reference\\\hline 
Interstellar gas correlated gamma-ray emission  & Hydrodynamical gas and dust maps divided in various rings: & \\
 & H{\tiny I} and H\boldmath$_{2}$ gas column density templates (4 rings each) plus & \\
& two dust correction templates & ~\cite{Macias:2016nev}\\
 &&\\
Inverse Compton emission & Seven IC maps are considered: a standard 2D IC map  & \\
& called ``Std-SA0'', in addition to six different 3D IC models: F98-,&\\
&SA0, F98-SA50, F98-SA100, R12-SA0, R12-SA50 and R12-SA100$^\dagger$& \cite{Porter:2017vaa}\\
&&\\
\textit{Fermi} bubbles& Three templates are tested: Catenary, structured \textit{Fermi} &\\
&bubbles (SFB) and an inpainted version of the latter (SFB (Inp.))&~\cite{Casandjian:andFermiLat2016,TheFermi-LAT:2017vmf}\\
&&\\
Loop I& Analytical model & \cite{Wolleben:2007}\\
Point sources & Second \textit{Fermi} Inner Galaxy Catalog (2FIG)& \cite{Fermi-LAT:2FIG}\\
Sun and Moon templates& Obtained with \textit{Fermi} science tools&\\
Isotropic emission& \texttt{iso$_{-}$P8R2$_{-}$ULTRACLEANVETO$_{-}$V6$_{-}$v06.txt}&\\\hline
&&\\
Nuclear bulge (NB)& Template constructed from star count data& ~\cite{Nishiyama2015}\\
Boxy bulge& Model derived from a fit to near-infrared data& ~\cite{Freudenreich:1997bx}\\
X-bulge& Residual image from WISE data& ~\cite{Ness:2016aaa}\\
Spherically symmetric template & Square of a Navarro-Frenk-White with a mild slope (NFW$^2$) &\\
\hline\hline
\end{tabular}
\begin{tablenotes}
\item  $^\dagger$We adopt two different Galaxy-wide dust and stellar distribution models as considered in Ref.~\cite{Porter:2017vaa}: Freudenreich (1998)~\cite{Freudenreich:1997bx} (F98) and Robitaille et al.~(2012)~\cite{R12} (R12). For each, three different CR propagation setups are also considered: SA0, SA50 and SA100. See Table 3 of Ref.~\cite{Porter:2017vaa}.
\end{tablenotes}
\end{threeparttable}
\end{adjustbox}
\end{table*}

\section{Data and Methods}\label{sec:setup}

We first describe our modeling procedure, starting with a description of data selection, the templates used in our analysis, and finally the analysis method. 

\subsection{Data selection}\label{subsec:dataselection}

We used $\sim 7$ years (August 4, 2008$-$September 4, 2015) of \textsc{Pass 8 ULTRACLEANVETO} class photons with reconstructed energy in the 667 MeV$-$158 GeV range. Photons detected at zenith angles larger than  90$^{\circ}$ were excised to limit the contamination from $\gamma$-rays generated by cosmic-ray interactions in the upper layers of the atmosphere. Moreover, we filtered the data using the recommended specifications (DATA$_{-}$QUAL$>$0)\&\&(LAT$_{-}$CONFIG==1). \textit{Fermi} Science Tools v10r0p5 and instrument response functions (IRFs) P8R2$_{-}$ULTRACLEANVETO$_{-}$V6 were used for this analysis. In addition, the analysis was restricted to a square region of $40^{\circ}\times 40^{\circ}$ centered at Galactic coordinates $(l,b) = (0,0)$ with a spatial binning of $0.2^\circ$.

\subsection{Baseline model templates}\label{subsec:baseline}

We started with a similar baseline model to that developed in Ref.~\cite{Macias:2016nev}. In summary, this consists of the point-like sources listed in the 2FIG catalog~\cite{Fermi-LAT:2FIG}, specialized templates for the \textit{Sun} and the \textit{Moon}, an isotropic component (\texttt{iso$_{-}$P8R2$_{-}$ULTRACLEANVETO$_{-}$V6$_{-}$v06.txt}), a standard 2D IC template generated with {\tt GALPROP} v56~\cite{Porter:2017vaa}, and an emission map for Loop I. In addition, the interstellar gas-correlated photons were modeled with a linear combination of atomic and molecular hydrogen gas templates divided into concentric rings (see Section \ref{subsec:analyspipelines}). Furthermore, these gas maps were obtained from a suite of hydrodynamical simulations of interstellar gas flow. This contrasts with gas maps in {\tt GALPROP} and the official \textit{Fermi} diffuse emission model that are constructed with an interpolation technique. Reference~\cite{Macias:2016nev} showed in detail that there are important morphological differences between the interpolated and hydrodynamic gas maps and that the latter provides a significantly better fit to the gamma-ray data in the Galactic Center region. See Table~\ref{Tab:definitions} for a summary of the model templates considered in this work.

\subsection{Bulge templates}\label{subsec:newtemplates}

Analyses~\cite{Wilandetal:1994,Binneyetal:1997,Freudenreich:1997bx} of the near-infrared emission measured by the DIRBE instrument on board of the COBE satellite were the first to uncover the non-spherical nature of the Galactic bulge. More recent studies of the Galactic bulge stars~\cite{Cao:2013dwa,WeggandGerhard:2013} have firmly established that the bulk of bulge stellar mass forms a Boxy/Peanut (B/P) spatial morphology (this geometry is also sometimes called an ``X-shape'' in the literature). Using $N$-body simulations of disk galaxy formation~\cite{Bland-Hawthorn2016}, it has been found that the bulge hosts a rapid initial starburst component, as well as a second component which forms after disk build-up through dynamical instabilities~\cite{Samland:2003}. The former is thought to be associated with a classical bulge structure\footnote{Classical bulges are known to have roughly spherically symmetric morphologies.} while the latter with the B/P structure. Studies~\cite{Shenetal:2010} of the internal kinematics of the stars of the Galactic bulge have determined that these are consistent with at least 90\% of the bulge mass being contained in the B/P structure. Interestingly, the detection of a classical bulge component in the Milky Way has been claimed in~\cite{Kunderetal:2016}. However, as noted above, this is expected to harbor a sub-dominant fraction of the Galactic bulge stellar mass.  

Further evidence for an X-shaped or B/P bulge morphology was deduced from the bifurcation in red clump giant counts in the 2MASS and OGLE-III surveys~\cite{Skrutskie:2006,McWilliam:2010,Nataf:2010}. The analysis in Ref.~\cite{Ness:2016aaa}  found, by constructing an image of the Milky Way bulge from an independent co-adding of publicly available WISE data, that the Milky Way bulge shows a distinct X-shaped structure in the projected stellar distribution that is different from previous works in that it has a more pronounced pinch-in effect at longitude ($l\sim 0^{\circ}$).

It has been posited in previous studies (e.g.~\cite{Abazajian:2010zy}) that the GCE could be due to a population of unresolved sources such as MSPs, which are known to have GeV-peaked gamma-ray spectra. Since MSPs can be generated from old stellar populations, here we test several different stellar templates for the Galactic bulge in the \textit{Fermi} data.

\subsubsection{Boxy/Peanut bulge}

Reference~\cite{Freudenreich:1997bx} derived a parametric model for the spatial morphology of the Galactic bulge by fitting to COBE/DIRBE near-infrared ($1.25$--$4.9$~$\mu$m) data. The best-fitting model obtained in Ref.~\cite{Freudenreich:1997bx} was \textit{Model S}, which is a $\sech$-squared function on the bar radial spatial profile. The density distribution of the boxy bulge for \textit{Model S} is written as\footnote{Note that there was a typo in the argument of the $\exp$ function in Eq.~(14) of Ref.~\cite{Freudenreich:1997bx} which has been corrected in our Eq.~(\ref{Eq:F98bardensity}). We have confirmed this in private communication with H. Freudenreich.}

\begin{equation}
\begin{aligned}
  \rho_{\rm bar} (R,\phi,z) &\propto  \sech^2 (R_s) \\
  &\quad\times
  \begin{cases}
    1, \,\,\,\mbox{$R \leq R_{\rm end}$}\\
    e^{-[(R - R_{\rm end})/h_{\rm end}]^2}, \,\,\,\mbox{$R > R_{\rm end}$},
  \end{cases}
  \end{aligned}
  \label{Eq:F98bardensity}
\end{equation}

\noindent
where $R_s$ is the effective radius:

\begin{eqnarray}
R_\perp ^{C_\perp} & = & \left(|X^\prime|/a_x\right)^{C_\perp} + \left(|Y^\prime|/a_y\right)^{C_\perp},\\
  R_s^{C_{||}} & = & R_\perp ^{C_{||}} + \left( |Z^\prime|/a_z \right)^{C_{||}},\label{Eq:F98barradius}
\end{eqnarray}

\noindent
and ($X^\prime$, $Y^\prime$, $Z^\prime$) is the Boxy bulge frame of reference. The bulge axis scale-lenghts are given by $a_x=1.696$ kpc, $a_y=0.6426$ kpc and $a_z=0.4425$ kpc, the bulge cut-off radius is $R_{\rm end}=3.128$ kpc and the scale height $h_{\rm end}=0.461$ kpc. In addition, the bulge face-on and edge-on shape parameters are
$C_{||}=3.501$ and $C_\perp=1.574$ respectively.

In order to transform from the Galactocentric frame\footnote{The Galactocentric $y$ axis is assumed to be positive in the Sun-Galactic Center direction and the $xy$-plane is parallel to the plane of the Galaxy.}  $(x,y,z)$, to the bulge frame ($X^\prime$, $Y^\prime$, $Z^\prime$), we need to perform two successive passive rotations of coordinates: the first is a clockwise rotation $\theta_0=13.79^{\circ}$ around the Galactocentric $z$ axis, and the second is a clockwise rotation of $\beta_{\rm bulge}=0.023^{\circ}$ around the new axis $y^{\prime}$. The Sun is assumed to be at $z_0=16.46$ pc and $R_0=8.5$ kpc in the Galactocentric frame. Lastly, we performed a line-of-sight integration of the density function in Eq.~(\ref{Eq:F98bardensity})
and normalized the map appropriately for analyses within the \textit{Fermi} Science Tools.

\subsubsection{X-shaped bulge}

We reproduced the method described in Ref.~\cite{Ness:2016aaa} and applied it to the raw infrared WISE image of the Milky Way made available in Ref.~\cite{Ness:2016aaa}.
We employed the 3.4 (W1) and 4.6 (W2) micron WISE data and masked out pixels with negative flux, as well as the top and bottom 5\% of pixels based on W1$-$W2 color.
We fitted an exponential disk model to the W1 and W2 bands and then subtracted the best fit models from the W1 and W2 data, respectively. Namely, we assumed
\begin{multline}
\mbox{bulge}\propto  \exp
\left(
	-
	\left\{
		\left(
        	\frac{  (b-b_0)\cos (\theta )+(l-l_0)\sin (\theta ) }{\beta}
        \right)^2 
    \right. 
\right.
    \\
\left.
    \left.
    \vphantom{\frac{  (b-b_0)\cos (\theta )+(l-l_0)\sin (\theta ) }{\beta}}
        +[(l-l_0)\cos (\theta ) -(b-b_0)\sin (\theta ) ]^2
	\right\}^{1/2}
    /\alpha
\right),
\end{multline}
where $\alpha=1.2$~{ kpc}, $\beta=0.35$, $l_0=-0.45^\circ$,  $b_0=0.11^\circ$, and $\theta=0.06^\circ$.
We then applied a median filter of radius $1.7^\circ$ to the masked versions of both residual templates. In order to include this template in our gamma-ray pipeline, all pixels values with negative fluxes were set to zero in each median filtered exponential subtracted map. Finally, we computed the average of the two resultant maps and normalized it to flux unity for analyses with the \textit{Fermi} Science Tools.

\subsubsection{Nuclear Bulge}

Another stellar structure that is evident from infrared observations corresponds to the so called Nuclear Bulge (NB); this is thought to consist of a spherical Nuclear Stellar Cluster centered at the position of the supermassive black hole and a Nuclear Stellar Disk with radius $\sim 230$ pc and scale height $\sim 45$ pc. The gamma-ray emission from this stellar component was also studied in Ref.~\cite{Macias:2016nev}.  This template  was constructed from a near-infrared stellar density measurement~\cite{Nishiyama2015} of the central region of our Galaxy ($|l|\geq 3^\circ$ and $|b|\geq 1^\circ$). More details about this template can be found in Ref.~\cite{Macias:2016nev}. 

\begin{figure}[t!]
\centering
\includegraphics[scale=0.2]{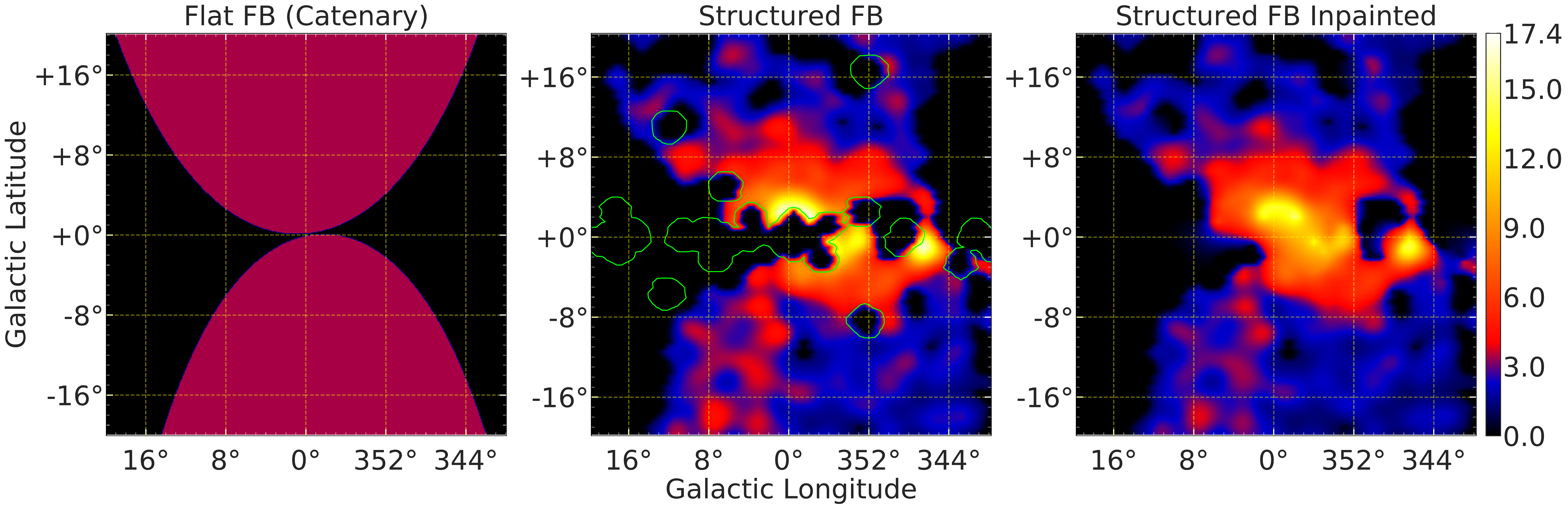}
\caption{\label{Fig:Fermibubbles} \textbf{\textit{Fermi} bubbles templates considered in this analysis.} \textit{Left panel:} Flat \textit{Fermi} bubbles map obtained in Ref.~\cite{Casandjian:andFermiLat2016}. This map consists of two catenary curves describing the edges of the bubbles. This was the \textit{Fermi} bubbles model considered in our previous GCE analysis~\cite{Macias:2016nev}. \textit{Center panel:} Structured \textit{Fermi} bubbles template recovered with a spectral component analysis in Ref.~\cite{TheFermi-LAT:2017vmf}. The green contours represent the point source mask applied in that work. \textit{Right panel:} Structured \textit{Fermi} bubbles template after application of a Laplace inpainting method \cite{Laplace} to correct for artifacts of the point source mask. The maps are normalized to unitary flux in the RoI and the colorbar shows the full range of variations in arbitrary units for the three panels.  }\label{fig:Fermibubbles}
\end{figure}

\begin{table*}[!b]\caption{{\bf Summary of the likelihood analyses for  \textit{Fermi} bubbles maps.\label{Tab:likelihoods}}}
\begin{adjustbox}{width=1.0\textwidth, center}
\centering\begin{threeparttable}
 \scriptsize 
\begin{tabular}{llllccc}
\hline\hline

Base  & Source &  $\log(\mathcal{L}_{\rm Base})$  & $\log(\mathcal{L}_{{\rm Base}+{\rm  Source}})$  &  $\mbox{TS}_{\rm Source}$& Number of & Reference\\ 
           &              &                                  &                         &                                                & source parameters & for FB template\\\hline
baseline  & Catenary& -2486188.1         & -2486753.1          & 1130      & 15 & \cite{Casandjian:andFermiLat2016}\\
baseline  & Structured FB& -2486188.1           &-2487322.3      & 2268    & 15 &\cite{TheFermi-LAT:2017vmf}\\ 
baseline  & Structured FB (Inpainted) & -2486188.1            & -2487802.1         & 3228  &  15& adapted from~\cite{TheFermi-LAT:2017vmf}\\ 
\hline\hline
\end{tabular}
\begin{tablenotes}
      \item  The \textit{baseline} model includes the 2FIG~\cite{Fermi-LAT:2FIG} point sources, Loop I, a standard 2D IC map predicted by {\tt GALPROP}, hydrodynamic interstellar gas and dust maps divided in several rings, an isotropic component, and a template for the Sun and the Moon that matches the filters and cuts applied to the data (see also Table~\ref{Tab:definitions} for details). Three different models for the \textit{Fermi} bubbles were considered. See Fig.~\ref{Fig:Fermibubbles} and Sec.~\ref{subsec:newtemplates} for details.
      The statistical significance of a new source is given by 
       TS$_{\rm Source}\equiv 2(\log(\mathcal{L}_{{\rm Base}+{\rm  Source}}) - \log(\mathcal{L}_{\rm Base}))$, where maximum likelihood ($\mathcal{L}$) values are computed independently for the Base and Base$+$Source models. Note that for both cases all parameters are maximized and so the $\mathcal{L}_{{\rm Base}+{\rm  Source}}$ will have additional parameters whose number is given in the second to last column of the table. 
       The conversion between TS$_{\mbox{Source} }$ and $\sigma$ is discussed in Ref.~\cite{Macias:2016nev}. 
    \end{tablenotes}
\end{threeparttable}
\end{adjustbox}
\end{table*}

\subsection{Spherically symmetric template}\label{subsec:DM}

In addition to the stellar bulge templates, we also consider a map describing the potential gamma-ray emission from dark matter annihilation or a spherically symmetric distribution of MSPs. We model this by the square of a NFW density profile with an inner slope of 1.2, which has been shown to describe the GCE well in previous works (see Ref.~\cite{Gordon:2013vta} for details on the precise parameters choices used in this work).

\subsection{\textit{Fermi} Bubbles}

Reference~\cite{Casandjian:andFermiLat2016} carried out a reanalysis of the low latitude counterpart of the \textit{Fermi} bubbles (FB). The residual map found in that study has a spatial morphology whose boundaries are well described by two catenary curves of the form $10.5^\circ \times (\cosh((l - 1^\circ)/10.5^\circ) - 1^\circ)$ and $8.7^\circ \times(\cosh((l+1.7^\circ)/8.7^\circ) - 1^\circ)$ for the Northern and the Southern bubbles, respectively. This flat FB template (hereafter simply called ``Catenary'') is displayed in the left panel of Fig.~\ref{fig:Fermibubbles}.  

Recently, Ref.~\cite{TheFermi-LAT:2017vmf} derived an all-sky template for the \textit{Fermi} bubbles using a spectral component analysis technique~\cite{Malyshev:2012mb}. This method assumes that the FB spectrum at high latitudes is the same as the spectrum at low latitudes in the energy range  $[1,10]$ GeV. We show their ``Structured FB''~\footnote{The Strucutured FB map used in this work corresponds to Fig.12 of Ref.~\cite{TheFermi-LAT:2017vmf}. The authors of that reference constructed the map with their 3 component spectral decomposition technique.} template for our RoI in the central panel of Fig.~\ref{fig:Fermibubbles}. it is evident that this FB template contains missing data (see green contours in Fig.~\ref{Fig:Fermibubbles}) due to the need of a point source mask template.

In order to ameliorate the artifacts present in the Structured FB map, we have applied an inpainting  method to the FB image. The inpainting algorithm utilizes Laplace Interpolation~\cite{Laplace} which is a specialized interpolation technique for restoring missing data on a grid. Specifically, this algorithm solves the Poisson equation in the masked pixels of an image using as boundary conditions the pixel values in the unmasked portion of the image. The inpainted FB template is shown in the right panel of Fig.~\ref{Fig:Fermibubbles}.

\subsection{Inverse Compton Models}\label{subsec:ICS}

\begin{figure}[t!]
\centering
\includegraphics[scale=0.4]{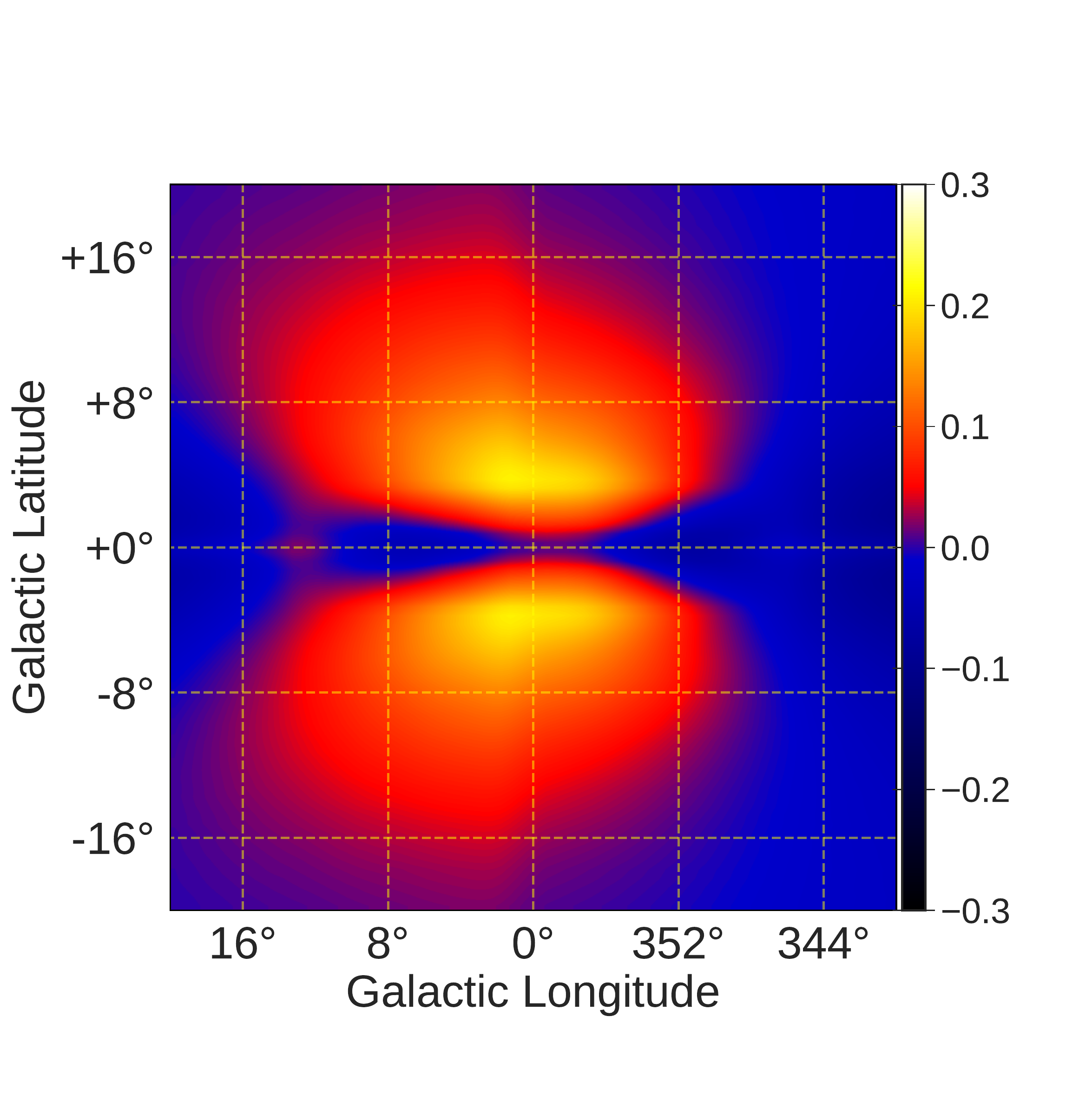} \caption{\label{Fig:ICS} \textbf{Fractional residual difference map for the 3D IC Model ``F98-SA0'' with respect to the Standard 2D IC Model ``Std-SA0'' in the inner $40^\circ\times 40^\circ$ of the Galactic Center $\left(\frac{(\mbox{F98-SA0})-(\mbox{Std-SA0})}{\mbox{Std-SA0}}\right)$.} The map is displayed  at an energy of $1.2$ GeV with a pixel size of $0.5^\circ$. The IC maps were computed with the most recent release of {\tt GALPROP} (v56)~\cite{Porter:2017vaa} and the new 3D ISRF data. No smoothing has been applied to the image. These maps reproduce the ones presented in Ref.~\cite{Porter:2017vaa}. See Sec.~\ref{subsec:ICS} for a description of the 3D IC models.}\label{fig:3Dinversecompton}
\end{figure}

Reference~\cite{Macias:2016nev} investigated the GCE systematics associated to the IC component with the use of the 2D Galactocentric cylindrically symmetric interstellar radiation field (ISRF) model attainable in {\tt GALPROP} v54~\cite{Galprop}. The authors of Ref.~\cite{ackermannajelloatwood2012} compared 128 different {\tt GALPROP} models with all sky \textit{Fermi}-LAT data finding 
a range of possible propagation parameter setups that are consistent with local CRs and gamma-ray measurements. Reference~\cite{Macias:2016nev} reconstructed a subset of the IC maps in Ref.~\cite{ackermannajelloatwood2012} that were found to have the largest morphological and spectral variations in order to estimate the impact of the IC component in the best-fit GCE emission. 

In this article we improve upon the pipeline of Ref.~\cite{Macias:2016nev} by using the two recently developed 3D ISRF models made publicly available in the most recent release of {\tt GALPROP} v56~\cite{Porter:2017vaa}. These are based in two different Galaxy wide dust and stellar distribution models labeled F98 and R12: the F98 model is based on the work by Freudenreich (1998)~\cite{Freudenreich:1997bx} and its bulge component is described in more detail in Sec.~\ref{subsec:newtemplates}, while the R12 is based on the study by Robitaille et al. (2012)~\cite{R12}. The corresponding spectral intensities for the ISRF were calculated by Ref.~\cite{Porter:2017vaa} with the Numerical Kalculation for Interstellar Emission (FRaNKIE) code. Although both models assume different stellar luminosities and dust densities, their predicted local intensities are consistent with near-infrared to far-infrared observations.

The IC maps investigated in \cite{Porter:2017vaa} utilized three different CR propagation setups (see Table 3 of that publication). These were labeled SA0, SA50 and SA100 according to the proportion of  CR luminosity injected by the spiral arms. We have reproduced the main IC skymaps presented in that study and included them in our step-wise statistical procedure described in the following section. In particular, for each Galaxy model (F98 and R12) we have utilized the three proposed CRs propagation setups. These are named: F98-SA0, F98-SA50, F98-SA100, R12-SA0, R12-SA50 and R12-SA100. Figure~\ref{fig:3Dinversecompton} displays the fractional residual differences between a standard 2D IC map extracted from Ref.~\cite{ackermannajelloatwood2012} (Std-SA0) and the 3D model F98-SA0.   

\subsection{Analysis pipeline}\label{subsec:analyspipelines}
In Ref.~\cite{Macias:2016nev} some of us showed that the GCE spatial morphology was better explained by the stellar nuclear bulge and Boxy/Peanut (or X-shaped) bulge templates than by a spherically symmetric excess map given by an NFW$^2$ profile.  This was done by performing a set of maximum-likelihood fits summarized in Table~I of that work~\cite{Macias:2016nev}. In the present analysis we use a similar step-wise statistical procedure to establish whether a certain extended template for the GCE is statistically preferred over another.

The analysis pipeline used here to study the \textit{Fermi}-LAT data utilizes the inner $40^{\circ}\times 40^{\circ}$ of the Galactic Center. Similar to Ref.~\cite{Macias:2016nev}, we employed a bin-by-bin analysis method in which the full \textit{Fermi}-LAT data were divided into 15 logarithmically spaced energy bins where, for each energy bin, we performed a separate maximum-likelihood fit with \textit{pyLikelihood} within the \textit{Fermi} Science Tools. In each of these maximum-likelihood runs we simultaneously fitted the normalization of all included diffuse emission components while keeping their spectral slope as a fixed parameter. Due to the small size of the bins, our results were not sensitive to the assumed spectral shapes of the extended and point-like sources. For simplicity we therefore used power-law spectra with fixed index of negative two. 

The point source modeling was done with the 2FIG catalog~\cite{Fermi-LAT:2FIG}. We considered only the point sources that were confidently found with the two different interstellar emission models considered in that work. There is a total of 288 such point sources inside our RoI. Due to limitations in the maximum number of parameters that can be reliably fitted in a given run with \textit{Fermi} Science Tools, we have opted to follow a hybrid fitting approach analogous to that employed in Ref.~\cite{TheFermi-LAT:2017vmf}. Namely, we independently varied the normalizations of the 100 brightest 2FIG point sources in the RoI, while for the remaining point sources we constructed a single point source template in each energy bin. This is a reasonable approximation given that the amount of data utilized in the present study is very similar to the one used by the authors of the 2FIG catalog. To construct the combined point source template, we used the best-fit spectra in the 2FIG catalog and convolved it with the \textit{Fermi} point spread function at each energy bin with the \textit{gtmodel} tool. The resulting point source template was appropriately normalized and included in the fit along with the other extended and point-like sources. Other extended sources (W28, W30, RXJ1713, HESSJ1825-137, HESSJ1632-478 and HESSJ1837-069) that exist in our RoI were taken from the 3FGL catalog~\cite{3FGL} and varied independently in the fits.

The significance of each source was evaluated using the test statistic $\mbox{TS} = 2(\ln \mathcal{L}_1 -\ln \mathcal{L}_0)$, where $\mathcal{L}_0$ is the likelihoods of the background (null hypothesis) and $\mathcal{L}_1$ is the likelihood of the source being tested plus background. The total TS of a source is given by $\mbox{TS}=\sum_{i=1}^n \mbox{TS}_i$ where $\mbox{TS}_i$ is the $\mbox{TS}$ for bin $i$ and there are $n=15$ bins. Since a source has now 15 degrees of freedom (non-negative normalization in each energy bin) instead of two in a standard broad-band analysis (non-negative normalization and index), the detection threshold condition has to be adjusted accordingly. As discussed in Ref.~\cite{Macias:2016nev}, this can be done by utilizing the formula given in case 9 of \cite{SelfLiang87}:

\begin{equation}
p({\rm TS})= 2^{-n}\left(\delta({\rm TS}) +\sum_{i=1}^{n} \binom{n}{i} \chi_{i}^2({\rm TS})\right)
\label{eq:pTS}
\end{equation}
where  $\delta$ is the Dirac delta function, $\binom{n}{i}$ is a binomial coefficient with $n=15$, and $\chi_{i}^2$ is a $\chi^2$ distribution with $i$ degrees of freedom. It follows  (see Ref.~\cite{Macias:2016nev}) that a threshold of $\mbox{TS} \ge 34.8$ corresponds to a 4$\sigma$ detection of a new extended source.

\begin{figure}[t!]
\centering
\includegraphics[scale=0.5]{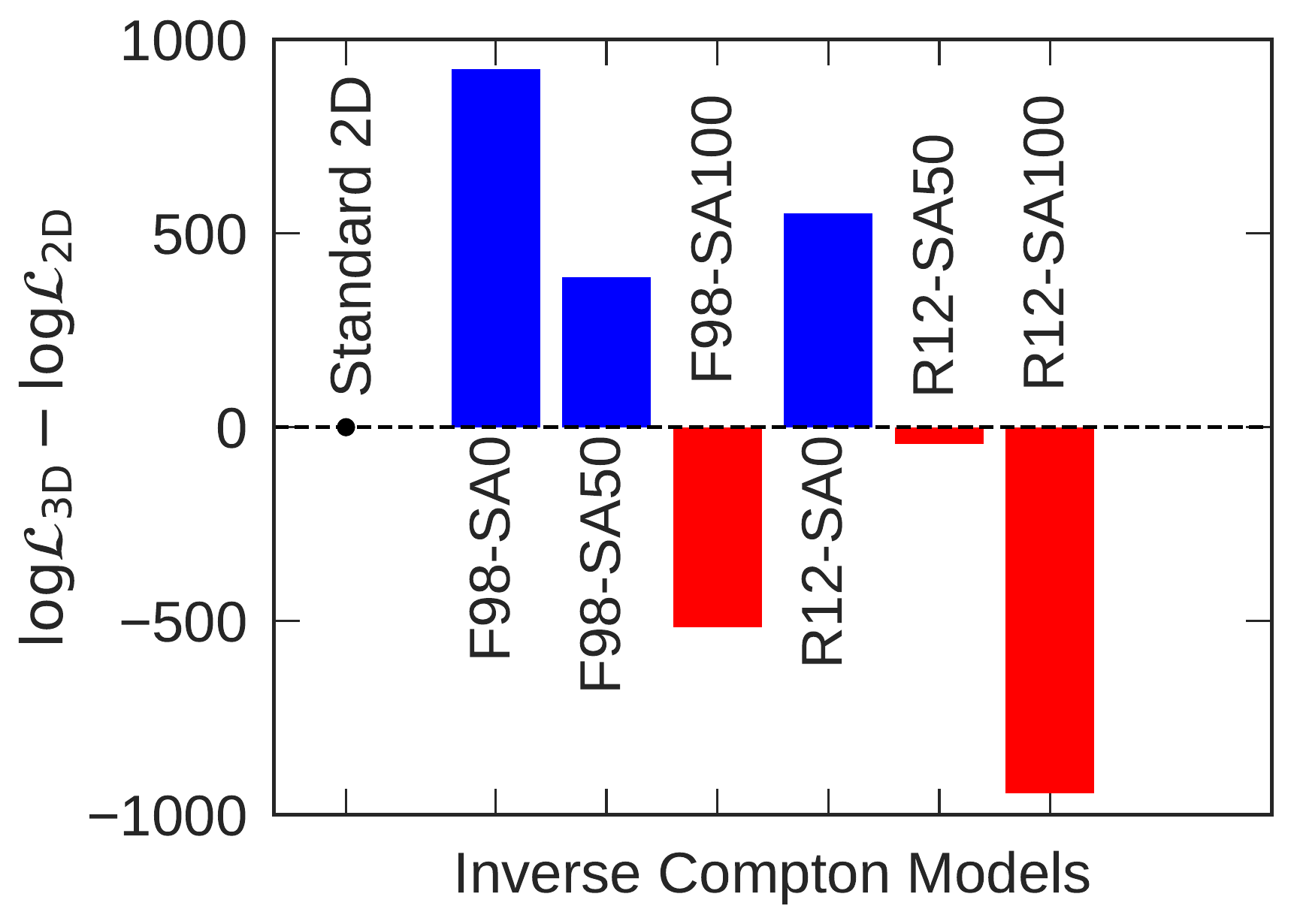} \caption{\label{Fig:bardiagram} \textbf{Comparison of the log-likelihood values obtained for our six 3D-IC-Models vs a Standard 2D-IC Model.} Blue (red) bars indicate which of the 3D-IC-Models considered improve (worsen) the fit to the inner $40^\circ \times 40^\circ$ region of the Galactic Center compared to a Standard 2D-IC Model (Std-SA0). The fits to the data include: the gas maps, 2FIG point source catalog, Isotropic, Loop I, Sun \& Moon and inpainted \textit{Fermi} bubbles template plus an alternative IC map at each different run. Notice that the different IC maps were not nested in the fits. See text for a description of the IC models evaluated in this work.   }\label{Fig:ICSDeltaLoglikes}
\end{figure}

\begin{figure}[t!]
\centering
\begin{tabular}{cc}
\includegraphics[scale=0.45]{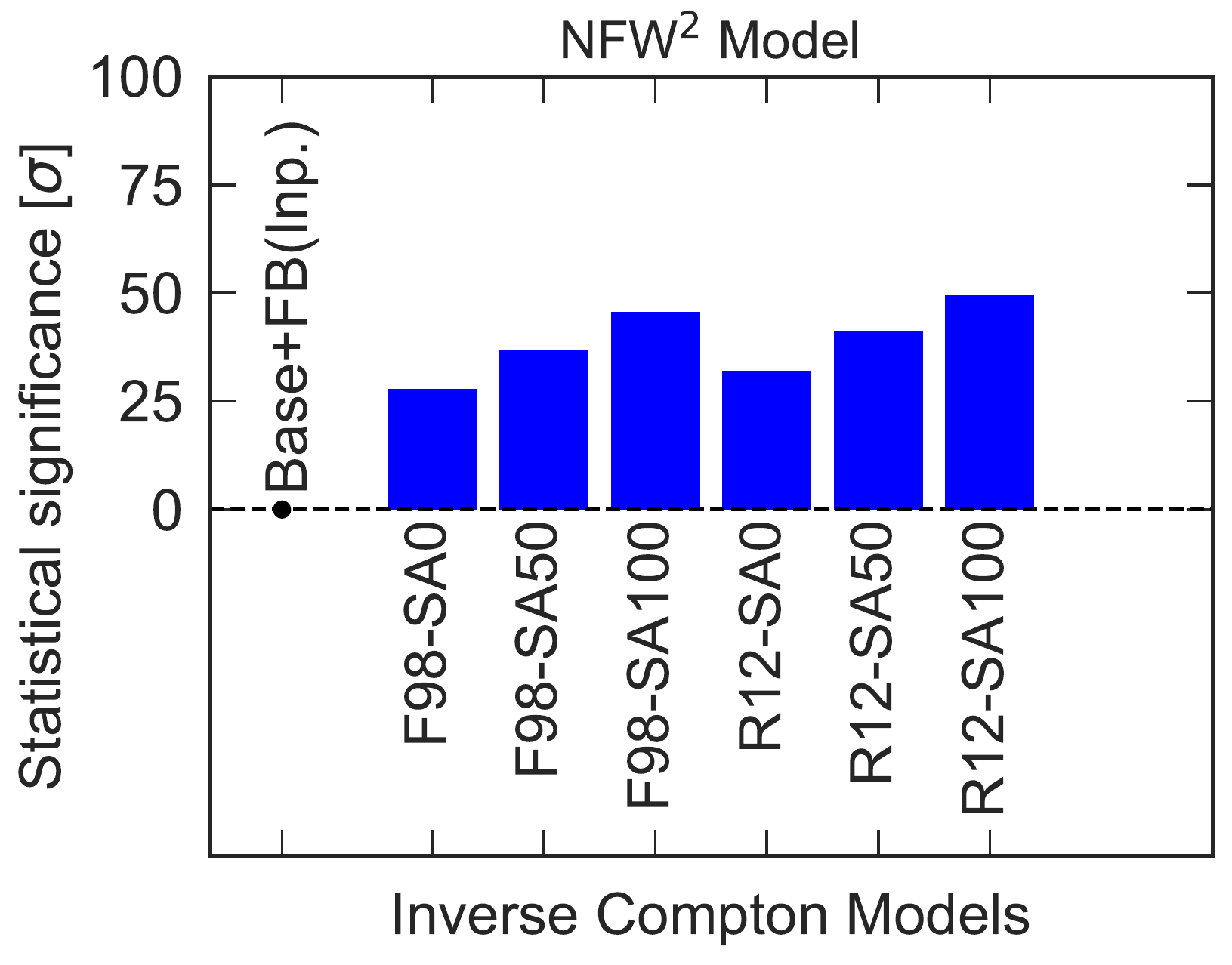} & \includegraphics[scale=0.45]{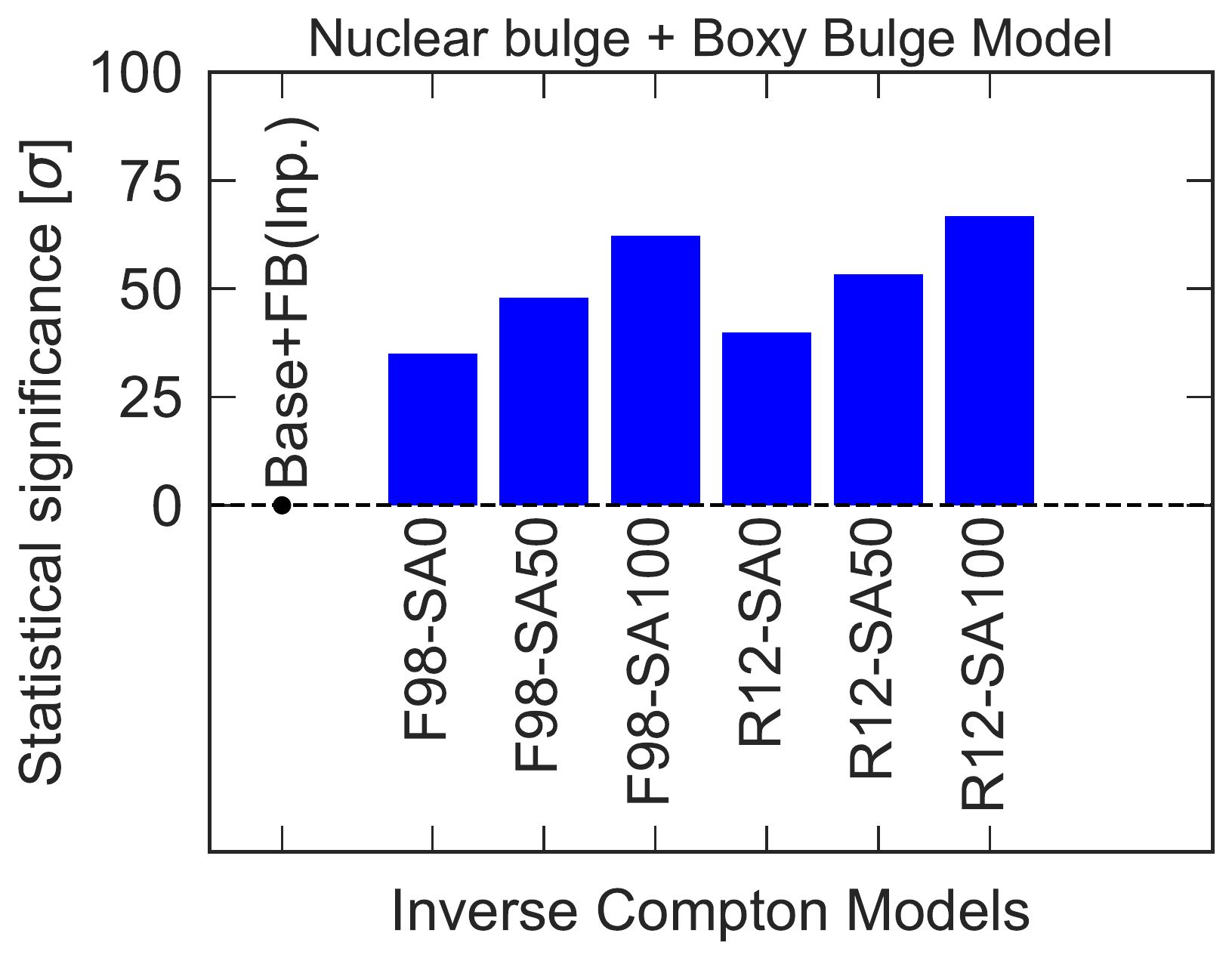} 
\end{tabular}
\caption{ \textbf{Statistical significance in $\sigma$ units of the two main Galactic bulge models considered in this work:} Navarro-Frenk-White squared map (NFW$^2$) (left) and nuclear bulge + boxy bulge (right). The reference model used in the computation of the statistical significance is similar to the baseline described in Sec~\ref{subsec:baseline}, except that use of the inpainted structured map (far left black circle). Each bar shows the replacement of the Standard 2D map in the baseline model by a 3D IC map. The computation of the $\sigma$ values is done with a maximum-likelihood routine in which both the reference model (Base + FB(Inp)) and the bulge model (NFW$^2$ or nuclear bulge + boxy bulge) are nested in the fit. More details can be found in Table~\ref{Tab:GCElikelihoods}. }\label{Fig:BulgeLoglikes}
\end{figure}

\begin{figure}[t!]
\centering
\includegraphics[scale=0.5]{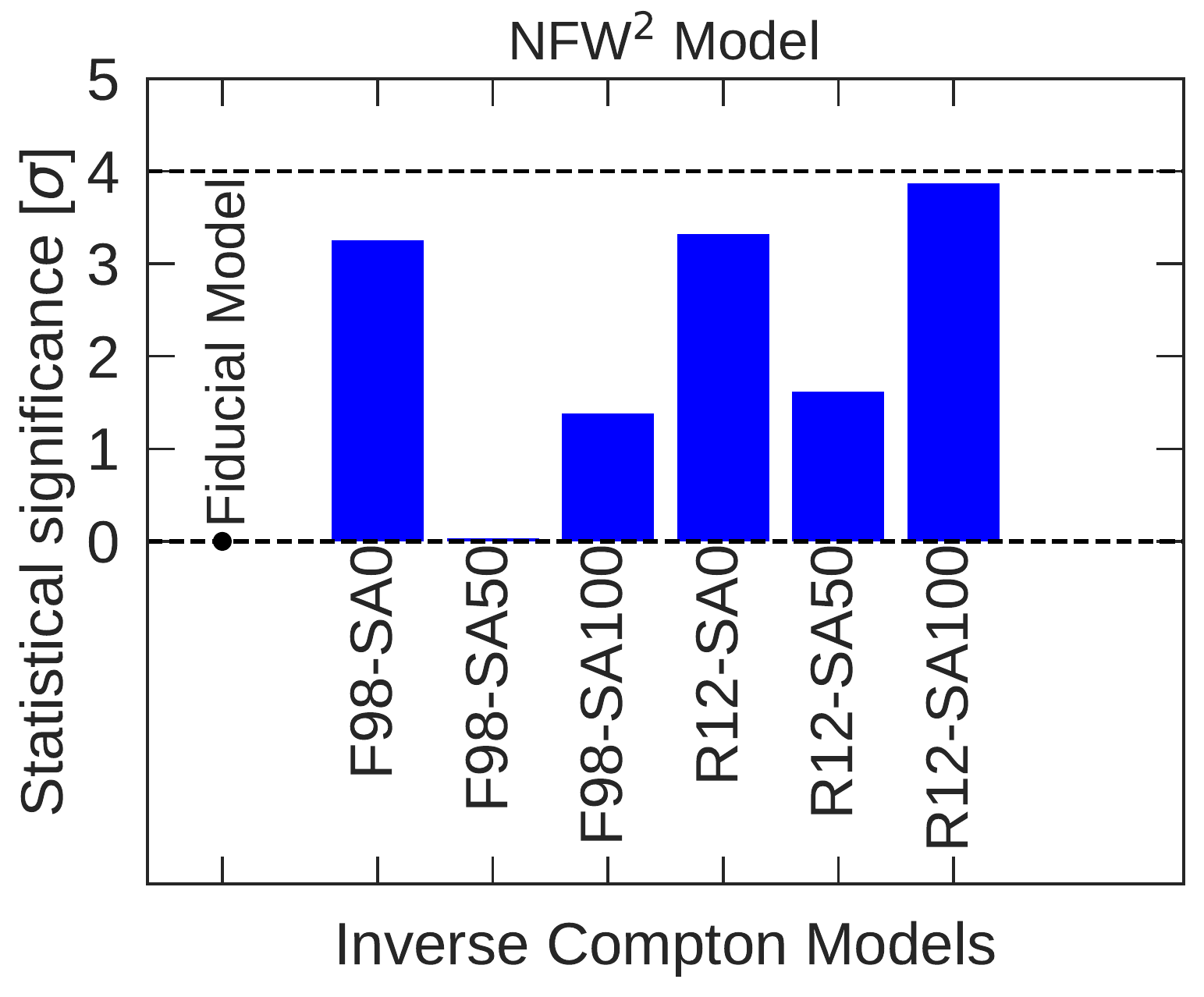} 
\caption{\label{Fig:NFWsignificance} \textbf{Significance of the NFW$^2$ template, in $\sigma$ units, after the Galactic bulge template has been included to the baseline model.} In other words, the fiducial model used in the computation of the NFW$^2$ significances consists of the baseline described in Sec~\ref{subsec:baseline}, the inpainted structured FB map, and nuclear bulge + boxy bulge stellar map. Each bar displays a different 3D IC map as described in Sec.~\ref{subsec:ICS}. The computation of the TS-values is done with a maximum-likelihood routine in which both the fiducial model and the NFW$^2$ are varied in the fit. We note that in none of the cases considered has the NFW$^2$ template  been significantly detected. More details can be seen in Table~\ref{Tab:GCElikelihoods}. }
\end{figure}

\section{Results}\label{sec:tests}

\begin{table*}[!htbp]\caption{{\bf Summary of the main maximum-Likelihood analyses considered in this study.\label{Tab:GCElikelihoods}}}
\begin{adjustbox}{width=1.0\textwidth, center}
\centering
\begin{threeparttable}
\scriptsize
\begin{tabular}{llllcrc}
\hline\hline
Base  & Source &  $\log(\mathcal{L}_{\rm Base})$  & $\log(\mathcal{L}_{{\rm Base}+{\rm  Source}})$  &  $\mbox{TS}_{\rm Source}$& $\sigma$ & Number of\\ 
           &              &                                  &                         &                                               & & source parameters\\\hline

baseline     &SFB (Inp.) & -2486188.1    &-2487802.1 & 3228& & 15 \\ 
baseline$+$SFB (Inp.)     & NFW$^2$ &-2487802.1     & -2488619.8  &1635 & & 15 \\
baseline$+$SFB (Inp.)     & X-bulge$+$NB & -2487802.1     &-2488839.6  &2075  & & $2\times 15$ \\
baseline$+$SFB (Inp.)     & Boxy bulge$+$NB & -2487802.1     & -2489230.1 & 2856 & & $2\times 15$ \\
  baseline$+$SFB (Inp.)$+$Boxy bulge$+$NB     & NFW$^2$ & -2489230.1    & -2489233.9  &8 & 0.9 & 15 \\ \hline
baseline$^1$     &SFB (Inp.) & -2487135.2     &-2488729.2 &3188 & & 15 \\ 
baseline$^1+$SFB (Inp.)     & NFW$^2$ & -2488729.2     & -2489119.4  &780 & & 15 \\
  baseline$^1+$SFB (Inp.)     & X-bulge$+$NB & -2488729.2    &-2489242.2  & 1026  & & $2\times 15$ \\
baseline$^1+$SFB (Inp.)     & Boxy bulge$+$NB & -2488729.2    & -2489341.3 & 1224 & & $2\times 15$ \\
  baseline$^1+$SFB (Inp.)$+$Boxy bulge$+$NB     & NFW$^2$ & -2489341.3    & -2489354.7 &27 & 3.3 & 15 \\ \hline
baseline$^2$     &SFB (Inp.) & -2486191.5    & -2488329.8& 4276& & 15 \\ 
baseline$^2+$SFB (Inp.)     & NFW$^2$ & -2488329.8    & -2489004.2 & 1349 & & 15 \\
  baseline$^2+$SFB (Inp.)     & X-bulge$+$NB & -2488329.8    & -2489120.2  & 1581& & $2\times 15$ \\
baseline$^2+$SFB (Inp.)     & Boxy bulge$+$NB & -2488329.8    & -2489479.9 &2300 & & $2\times 15$ \\
  baseline$^2+$SFB (Inp.)$+$Boxy bulge$+$NB     & NFW$^2$ & -2489479.9    & -2489480.6 &1 & 0.0 & 15 \\ \hline
baseline$^3$     &SFB (Inp.) & -2484807.7    &-2487555.6 & 5496 & & 15 \\ 
baseline$^3+$SFB (Inp.)     & NFW$^2$ & -2487555.6    & -2488596.9 & 2083& & 15 \\
  baseline$^3+$SFB (Inp.)     & X-bulge$+$NB &  -2487555.6   &-2488700.0  &2289 & & $2\times 15$ \\
baseline$^3+$SFB (Inp.)     & Boxy bulge$+$NB &  -2487555.6   & -2489489.4 &3868 & & $2\times 15$ \\
  baseline$^3+$SFB (Inp.)$+$Boxy bulge$+$NB     & NFW$^2$ & -2489489.4  &-2489495.1  &11 &1.3 & 15 \\ \hline
baseline$^4$     &SFB (Inp.) & -2486304.7    & -2488491.0&4373 & & 15 \\ 
baseline$^4+$SFB (Inp.)     & NFW$^2$ &-2488491.0  & -2489006.0 & 1030& & 15 \\
  baseline$^4+$SFB (Inp.)     & X-bulge$+$NB & -2488491.0   & -2489093.6 & 1205& & $2\times 15$ \\
baseline$^4+$SFB (Inp.)     & Boxy bulge$+$NB & -2488491.0   & -2489285.1 &1588 & & $2\times 15$ \\
  baseline$^4+$SFB (Inp.)$+$Boxy bulge$+$NB     & NFW$^2$ &  -2489285.1   & -2489298.7 &27 &3.3 & 15 \\ \hline
baseline$^5$     &SFB (Inp.) & -2485197.9    &-2488049.6 &5703 & & 15 \\ 
baseline$^5+$SFB (Inp.)     & NFW$^2$ &-2488049.6  & -2488901.7 &1704 & & 15 \\
  baseline$^5+$SFB (Inp.)     & X-bulge$+$NB & -2488049.6   &-2488927.8   &1756 & & $2\times 15$ \\
baseline$^5+$SFB (Inp.)     & Boxy bulge$+$NB & -2488049.6   & -2489474.6  & 2850& & $2\times 15$ \\
  baseline$^5+$SFB (Inp.)$+$Boxy bulge$+$NB     & NFW$^2$ & -2489474.6    & -2489481.2 & 13 & 1.6& 15 \\ \hline
baseline$^6$     &SFB (Inp.) & -2483983.0    &-2487246.5 & 6527 & & 15 \\ 
baseline$^6+$SFB (Inp.)     & NFW$^2$ & -2487246.5    & -2488472.5 & 2452& & 15 \\
baseline$^6+$SFB (Inp.)     & X-bulge$+$NB & -2487246.5    &-2488479.3  & 2466 & & $2\times 15$ \\
baseline$^6+$SFB (Inp.)     & Boxy bulge$+$NB & -2487246.5    & -2489479.6 & 4466 & & $2\times 15$ \\
  baseline$^6+$SFB (Inp.)$+$Boxy bulge$+$NB     & NFW$^2$ & -2489479.6    & -2489496.1 &33 & 3.9& 15 \\ \hline

\hline\hline
\end{tabular}
\begin{tablenotes}
\item The \textit{baseline} model in the first row set is described in Sec.~\ref{subsec:baseline} and assumes the Standard 2D IC map. In the next row sets we change the 2D IC component by alternative 3D IC maps. This is denoted by \textit{baseline$^i$} where $i=1,2,...,6$ with $i$ running through models \{F98-SA0, F98-SA50, F98-SA100, R12-SA0, R12-SA50, R12-SA100\} respectively. The Fermi bubbles model assumed corresponds to the Structured FB inpainted [SFB (Inp.)], see right panel of Fig.~\ref{Fig:Fermibubbles} and Sec.~\ref{subsec:newtemplates} for details. Other bulge templates are: the Nuclear Bulge (NB)~\cite{Nishiyama2015}, boxy bulge by Freudenreich (1998)~\cite{Freudenreich:1997bx}, Ness\&Lang X-bulge~\cite{Ness:2016aaa} and a Navarro-Frenk-White profile with $\gamma=1.2$ (NFW$^2$). See Sec.~\ref{subsec:newtemplates} for descriptions. The maximized likelihoods ($\mathcal{L}$) are given for the Base and Base$+$Source models and the significance of the new source is given by 
       TS$_{\rm Source}\equiv 2(\log(\mathcal{L}_{{\rm Base}+{\rm  Source}}) - \log(\mathcal{L}_{\rm Base}))$. See also Table~\ref{Tab:likelihoods} for definitions. 
    \end{tablenotes}
\end{threeparttable}
\end{adjustbox}
\end{table*}

\begin{figure}[t!]
\centering
\includegraphics[scale=0.15]{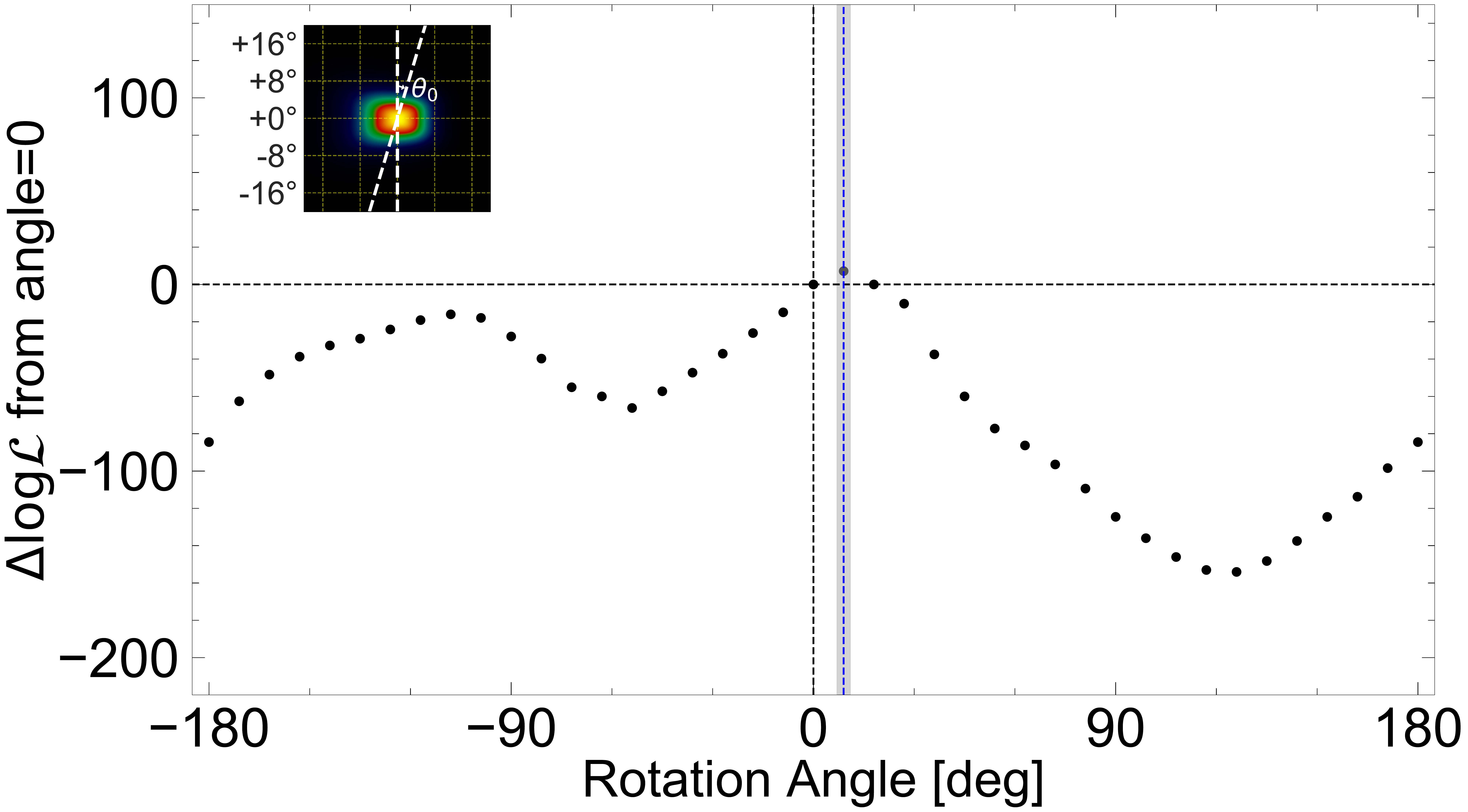} 
\caption{\label{Fig:Rotation} \textbf{Results of the rotation analysis.}  $\Delta \log (\mathcal{L})$ between the non-rotated boxy bulge and the boxy bulge template rotated with respect to the $l=0^{\circ}$ line and a best fit of $9^{\circ}\pm 2^{\circ}$ ($1 \sigma$) was found. This has a $3.8\sigma$ significance ($\Delta \log(\mathcal{L})=7.2$ for one extra parameter). This indicates that there is no significant evidence for rotating the infrared boxy bulge morphology. The gamma-ray background and foreground model used to evaluate the maximized log-likelihood values of rotated boxy bulge templates consists of the \textit{baseline$^1$} (for which the IC map corresponds to F98-SA0) described in Table~\ref{Tab:GCElikelihoods}, the inpainted structured FB and nuclear bulge map.}
\end{figure}

Using the initial baseline model described in Sec.~\ref{subsec:baseline}, the first step of our procedure is to find the best-fit normalization parameters for all free extended and point-like sources\footnote{The fitting of the 2FIG point sources in carried out with a hybrid approach described in Sec.~\ref{subsec:analyspipelines}.} in the RoI. The results are shown in Table~\ref{Tab:likelihoods}. In this first iteration we seek to find out which of the three alternative \textit{Fermi} bubbles maps (see Fig.~\ref{Fig:Fermibubbles}) is preferred by the data. Three fits were done, each time adding a different FB template: Catenary~\cite{Casandjian:andFermiLat2016}, Structured FB~\cite{TheFermi-LAT:2017vmf} and Structured FB corrected with the inpainting method. As summarized in this table, the Structured FB (Inpainted) map was the one found to improve the fit the most. Therefore, for the next iterations we add this FB template and re-optimize the baseline model.

In the next stage of our procedure we perform six fits in which we replace the standard 2D IC map by one of the new 3D IC templates: F98-SA0, F98-SA50, F98-SA100, R12-SA0, R12-SA50 or R12-SA100 (which were recently proposed in Ref.~\cite{Porter:2017vaa}; see Sec.~\ref{subsec:ICS}). A comparison of the log-likelihood values obtained for each 3D IC template versus the standard 2D IC is displayed in Fig.~\ref{Fig:bardiagram} (larger is better). Interestingly, gamma-ray data from the inner  $40^\circ \times 40^\circ$ prefer some of the 3D IC models to the Standard 2D one, with F98-SA0 being particularly favored. However, we note that although the new 3D IC maps are more physically motivated and much more realistic than the Standard 2D IC maps, the assumed CR propagation models and CR luminosities have not been tuned to match the gamma-ray data. Rather, they were obtained from a maximum-likelihood fit to local CR data~\cite{Porter:2017vaa}. So, in the remainder of this study we use the 3D IC maps in two different ways: first, we evaluate the impact of the new 3D IC maps on the results obtained in Ref.~\cite{Macias:2016nev}. And second, we replace the Standard 2D IC model by F98-SA0 in our baseline when performing empirical tests of the bulge-correlated gamma rays (described following sections).

Table~\ref{Tab:GCElikelihoods} shows the log-likelihood values obtained by adding to the fits the new extended templates (bulge models and preferred \textit{Fermi} bubbles map). There are seven row sets in which for each different set an alternative 3D IC map is included in the baseline instead of the Standard 2D IC map: the first row set considers the Standard 2D IC model, the second F98-SA0, the third  F98-SA50, and so forth, until the last row set that uses R12-SA100. Regardless of the IC map assumed, we find that a bulge model is still needed by the data. As can be seen, this holds even after inclusion of the inpainted structured FB map. Figure~\ref{Fig:BulgeLoglikes} shows a comparison of the TS values obtained for the boxy bulge + NB and NFW$^2$ models respectively. 

In each row set shown in Table~\ref{Tab:GCElikelihoods}, the methodology is to keep the component with highest log-likelihood value and move to the next iteration. In all the cases considered in this study we find that the stellar models have larger TS values than the NFW$^2$ map. Remarkably, when the boxy bulge + NB is included in the base model, there is no statistically significant need for an NFW$^2$ template (the significance of the NFW$^2$ map is always $\lesssim 4\sigma$). The stellar templates evidently absorb much of the residual photons and the data no longer need a map describing a dark matter distribution. Finally, even when the NFW$^2$ template is added to the base model, we find very strong evidence for the nuclear bulge + boxy bulge template. We therefore conclude that the data most prefer the nuclear bulge + boxy bulge. The low statistical significances obtained for the NFW$^2$ model when different IC maps were assumed can be seen in Fig.~\ref{Fig:NFWsignificance}. We note that this is consistent with our previous work~\cite{Macias:2016nev} where a smaller RoI and only the standard 2D IC maps were used.

\subsection{Rotation}\label{sec:tests:rotation}

Given the statistical preference for the stellar bulge templates under consideration, we continue to investigate whether the likelihoods can be improved by allowing more flexibility to the templates. In particular, we consider  rotations and  translations of the boxy bulge map. We even allow changes that are inconsistent with stellar data, which serve to test the interpretation of the gamma-ray correlations. Although the nuclear bulge is also robustly detected, we only perform these tests on the boxy bulge as this template is crucial to explain the GCE at higher latitudes. 
In order to test the rotational symmetry of the boxy bulge morphology we rotate the template about a line defined by $l=0^\circ$ (see inset image of Fig.~\ref{Fig:Rotation}) in increments of $9^{\circ}$ and compute the maximum log-likelihood at each turn. The results from this test are shown in Fig.~\ref{Fig:Rotation} and indicate that the template in its original orientation is preferred. Note that the boxy bulge is East-West asymmetric (this is elegantly illustrated in Fig.~9 of Ref.~\cite{Bland-Hawthorn2016}), and thus the $\Delta \log(\mathcal{L})$ profile looks different for negative and positive rotations. This indicates that the data has constraining power to statistically distinguish the East-West asymmetries present in the boxy bulge template. The maximum likelihood rotation was found to be $\sim 9^\circ \pm 2^\circ$ with significance $3.8\sigma$ ($\Delta \log(\mathcal{L})=7.2$ for one extra parameter). Therefore the data is consistent with the original orientation of the Boxy bulge image.

\begin{figure}[t!]
\centering
\includegraphics[scale=0.15]{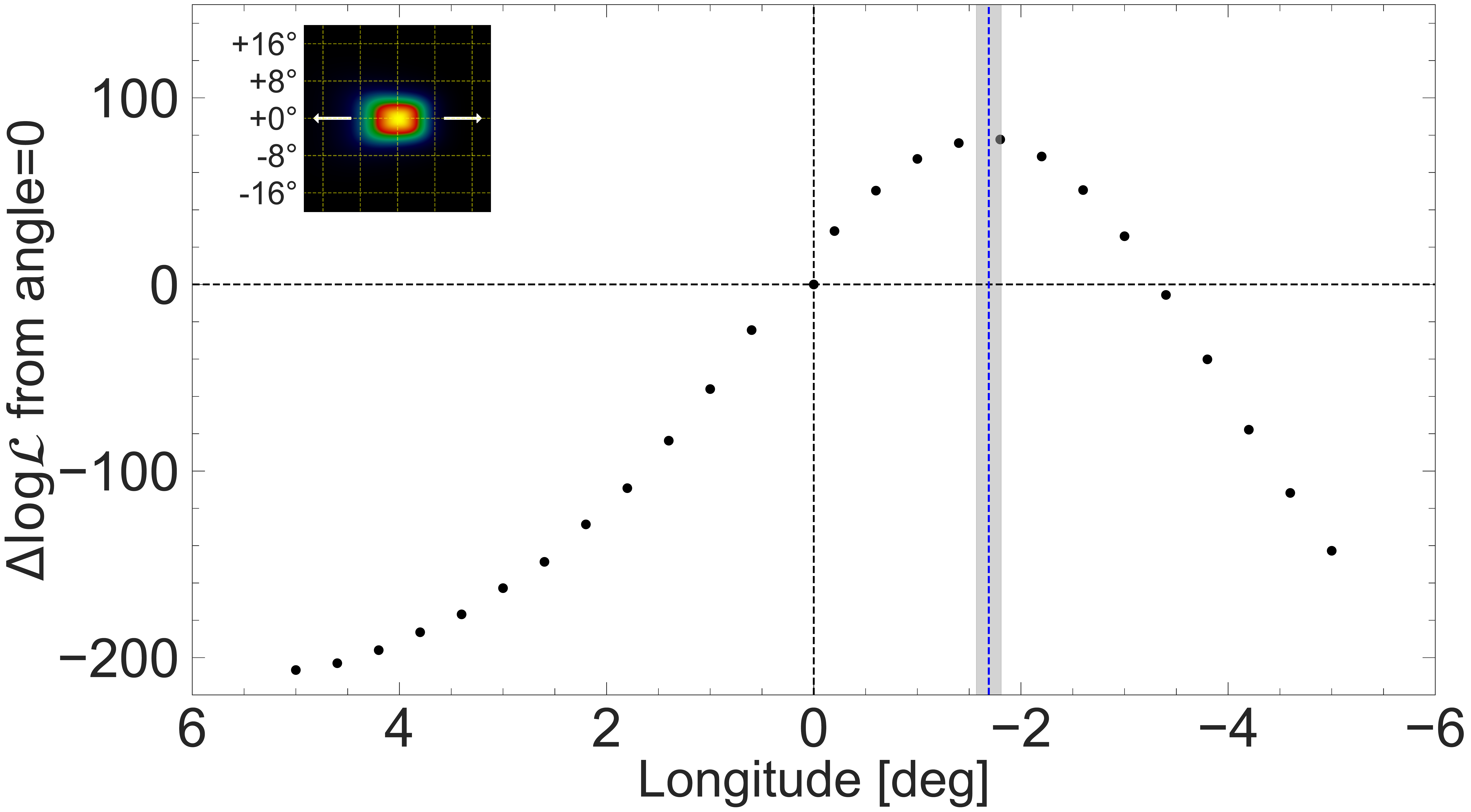} 
\caption{\label{Fig:translationLon} \textbf{Results of the translation analysis along the longitudinal direction.} $\Delta \log (\mathcal{L})$ values between the non-translated Boxy bulge template~\cite{Freudenreich:1997bx}  and the template artificially translated along the Galactic plane in steps of $0.4^{\circ}$. There is evidence for a shifted center at 
$l=-1.69^{\circ} \pm 0.12^{\circ}$ ($1 \sigma$). The statistical significance for the shift is $12.5\sigma$ ($\Delta \log(\mathcal{L})=77.7$ for one extra parameter), but systematic uncertainties (not included here) dwarf the statistical errors. See the caption of Fig.~\ref{Fig:Rotation} for a description of the gamma-ray background model used.}
\end{figure}

\begin{figure}[t!]
\centering
\includegraphics[scale=0.15]{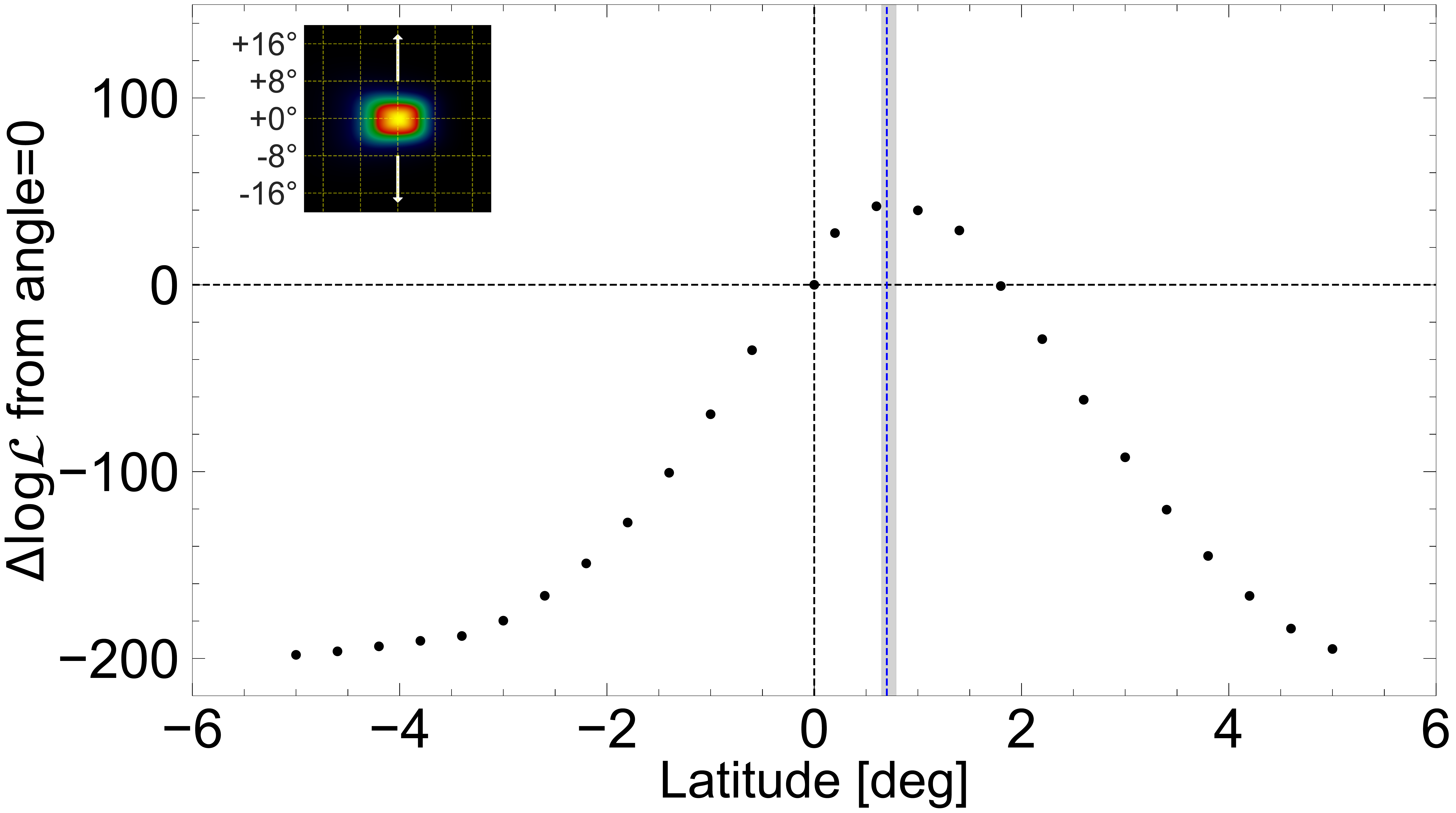}
\caption{\label{Fig:translationLat} \textbf{Results of the translation analysis along the latitudinal direction.}  Same as Fig.~\ref{Fig:translationLon}, except that this time the translation analysis is applied in the latitudinal direction. There is evidence for a small offset in position of $b=0.72^{\circ}\pm 0.07^\circ$ ($1\sigma$) with a $9.2\sigma$ significance ($\Delta \log(\mathcal{L})=42.0$ for one extra parameter). See the caption of Fig.~\ref{Fig:Rotation} for a description of the background model used.}
\end{figure}

It is instructive to compare the maximum likelihoods of the original Boxy bulge image for $\pm 90^\circ$ rotations. In the case of a clockwise rotation of $\theta_0 =90^\circ$, we find that the original orientation is a better model than the rotated one with a confidence of $15.8\sigma$ ($\Delta \log(\mathcal{L})=124.6$ for one additional parameter). For the counterclockwise rotation of the same angle ( $\theta_0=-90^\circ$), we find that the original orientation is preferred with a statistical significance of $7.5\sigma$ ($\Delta \log(\mathcal{L})=27.9$ for one extra parameter). Additionally, the profile displayed in Fig.~\ref{Fig:Rotation} shows that a rotation around $180^\circ$ is disfavored at $13.0\sigma$ ($\Delta \log(\mathcal{L})=84.5$ for one extra parameter). This demonstrates that the data can also statistically differentiate the North-South bulge asymmetries.

\subsection{Translation}\label{sec:tests:translation}


Next, we translate the boxy bulge template along the longitudinal and latitudinal directions in increments of $0.2^{\circ}$ and compute the log-likelihood value at each translated position respectively. The results from these tests are shown in Figs.~\ref{Fig:translationLon} and~\ref{Fig:translationLat}. 

In the longitudinal direction case (Fig.~\ref{Fig:translationLon}), we find evidence for a translation of $l \sim -1.7^{\circ}$, with a broad $\Delta \log(\mathcal{L})$  curve encompassing the range $l\sim [0^\circ, -3.5^{\circ}]$, while in the latitudinal case (Fig.~\ref{Fig:translationLat}) we find a translation at $b\sim 0.7^\circ$ with a $\Delta \log(\mathcal{L})$ curve that is comparatively smaller ($b\sim [0^\circ, 1.7^{\circ}]$). The boxy bulge images translated by $l \sim -1.7^{\circ}$ and $b\sim 0.7^\circ$ are preferred with $12.5\sigma$ ($\Delta \log(\mathcal{L})=77.7$) and $9.2\sigma$ ($\Delta \log(\mathcal{L})=42.0$ for one additional parameter) respectively. The preference for a shifted central position arises due to the presence of residuals on the Galactic west and north  sides of the Galactic Center that are able to be absorbed by the boxy bulge template when it is translated along the latitude and longitude directions. Therefore, the translations appear real in the sense it is warranted by the gamma-ray data. However, in the construction of the 2FIG catalog~\cite{Fermi-LAT:2FIG}, a group of several new point sources separated by less than $0.6^\circ$ were found around the region $l\sim 0^\circ$ and $b\sim [0^\circ, 2^\circ]$. Due to those new point sources not being easily individually resolved, the procedure of the authors of Ref.~~\cite{Fermi-LAT:2FIG} excluded that group of new point source candidates from the 2FIG catalog. It is likely that this exclusion drives the bulge translation shift and that with their inclusion in the form of, for example, an empirical patch of positive residual emission at $l\sim 0^\circ$ and $b\sim [0^\circ, 2^\circ]$, the boxy bulge latitudinal shift could be ameliorated. This is an interesting possibility that should be pursued in future work since it requires a dedicated focus along the Galactic plane, rather than the larger-scale Bulge emission of this study. 

With regards to the longitudinal shift, recent gamma-ray analyses of the inner Galaxy~\cite{Casandjian:andFermiLat2016,TheFermi-LAT:2017vmf,Cholis:2018} have also found excesses of gamma radiation off the Galactic Center at negative longitudes. Specifically, Refs.~\cite{Casandjian:andFermiLat2016,TheFermi-LAT:2017vmf} have both found that the base of the FB is a few degrees off the Galactic Center toward negative longitudes (see also central panel of Fig.~\ref{Fig:Fermibubbles}). In addition, Ref.~\cite{Cholis:2018} has pointed out that the GCE is potentially offset by $-4^\circ$. However, when the authors run their pipeline with the Galactic plane masked, the offset becomes $-1^\circ$.  The fact that this offset is sensitive to the applied masks indicates that the excess emission is mainly due to excess gamma-ray emission at very low latitudes ($b\sim 0^\circ$) but that there is also some evidence for a $\sim \mathcal{O}(1^\circ)$ shift of the gamma-ray bulge from high latitude data.

In this context, we note that the distribution of molecular clouds and young stars in the so-called ``central molecular zone'' (CMZ) is highly asymmetric~\cite{Yusef-Zadeh:2009,Immeretal:2012,Koepferletal:2015,Longmore:2018}. Close to 2/3 of the molecular gas is at positive longitude---Galactic East near Sgr A$^\star$---while about 2/3 of the young stars are at negative longitudes---Galactic West near Sgr A$^\star$~\footnote{See for example the bright 24 $\mu$m emission shown in Fig.1 of Ref.~\cite{Longmore:2018} Represented by blue color in that figure, this channel shows many bright point sources at $l\sim [0^\circ,-2^\circ]$ which correspond to a population of massive young stellar objects and evolved high mass stars.}. The asymmetric nature of the disposition of the molecular gas and young stars in the CMZ implies that the ISRF density distribution and the locations of CRs accelerators like supernovae and young massive stars, should be modeled with templates capable of accounting for these asymmetries. However, the triaxial function used in the derivation of the boxy bulge template adopted in this work~\cite{Freudenreich:1997bx} is not able to account for deviations of the triaxial symmetry in the stellar bulge data and, by construction, traces mostly the oldest bulge stars~\cite{Freudenreich:1997bx}. Nonetheless, Ref.~\cite{Cao:2013dwa} performed  a translation analysis in their fits to red giant clump stars and found that the data always preferred an offset toward negative longitudes (see Table 1 in Ref.~\cite{Cao:2013dwa}). While there are still quantitative differences, with the best-fit offsets of Ref.~\cite{Cao:2013dwa} being smaller than those we find in the gamma rays, it is very likely that these are mostly due to gamma-ray correlations with star formation sites along the disk or in the far Galactic East region of the nuclear bulge. As a consequence, high level emission from these sites might be biasing our fits to larger shift values than those found by Ref.~\cite{Cao:2013dwa}. The asymmetry of the gamma-ray bulge is a very interesting topic that should be more carefully investigated in future work, using bulge templates that go beyond the functional forms adopted in this study~\footnote{ We observe that the 3D IC templates~\cite{Porter:2017vaa} used in this work do not yet account for the stellar populations of the nuclear bulge and assume triaxial geometry (F98 or R12) for the bulge/bar structure. }.

\section{Gamma-ray properties of the bulge}\label{sec:bulge}

\begin{figure}[ht!]
\centering
\begin{tabular}{cc}
\includegraphics[scale=0.45]{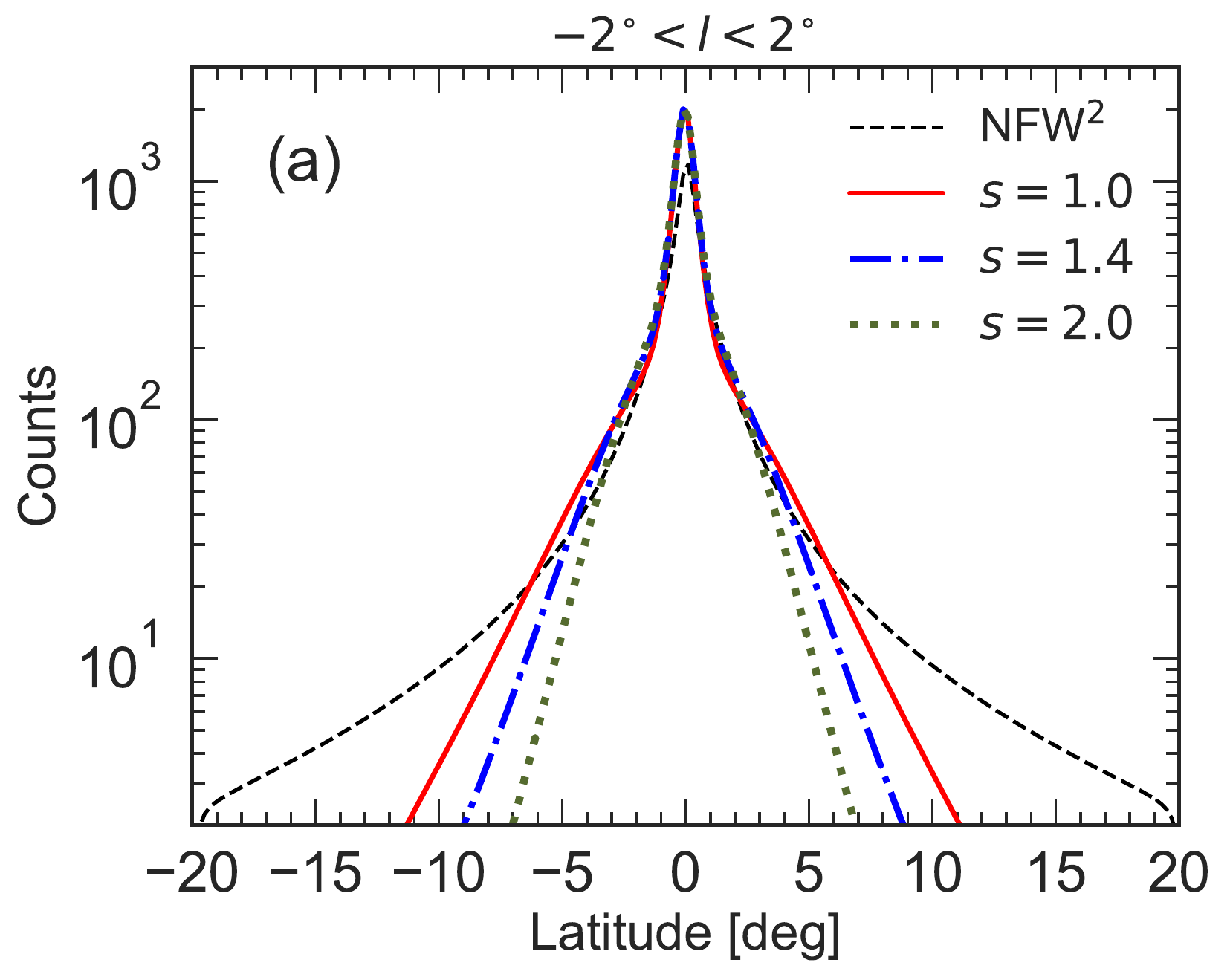} & \includegraphics[scale=0.45]{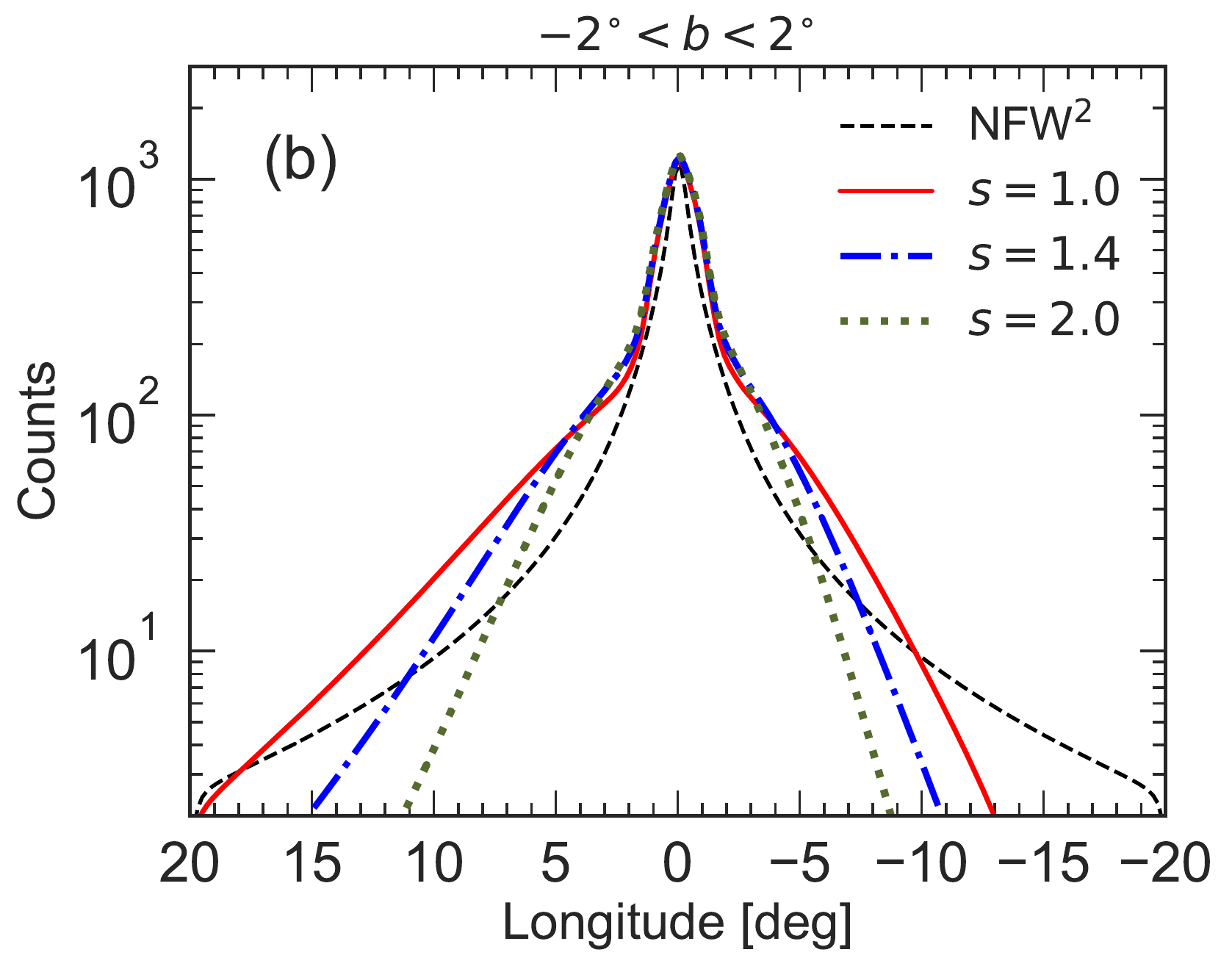} 
\end{tabular}
\includegraphics[scale=0.45]{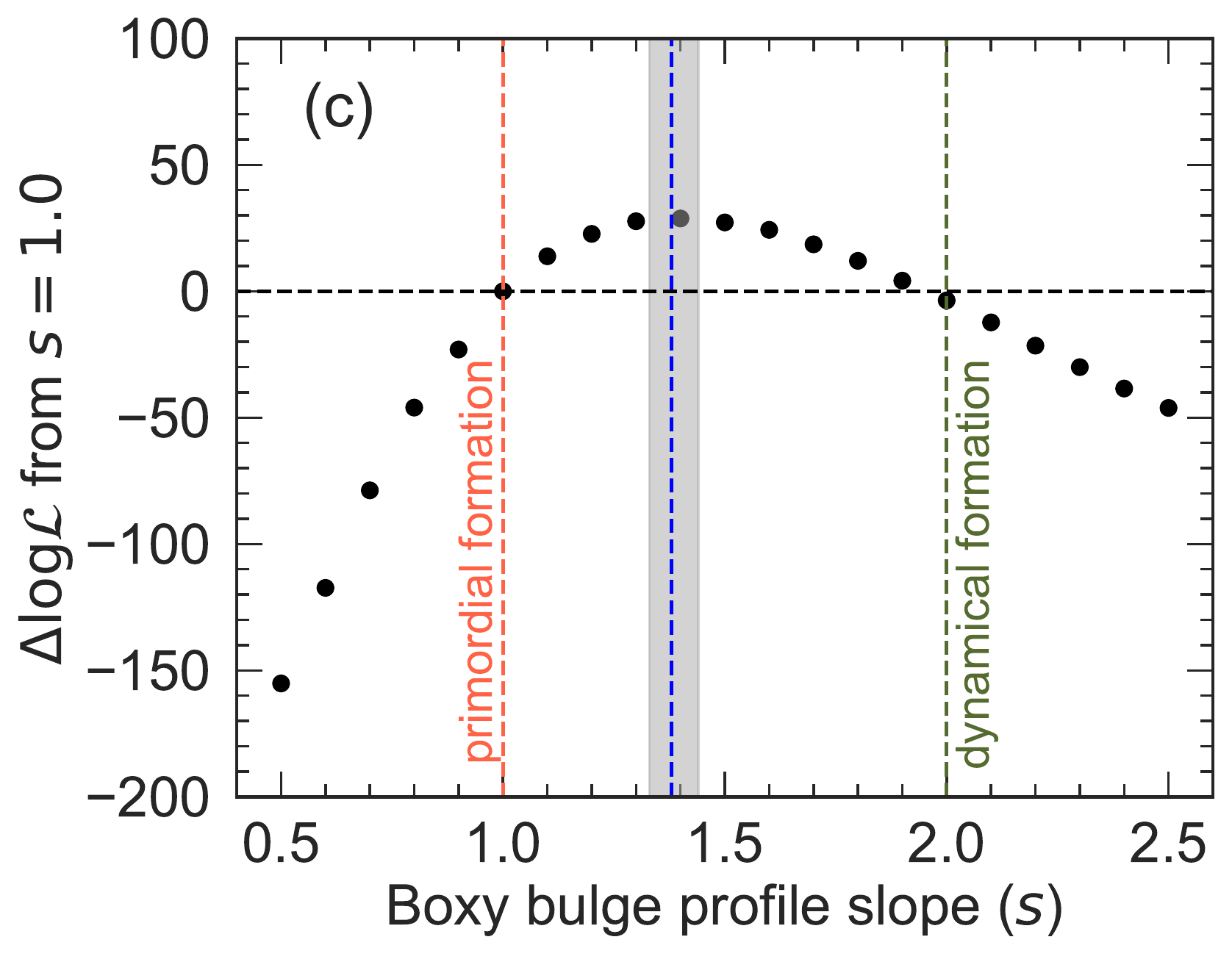} \caption{\label{Fig:BPslopeprofile} \textbf{Results of the morphological analysis:} Panels (a) and (b) show the latitudinal and longitudinal profiles of the boxy bulge + nuclear bulge best-fitting models summed over three energy bins between 1185.69 MeV and 2811.707 MeV. Note that in all cases the boxy bulge stellar density in Eq.~(\ref{Eq:F98bardensity}) has been modified to include an additional slope $s$ parameter of the form $\rho^s_{\rm bar} (R,\phi,z)$. The red solid line ($s=1$) represents the MSPs distribution in the primordial formation scenario while the green dotted line ($s=2$) represents the dynamical formation scenario. Panel (c) displays the $\Delta \log(\mathcal{L})$ profile for different stellar density slopes ($s$) with respect to the primordial formation model ($s=1$). The blue dash-dot line shows the best-fit slope $s=1.38^{+0.06}_{-0.05}$ which was found with a significance of $7.6\sigma$ ($\Delta \log(\mathcal{L})=28.8$ for one extra parameter). The grey band corresponds to the statistical only errors. }
\end{figure}

\subsection{Gamma-ray bulge morphology}\label{subsec:bulgeslope}

Globular clusters are old and extremely dense stellar agglomerations that might provide important clues about the formation mechanisms of MSPs in the Galactic bulge~\cite{Abazajian:2012pn}. There is strong observational evidence that some of the most massive globular clusters in the Milky Way contain large numbers of MSPs~\cite{TheFermi-LAT:2PC}. However, given the relatively low escape velocities in globular clusters ($\bar{v}_{\rm escape}\lesssim 60$ km/s)~\cite{Gnedin:2002un},  it is not straightforward to reconcile the presence of MSPs in these objects with empirical estimates of typical neutron star (NS) kick velocities ($\bar{v}_{\rm kick}\sim 90$ km/s)~\cite{Hobbs:2005yx}. 

In order to solve this ``NS retention problem''~\cite{Pfahl:2001df} in globular clusters, two scenarios have been proposed: First is the \textit{primordial formation scenario}~\cite{Brandt:1994rr}, which posits that NSs form in massive binaries making them more easy to retain. This is because the kick-momentum transferred to the NS is shared with a massive companion, which leads to a reduction in the velocity of the post-supernova compact object. The number of MSPs is expected to scale linearly with the total stellar mass in this scenario. Second is the \textit{dynamical formation} scenario, which rests on the observed positive correlations between the stellar encounter rate $\Gamma_c$ and the number of low-mass X-ray binaries (thought to be the progenitors of MSPs) in globular clusters~\cite{Huietal:2010}. Such observations indicate that after NSs are born, they can be captured by massive stars to form binaries which in turn can evolve to become MSPs. Since $\Gamma_c\propto \rho^2$, where $\rho$ is the stellar density, it follows that in this picture the number of MSPs scales with the square of the stellar density.

In spite of the very different stellar densities and dynamical histories of globular clusters and the Galactic bulge, similar MSPs formation mechanisms might be operating in the Galactic Center~\cite{Abazajian:2012pn,Abazajian:2014fta,Gonthier:2018ymi}.
The spatial distribution of low-mass X-ray binaries in M31 provides evidence for the view that both dynamical and primordial components exist~\cite{Voss:2006az}, with the dynamical component dominating in the center. The close match between M31 X-ray binaries and the GCE morphology was previously used by Ref.~\cite{Abazajian:2014fta} to argue for unresolved MSPs being the dominant source of the GCE. 

 Here, we perform a morphological analysis of the GCE data to find out whether the spatial distribution of photons is better explained by the primordial or dynamical formation scenario. To this end, we modify the boxy bulge density function in Eq.~(\ref{Eq:F98bardensity}) to include an additional slope parameter $s$ in the form $\rho^s_{\rm bar} (R,\phi,z)$. By varying $s$ in the range $[0.5,2.5]$ in steps of 0.1 and integrating the resulting density function along the line-of-sight, we create additional stellar maps which we individually pass through our pipeline routine. We note that in doing so the nuclear bulge map is left unmodified as we are focusing on the larger scale bulge component.

The results of the morphological analysis are presented in Fig.~\ref{Fig:BPslopeprofile}. Panels (a) and (b) show the longitudinal and latitudinal dependencies of the boxy bulge + NB models for various slope values: the primordial formation ($s=1.0$), dynamical formation ($s=2.0$) and the best-fit slope ($s=1.4$). There are appreciable differences between the modified boxy bulge + NB profiles and the NFW$^2$, which helps to explain why the maximum-likelihood analysis assigns largely different log-likelihood values to bulge models compared to spherically symmetric ones (\textit{e.g.}, Table~\ref{Tab:GCElikelihoods}). Interestingly, we find that the density slope that best fits the data is $s=1.38^{+0.06}_{-0.05}$, which is preferred over the primordial scenario with a statistical significance of $7.6 \sigma$ ($\Delta \log(\mathcal{L})=28.8$ for one additional parameter), see panel (c) of the same figure. Therefore, we find significant evidence for an \textit{admixture formation} scenario in which a fraction of the MSPs are formed through the primordial channel and the remainder are formed dynamically. Figure~\ref{Fig:BPslopeprofile}-(c) also shows that the fit worsens when stellar density models are assumed with slopes $s<1.0$ or $s>2.0$.
For instance, a slope $s=0.9$ is statistically disfavored with $6.8\sigma$ ($\Delta \log(\mathcal{L})=23.1$ for one additional parameter) significance. Likewise, slopes of $s=2.1$ and $s=2.2$ are disfavored with $5.0\sigma$ ($\Delta \log(\mathcal{L})=12.3$) and $6.6\sigma$ ($\Delta \log(\mathcal{L})=21.5$) significance, respectively. This is an important consistency check that gives further support to the MSPs hypothesis for the GCE.

It has been argued that the putative population of Galactic bulge MSPs could have been the result of depositions from tidally disrupted globular clusters~\cite{Gnedin:2013cda,Brandt:2015ula,Fragione:2017rsp}. However, a fundamental prediction of this model seems to be a spherically symmetric distribution of Galactic bulge MSPs~\cite{Fragione:2017rsp}, which we have demonstrated to be highly disfavored by the data. There is a growing consensus that the origin of the Boxy/Peanut bulge morphology is the result of the dynamical evolution of a disk stars via buckling instabilities~\cite{Nataf:2017,Bland-Hawthorn2016}. In this picture, the Galactic bulge is predominantly composed of disk stars now on bar orbits. It is possible that by changing the initial conditions of their $N$-body simulations, the disrupted globular cluster scenario could reproduce the Boxy/Peanut bulge morphology that the gamma-ray data requires. Nevertheless, such simulations would still need to explain the nearly isotropic distribution of the surviving population of Milky Way globular clusters. We note that accounting for these two seemingly irreconcilable observations in the disrupted globular cluster scenario~\cite{Gnedin:2013cda,Brandt:2015ula,Fragione:2017rsp} appears to be a difficult task, thus disfavoring current versions of this model for the GCE.

\subsection{Gamma-ray bulge spectrum}\label{subsec:bulgespectrum}

\begin{figure}[t!]
\centering
\includegraphics[scale=0.6]{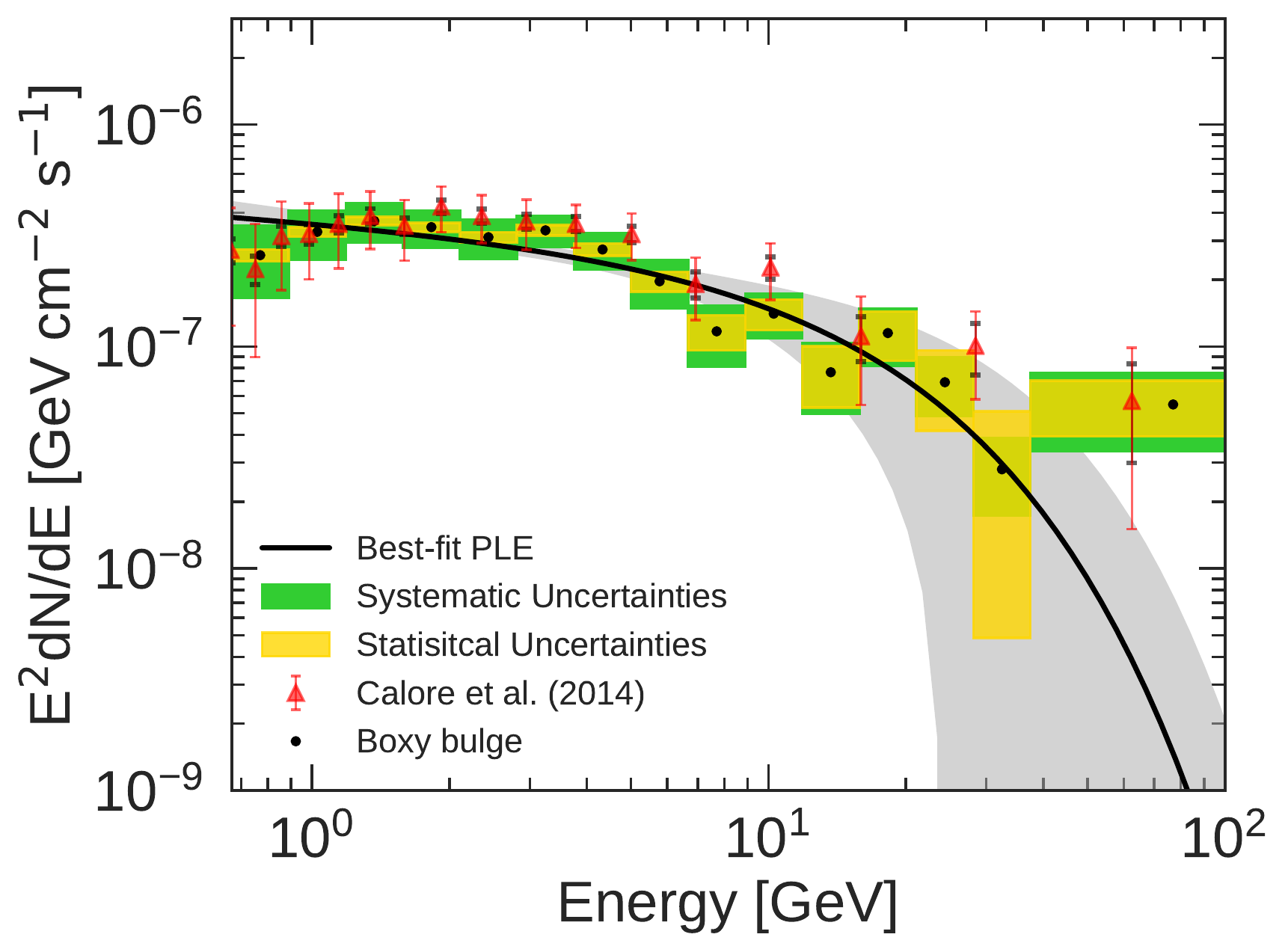} \caption{\label{Fig:BoxybulgeSpectrum} \textbf{Boxy bulge spectrum:} Shown are the boxy bulge fluxes (black points) and corresponding statistical (yellow) and systematic (green) errors. Systematic errors were estimated by calculating the dispersion of the best-fit bin fluxes obtained from fits that assumed the alternative 3D IC maps or the standard-2D IC model, respectively. Best-fit spectrum (black line) is given by $dN/dE = N_0\; E^{-\Gamma}\exp(-E/E_{\rm cut})$ with $N_0=(3.8\pm 0.4)\times 10^{-10}$ MeV$^{-1}$ cm$^{-2}$ s$^{-1}$, $\Gamma=2.1\pm 0.2$ and $E_{\rm cut}=15\pm 14$ GeV ($\chi_{\rm PL}^2=11.12$ for 15-3 dof, that is, $\mbox{p-value}=0.5$). This functional form was preferred over a simple power law (PL) with a $2.6\sigma$  statistical significance ($\Delta \chi^2=\chi^2_{\rm PL}-\chi^2_{\rm PLE}$=5.3 for one extra parameter. The grey shaded area displays the uncertainties on the PL with exponential cut-off spectral parameters. We compare the boxy bulge spectrum to the GCE obtained by assuming an NFW$^2$ template in Ref.~\cite{Calore:2014xka} (red data points). Red (black) error bars correspond to their systematic (statistical) uncertainties.  The data points were adapted by accounting for the solid angle of our RoI. }
\end{figure}

\begin{figure}[t!]
\centering
\begin{tabular}{ccc}
\includegraphics[width=0.26\textwidth]{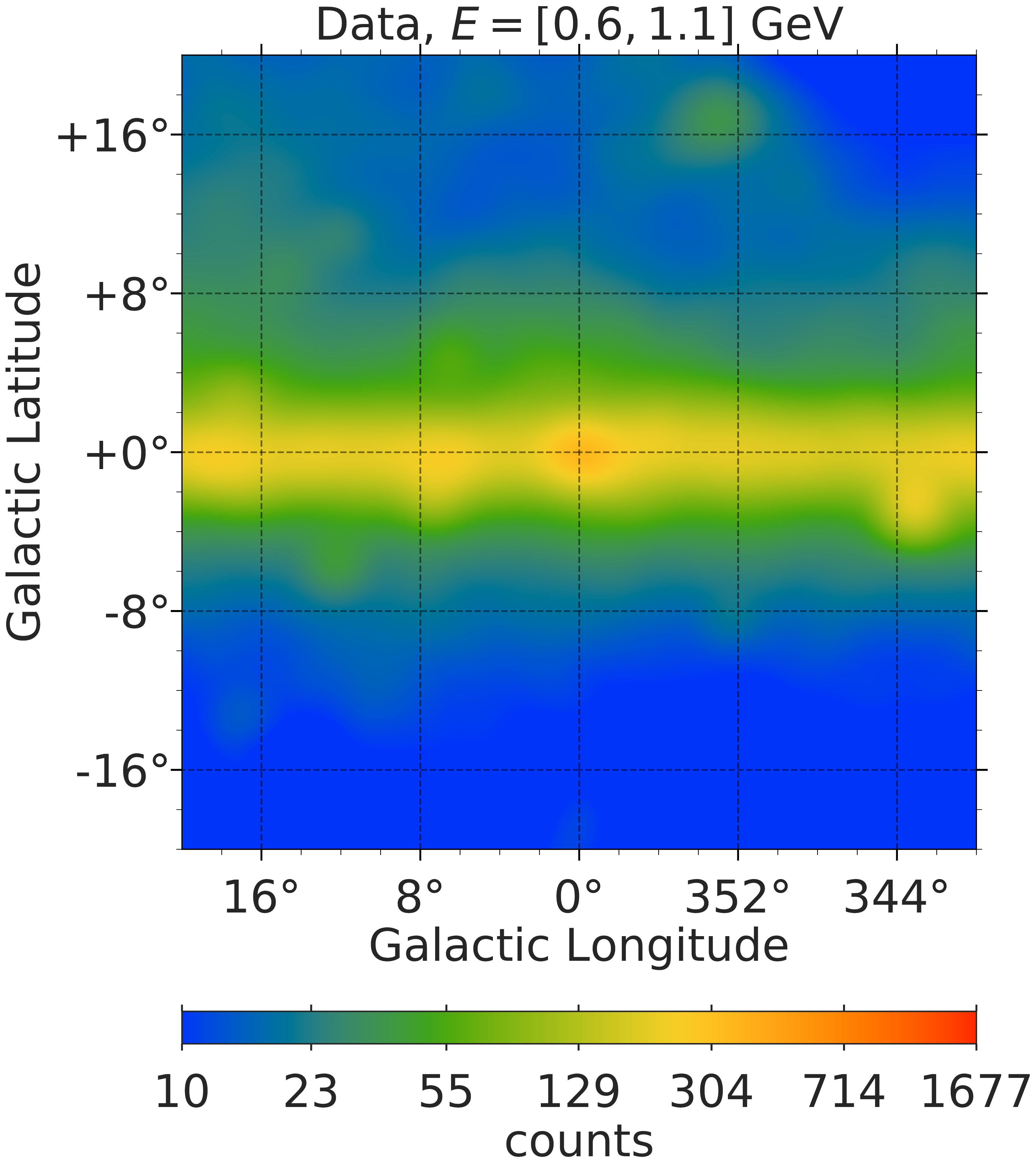} & \includegraphics[width=0.26\textwidth]{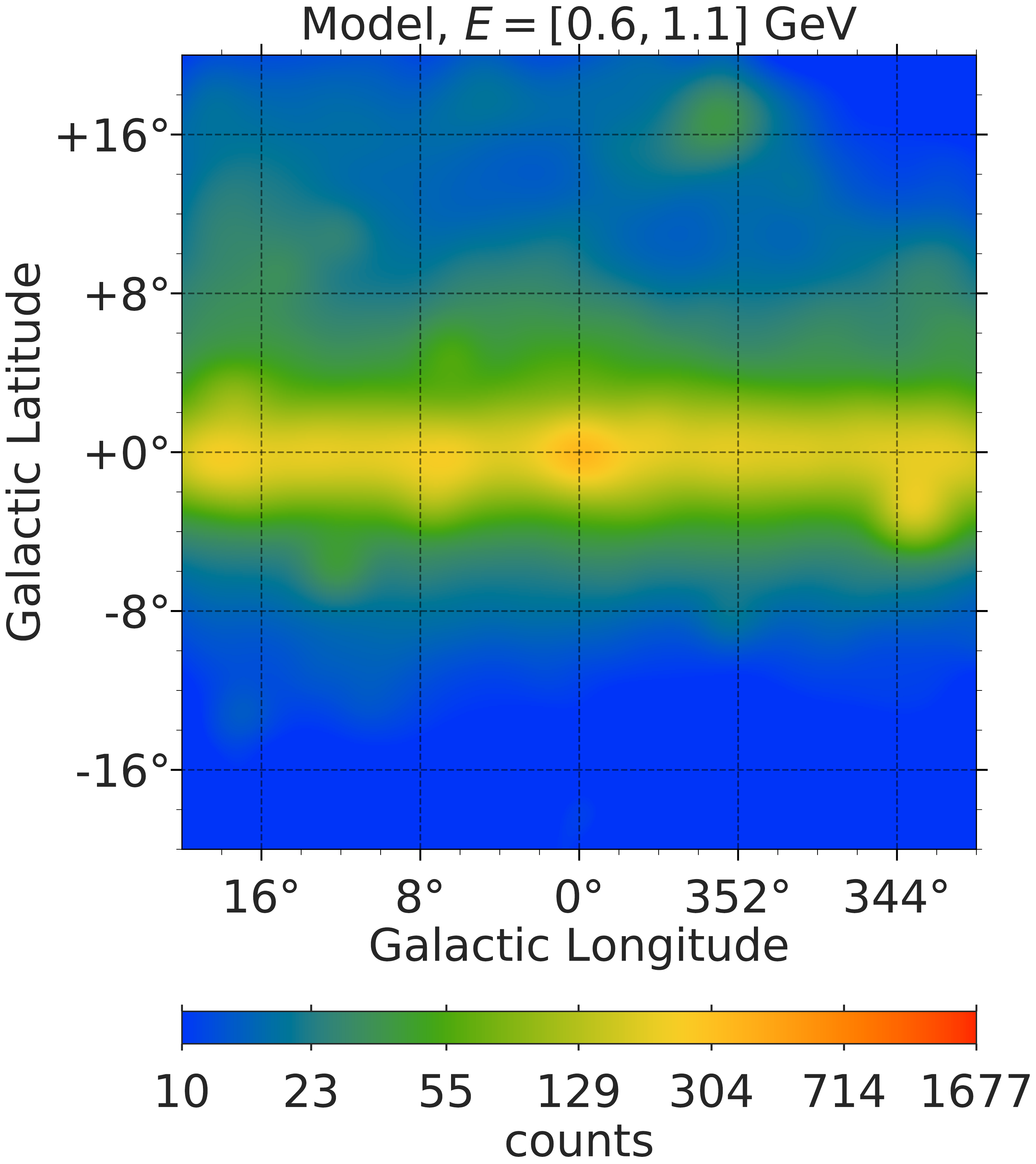}  & \includegraphics[width=0.258\textwidth]{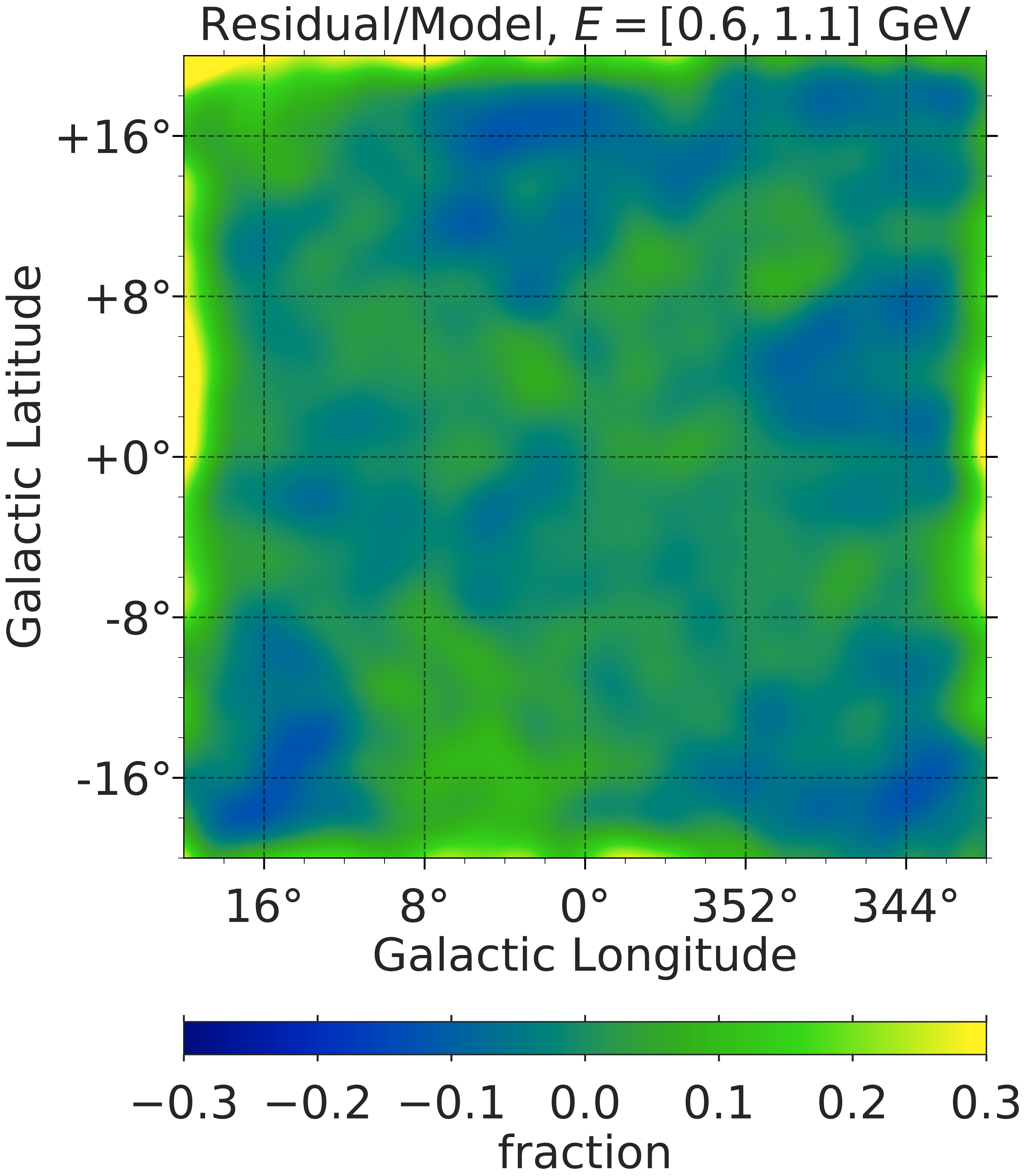}\\
\includegraphics[width=0.26\textwidth]{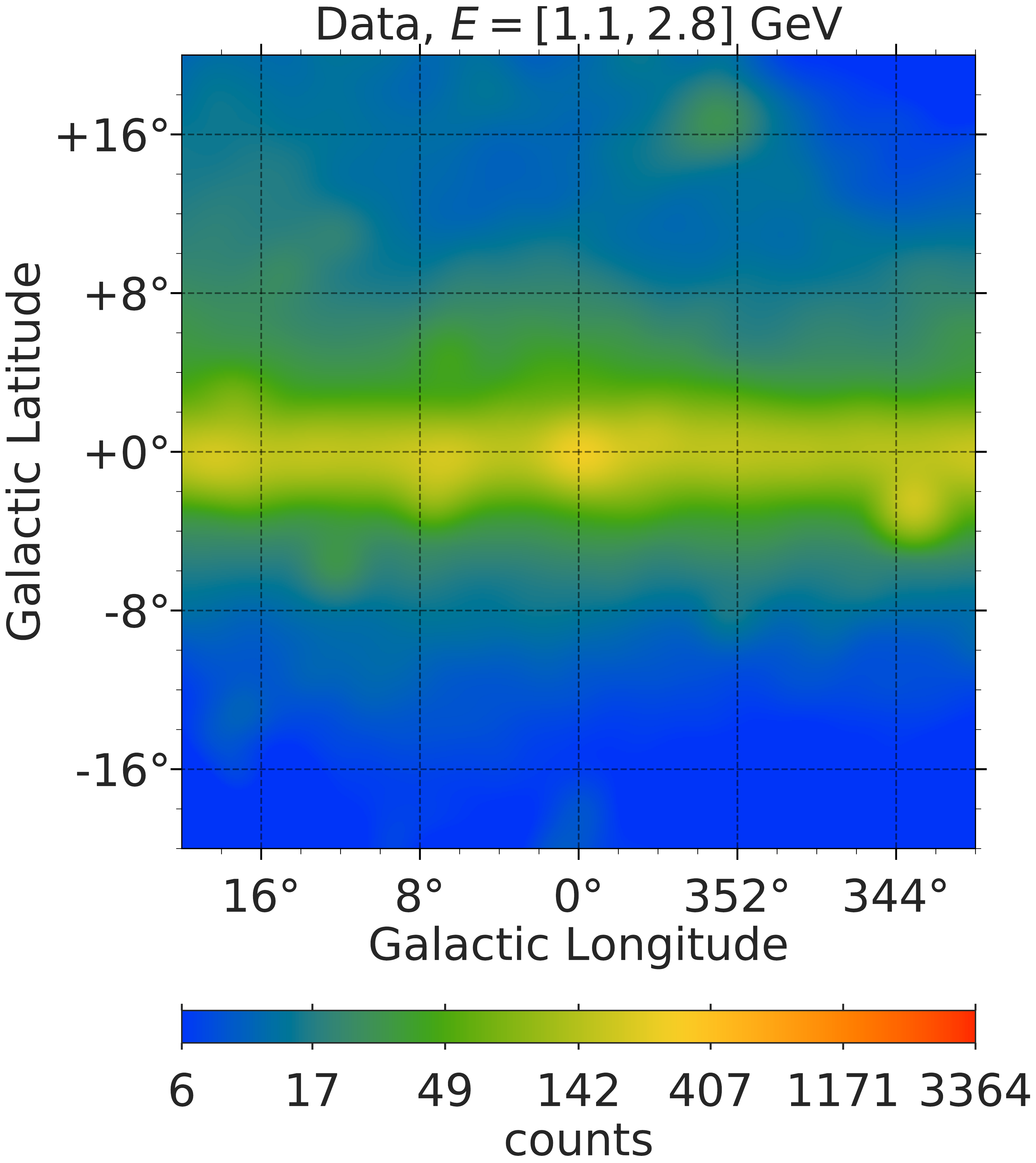} & \includegraphics[width=0.26\textwidth]{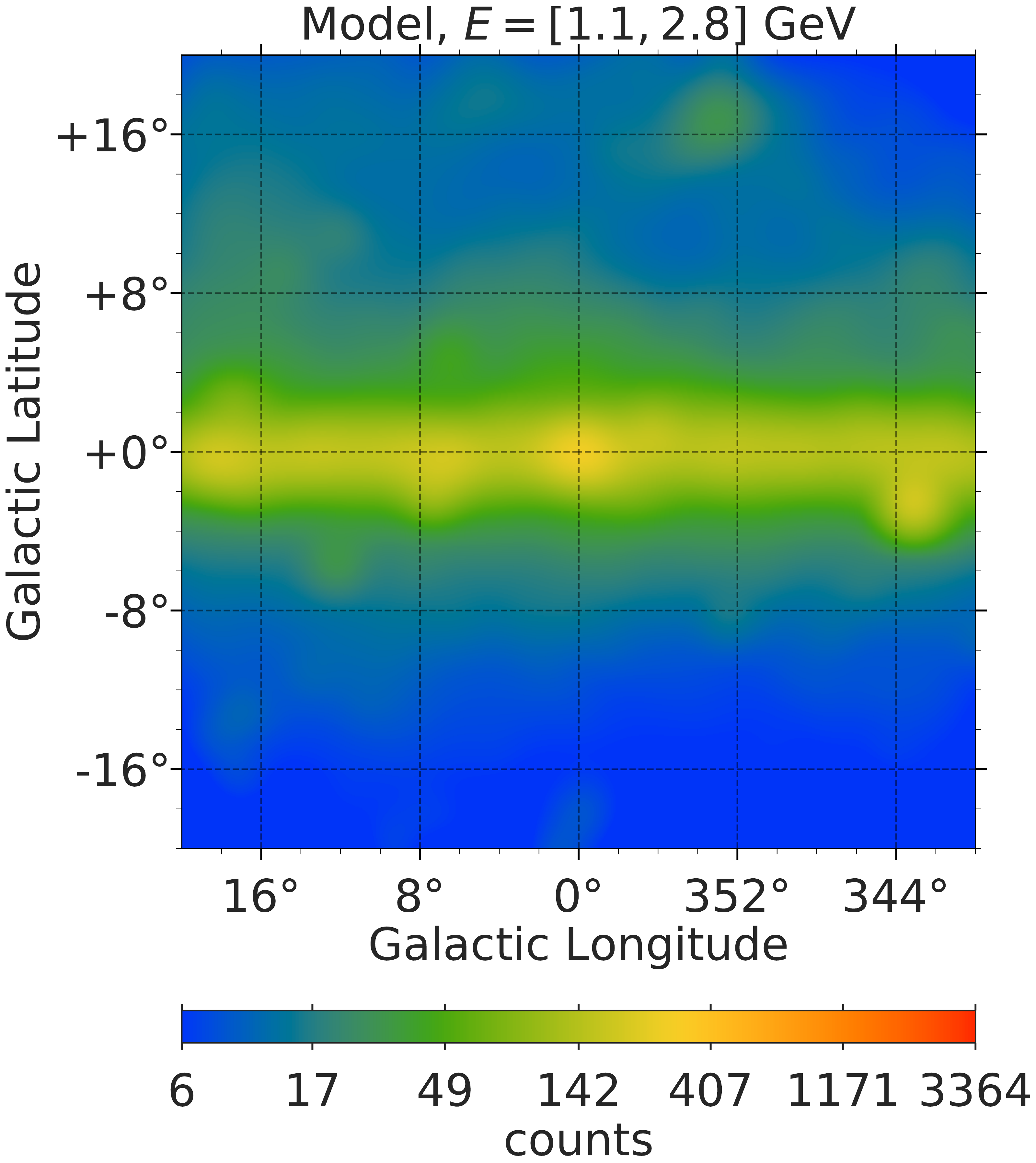}  & \includegraphics[width=0.258\textwidth]{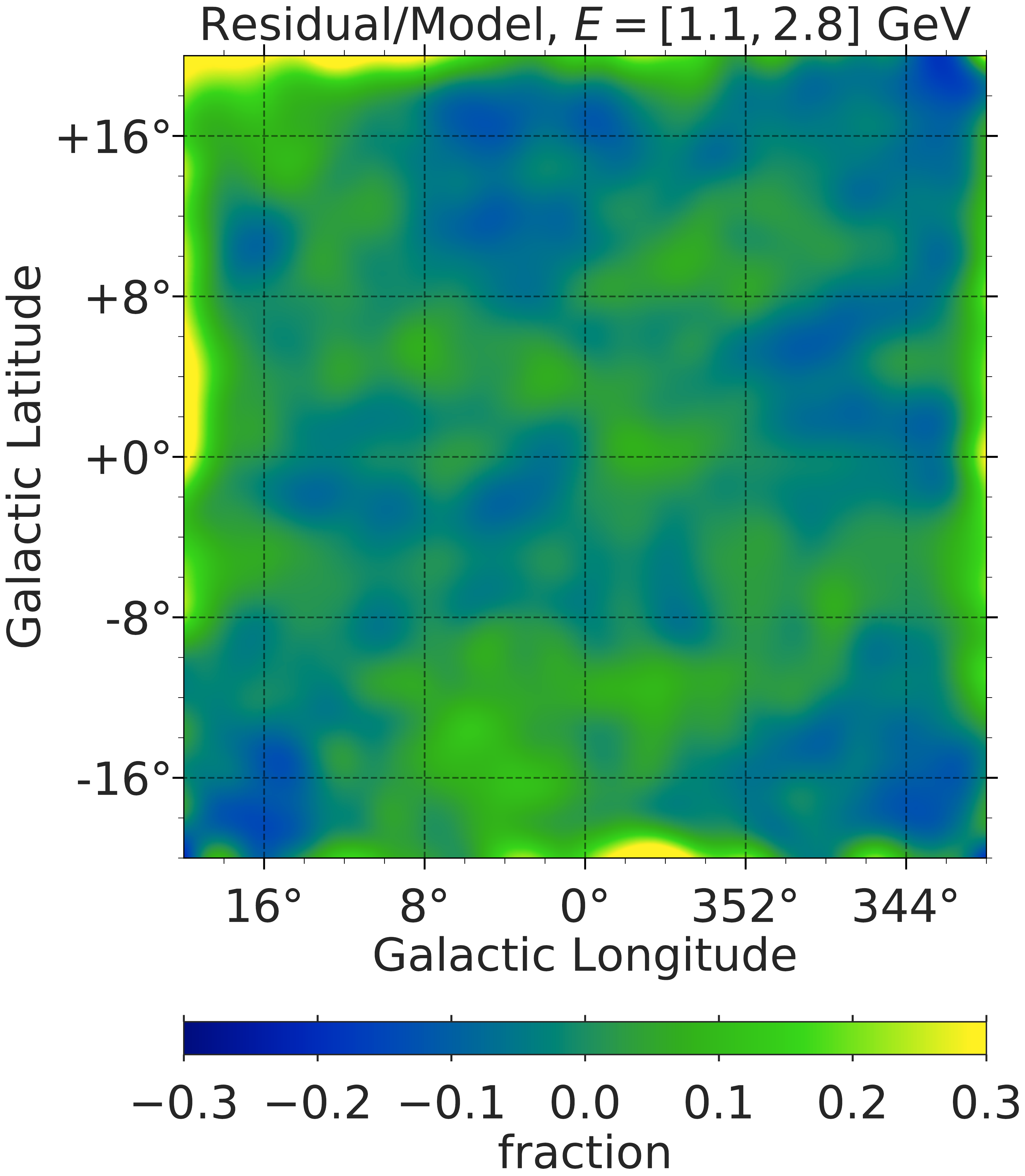}\\
\includegraphics[width=0.26\textwidth]{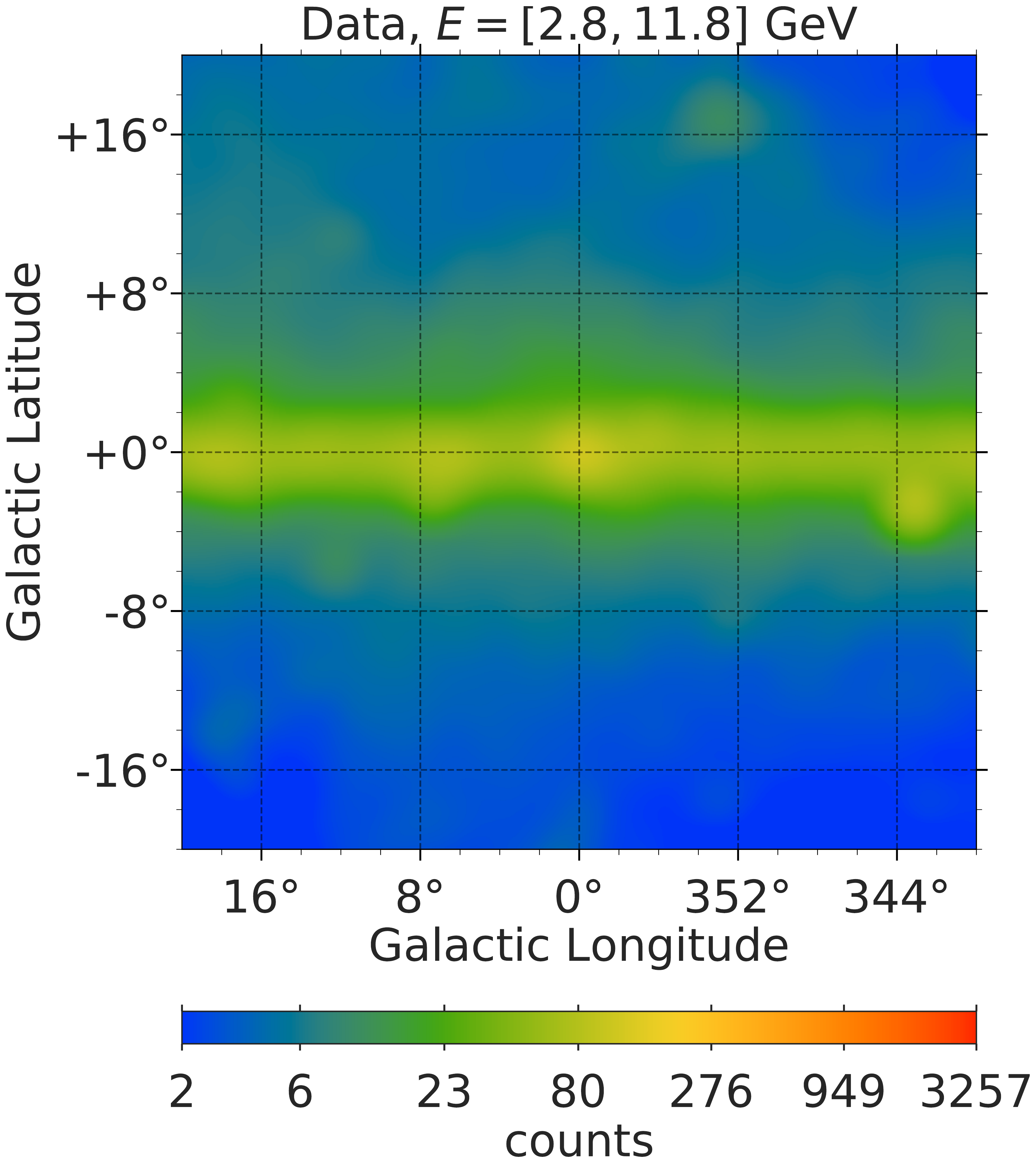} & \includegraphics[width=0.26\textwidth]{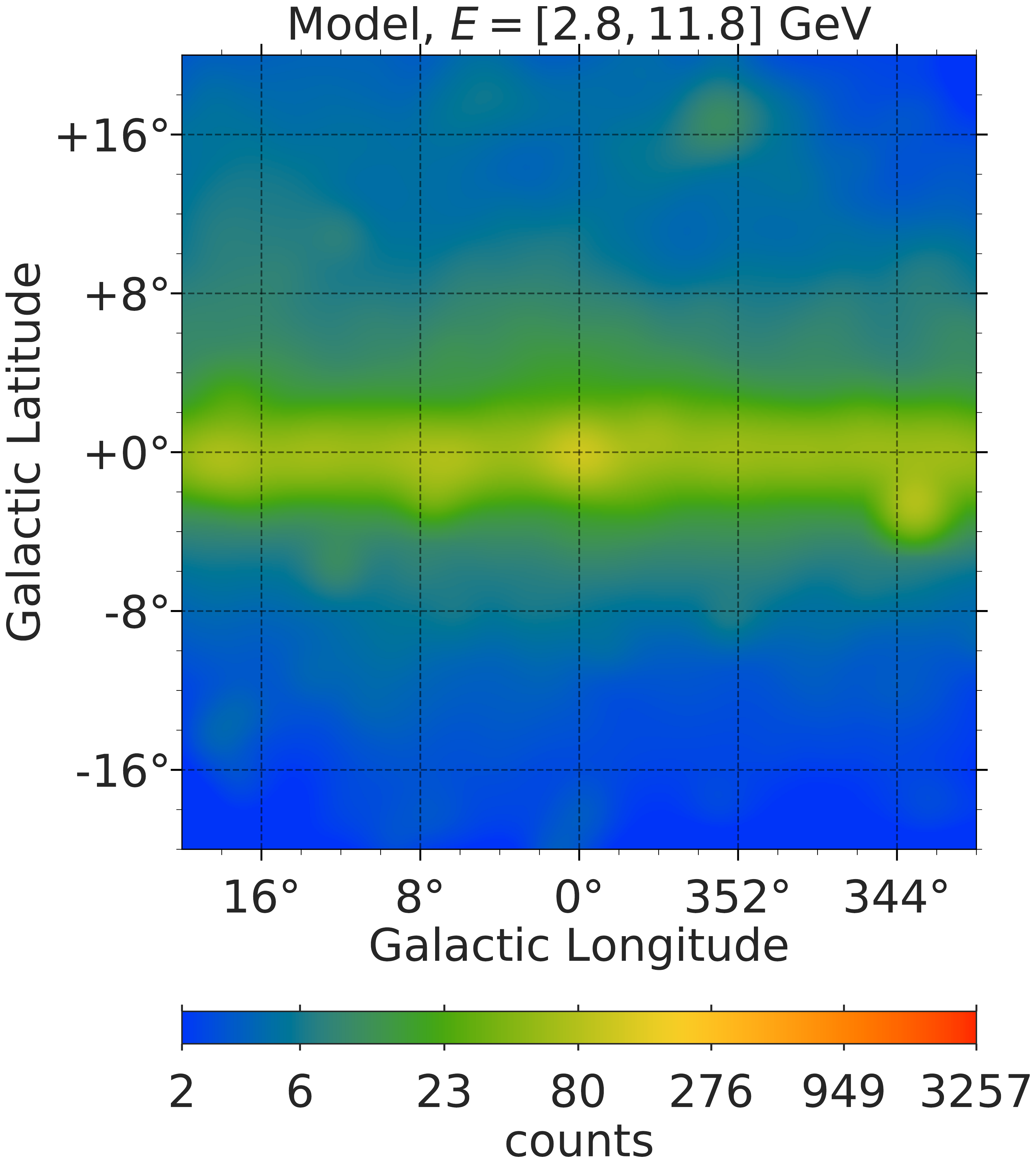}  & \includegraphics[width=0.258\textwidth]{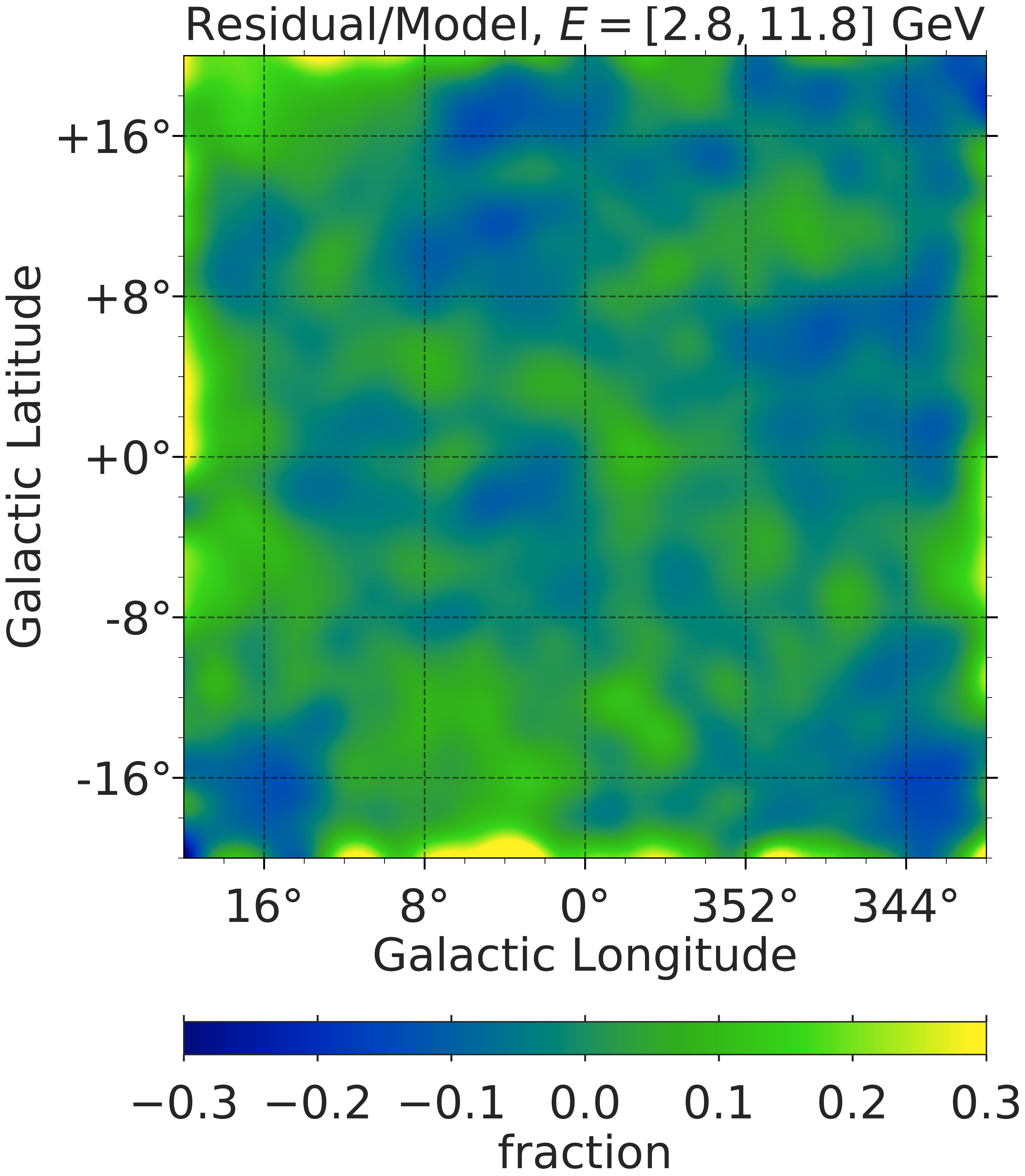}\\
\includegraphics[width=0.26\textwidth]{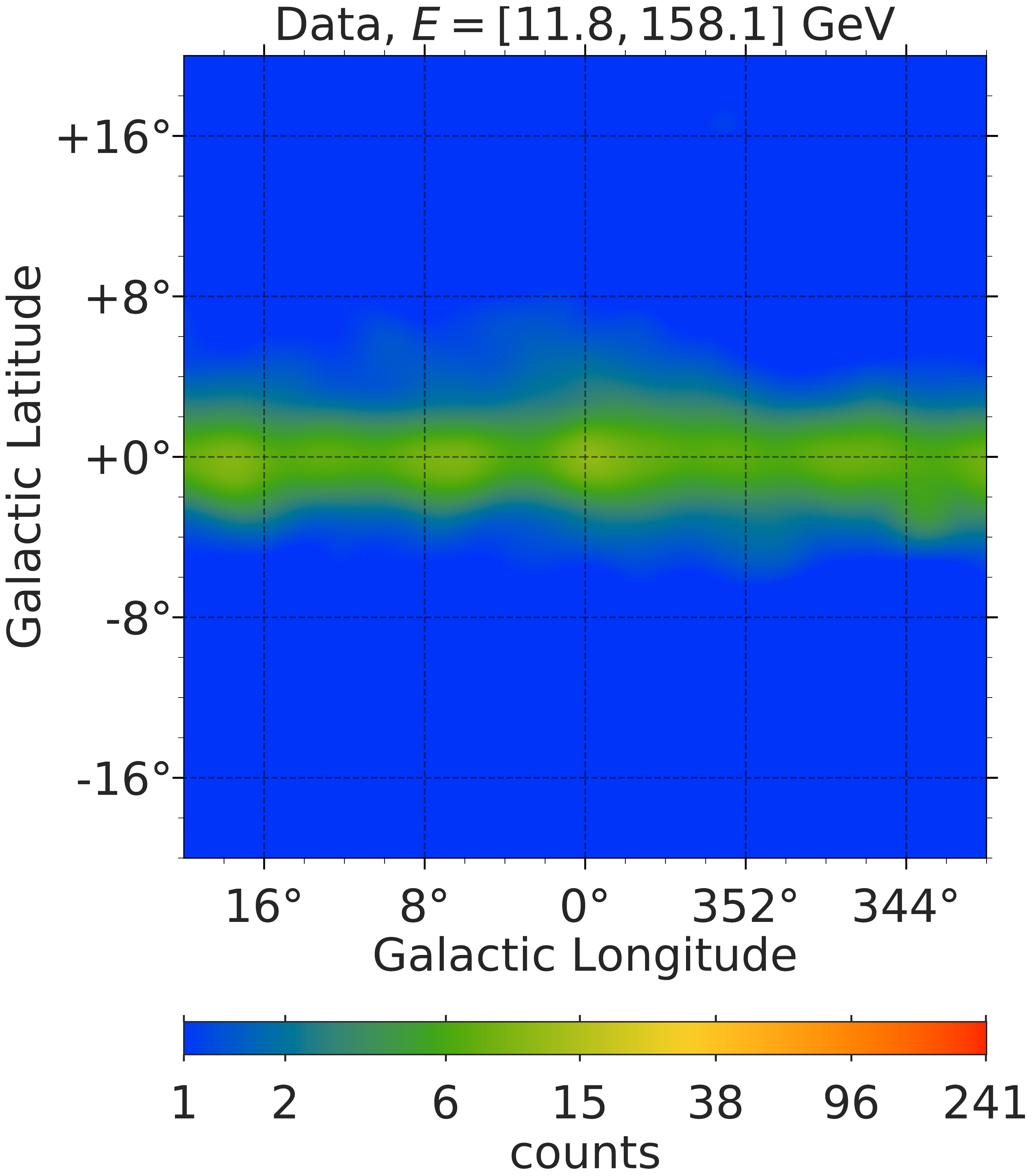} & \includegraphics[width=0.26\textwidth]{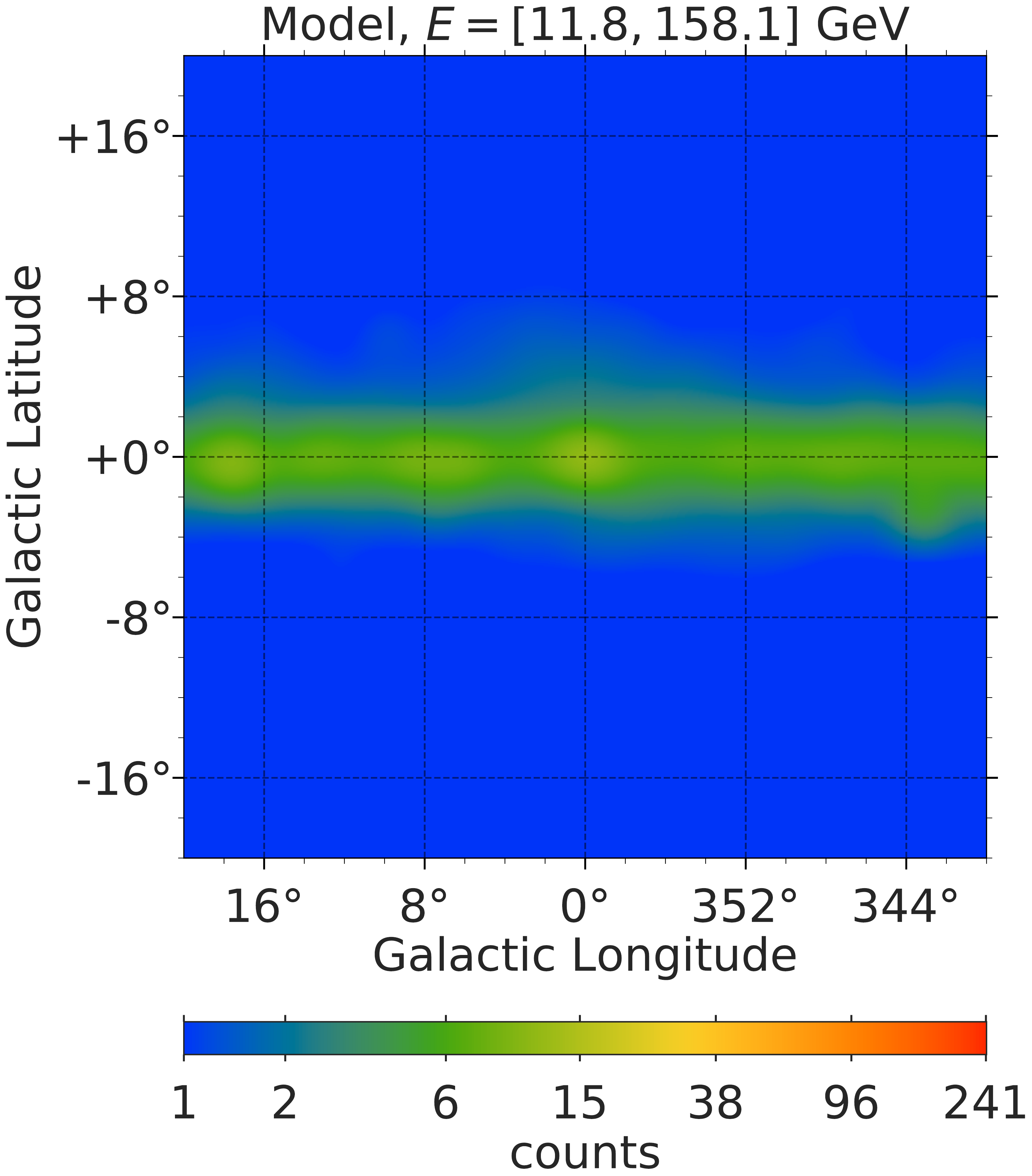}  & \includegraphics[width=0.258\textwidth]{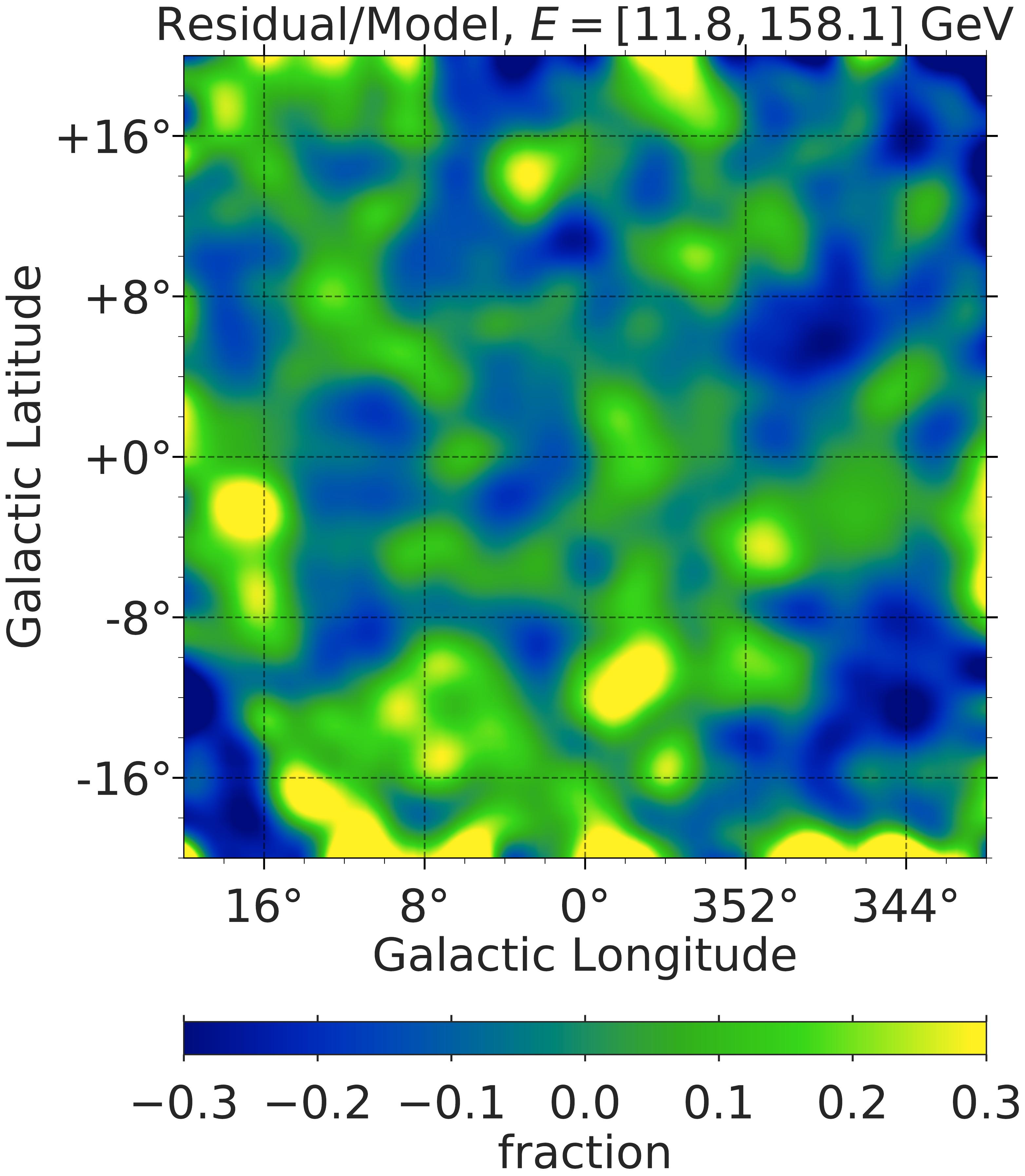}
\end{tabular}\caption{\textbf{Gamma-ray data (left), best-fitting model (middle) and fractional residuals $(Data-Model)/Model$ (right).} The best-fit model shown in the second column corresponds to baseline$+$SFB (Inp.)$+$Boxy bulge$+$NB (see Table~\ref{Tab:GCElikelihoods} for the model details). In the baseline model we included the 3D ICS map F98-SA50. The maps are smoothed with a Gaussian filter of radius $0.5^\circ$. } 
\label{fig:Residuals}
\end{figure}

\begin{figure*}[ht!]
    \centering
    \includegraphics[width=0.48\textwidth]{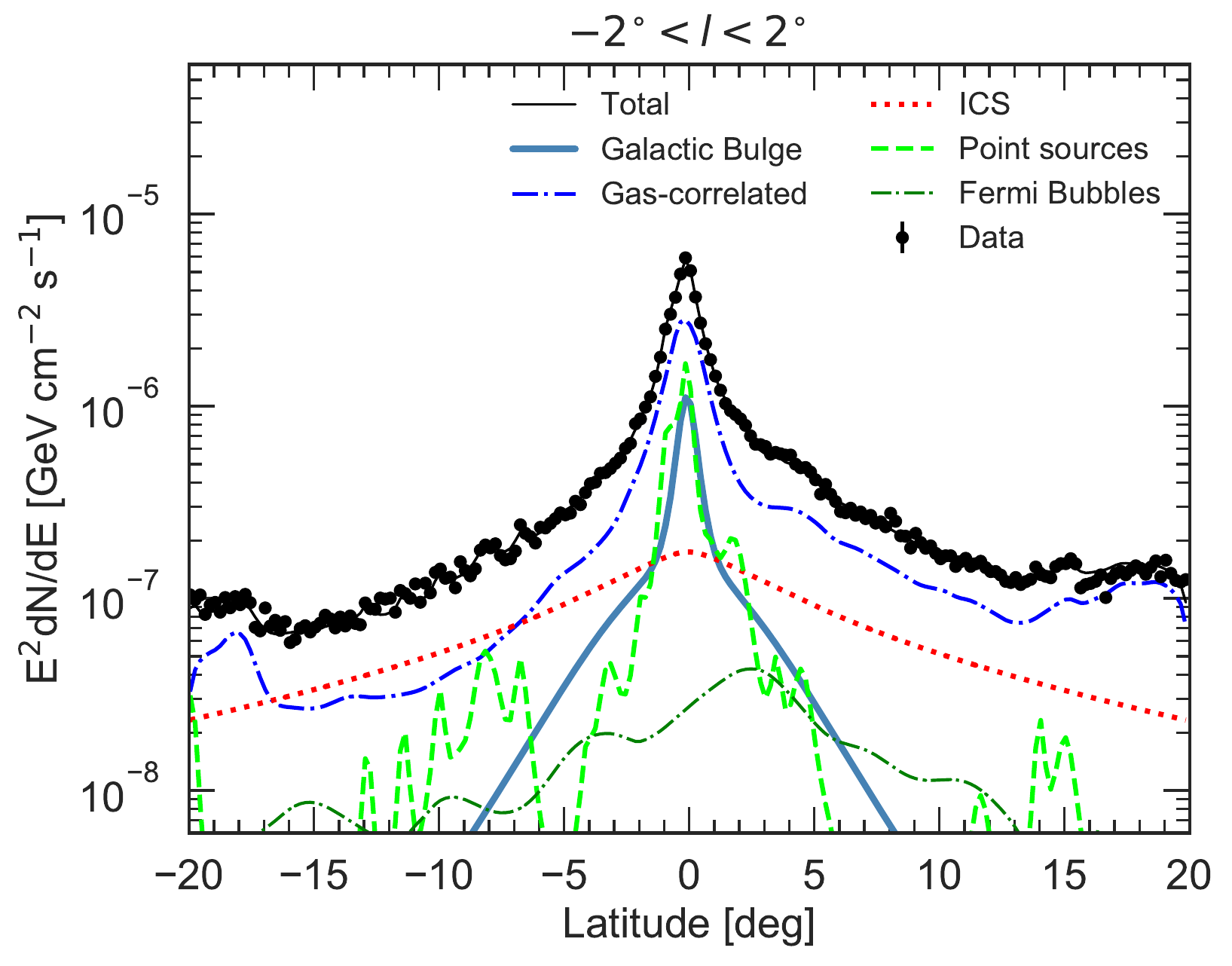}
    \includegraphics[width=0.48\textwidth]{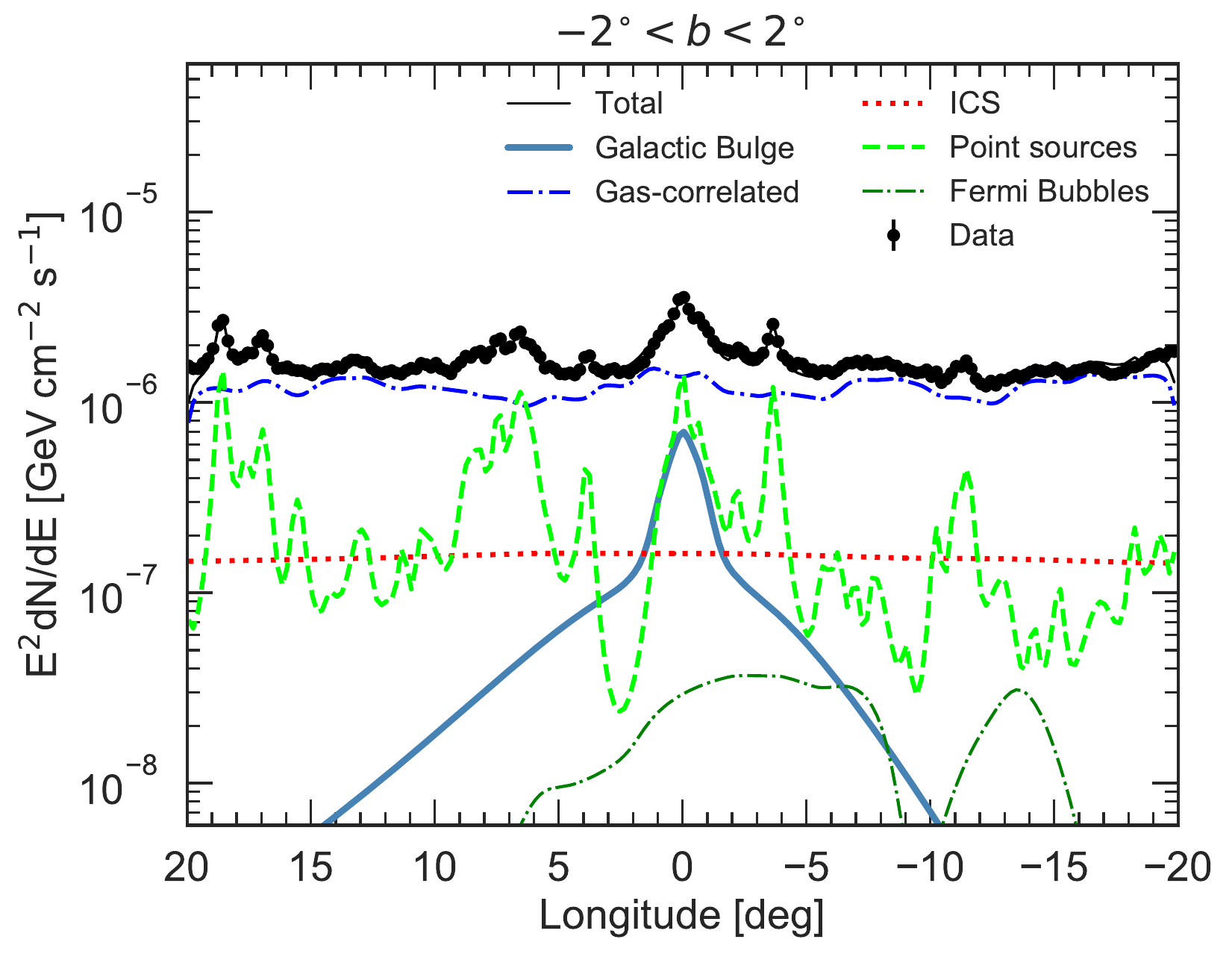}
    \caption{Spatial distribution of the main model components included in the fit. A description of the model chosen for display can be seen in the caption of Fig.~\ref{fig:Residuals}. Both panels display the flux profiles in the energy range $[1.1,2.8]$ GeV and in the regions $|l|<2^\circ$ and $|b|<2^\circ$. Black dots represent the data and the continuous black line the total best-fitting model. Other components not shown here (\textit{e.g.,} isotropic, Sun, Moon and Loop I) are $\sim \mathcal{O}(1)$ less bright in the region used to construct the profile. }
    \label{fig:fluxprofiles}
\end{figure*}

The gamma-ray luminosity\footnote{In the computation of the luminosities we assumed 8.25 kpc for the distance to the Galactic Center.} of the boxy bulge model ($s=1$) was found to be $(2.2 \pm 0.4)\times10^{37}$ erg s$^{-1}$ for $E\geq100$ MeV, as estimated for the whole $40^\circ \times 40^\circ$ region of interest. A power-law with an exponential cut-off (PLE) model was preferred to a simple power law (PL) with a confidence of $2.6\sigma$. As the amplitude is restricted to be non-negative, a $\chi^2/2$ distribution rather than the $\chi^2$ distribution is needed. We computed the best-fit spectral parameters with a $\chi^2$ fitting method to the inferred fluxes at each energy bin. We found $E_{\rm cut}= 15\pm 14$ GeV and $\Gamma=2.1\pm 0.2$, which as can be seen from Fig.~5 of Ref.~\cite{Macias:2013vya}, are consistent with the resolved MSPs and also the GCE obtained using an NFW$^2$ template. In Fig.~\ref{Fig:BoxybulgeSpectrum}, we show the spectra of the boxy bulge component in the inner $40^\circ \times 40^\circ$ around the Galactic Center. The yellow (green) error bars show the statistical (systematic) uncertainties. We have estimated the systematic uncertainties by computing the dispersion of the best-fit bin fluxes obtained when all the alternative 3D and standard-2D IC models are included in the maximum-likelihood analyses respectively. Our approach is consistent with earlier analyses by the \textit{Fermi}-LAT collaboration~\cite{TheFermi-LAT:2015kwa} that argued for a large impact in the GCE spectrum due to IC modeling. After rescaling the GCE data from Ref.~\cite{Calore:2014xka} to our RoI and comparing to our boxy bulge spectrum (red data points), we find very good agreement with their results. We note that in that work the GCE was obtained by masking the inner $|b|<2^\circ$ region of the Galactic plane and assuming an NFW$^2$ spatial model.

\begin{figure}[t!]
\centering
\begin{tabular}{cc}
\includegraphics[scale=0.43]{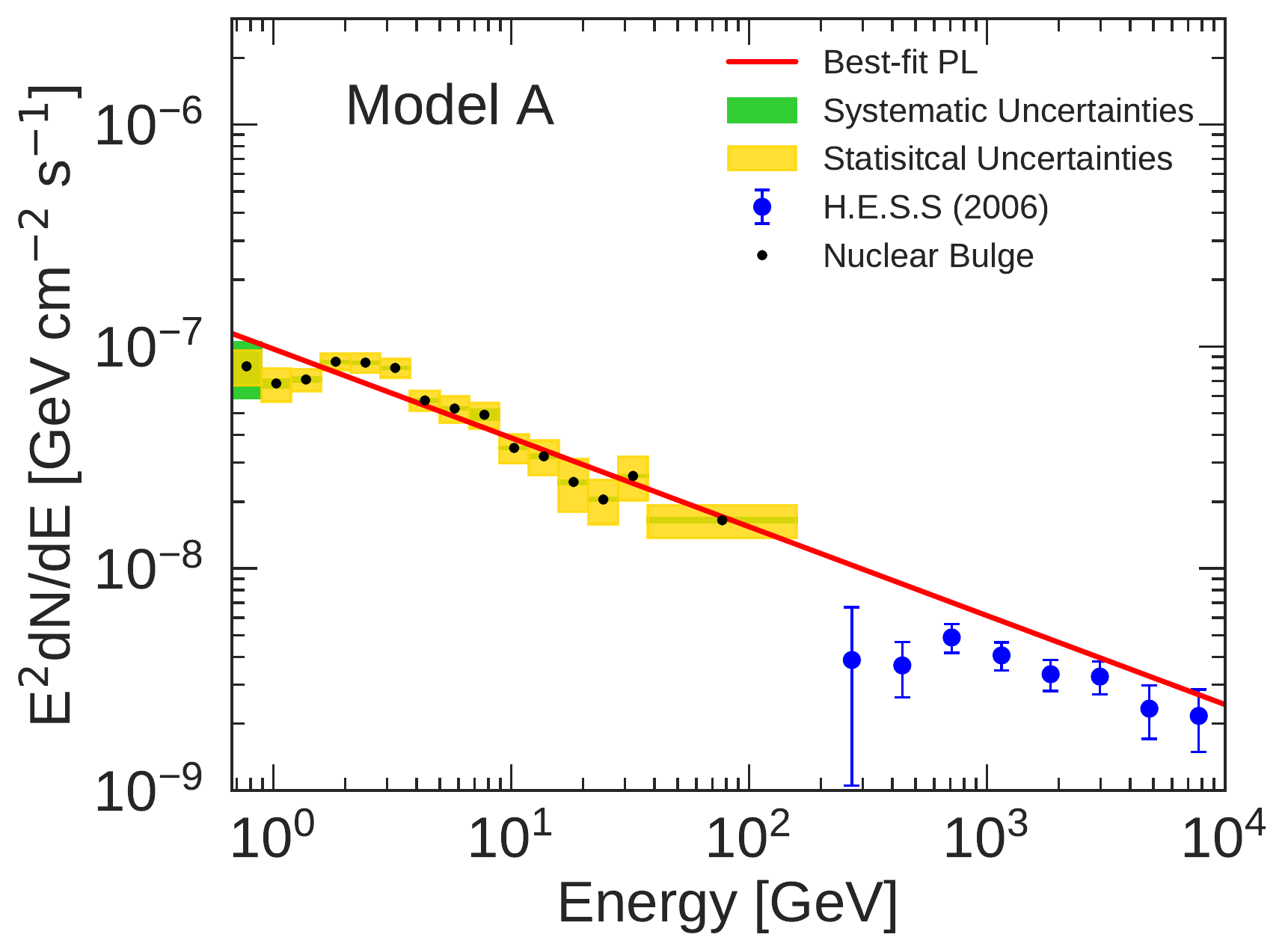} & \includegraphics[scale=0.43]{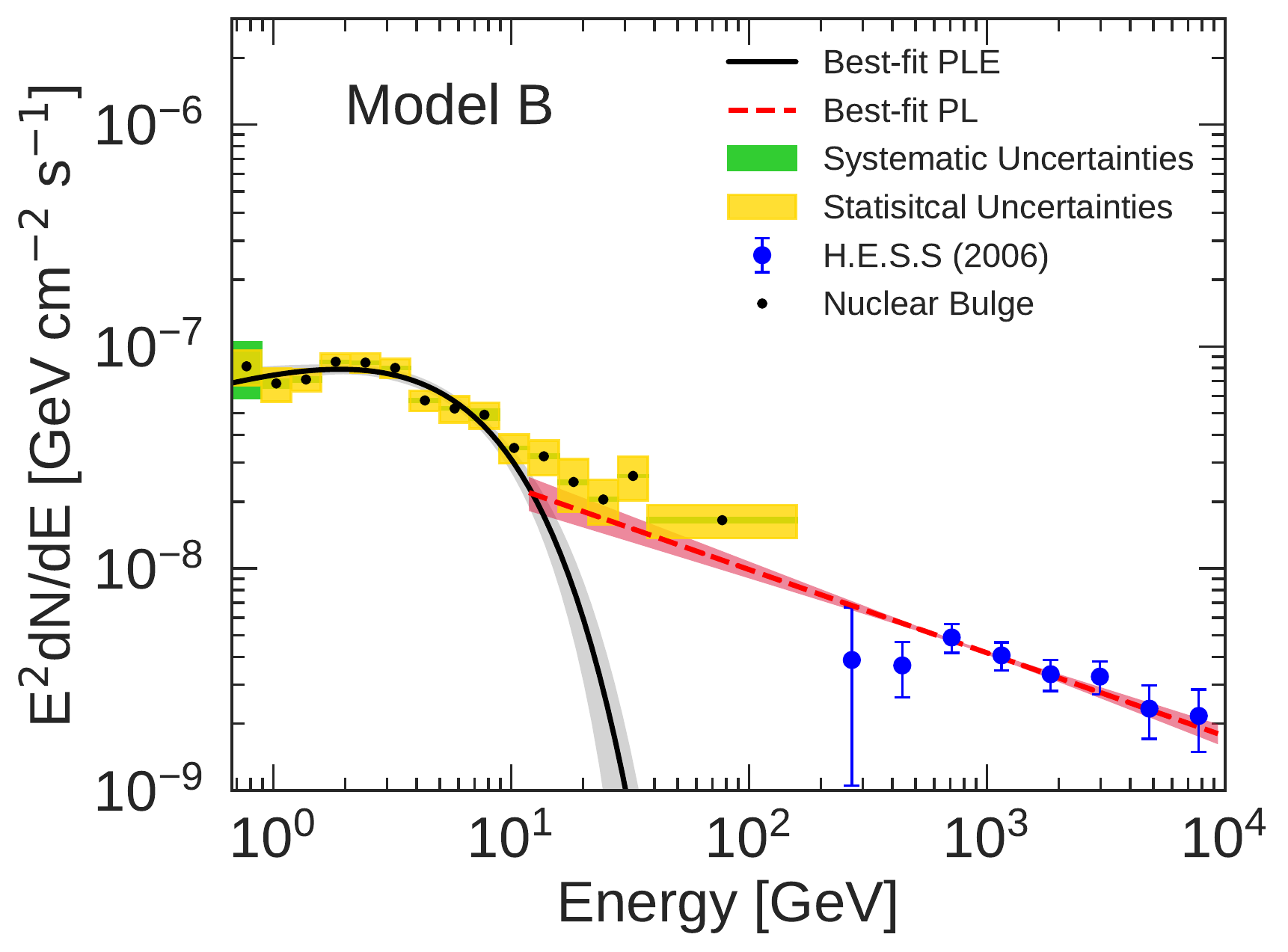}
\end{tabular}
\caption{\label{Fig:NuclearbulgeSpectrum} \textbf{Nuclear bulge spectrum:} Shown are the nuclear bulge fluxes (black points) and corresponding statistical (yellow) and systematic (green) errors bars. The blue data points correspond to H.E.S.S.~\cite{Aharonian:2006au} data from the Galactic ridge region. \textit{Model A:} A single power-law (PL) fit to the \textit{Fermi} gamma-ray data from the nuclear bulge gives an acceptable fit ($\chi_{\rm PL}^2=22.7$ for 15-2 dof, that is, $\mbox{p-value}=0.045$) with best-fit slope of $\alpha=2.40\pm 0.03$ and norm $N_0=(9.8\pm 2.4)\times 10^{-11}$ MeV$^{-1}$ cm$^{-2}$ s$^{-1}$. However, we observe that when this PL fit is extrapolated to higher energies it overshoots the H.E.S.S.~measurements from the same region. \textit{Model B:} We have thus included the H.E.S.S.~data points to the $\chi^2$ test, and obtained $\chi^2_{\rm PL+PLE}=19.2$ for 15+8-2 dof, that is, $\mbox{p-value}=0.58$.  A two-component model was preferred to a single power-law with a confidence of $3.4\sigma$ ($\Delta \chi^2=\chi^2_{\rm PL}-\chi^2_{\rm PL+PLE}$=17.7 for four extra parameters).  The best-fit spectrum is given by $dN/dE = N_0\; E^{-\Gamma}\exp(-E/E_{\rm cut})$ with $N_0=(9.0\pm 0.6)\times 10^{-11}$ MeV$^{-1}$ cm$^{-2}$ s$^{-1}$, $\Gamma=1.6\pm 0.2$ and $E_{\rm cut}=5.3\pm 1.4$ GeV in addition to $dN/dE = N_0\; E^{-\alpha}$ with $N_0=(5.6\pm 1.5)\times 10^{-11}$ MeV$^{-1}$ cm$^{-2}$ s$^{-1}$, $\alpha=2.4\pm 0.1$ for $E\geq 11$ GeV.    Shaded regions represent the uncertainties on the best-fit model parameters.}
\end{figure}

The quality of the fit to the inner $40^\circ\times 40^\circ$ of the GC is illustrated in Fig.~\ref{fig:Residuals}. The different columns show the data, best-fitting model and fractional residuals $(Data-Model)/Model$, respectively. 
The rows show different energy ranges, as labeled above the panels. All maps were 
smoothed using a gaussian filter of $0.5^\circ$ radius for display purposes. The first two columns demonstrate that there is in general a good agreement between the model and the data. As can be seen from the third column, our model underpredicts the data at the $\lesssim 20\%$ level for energy bins  $[0.6,1.1]$, $[1.1,2.8]$ and $[2.8,11.8]$ GeV. The level of agreement of the model with the data is downgraded to the $\sim 30\%$ level for energies $[11.8,158.1]$ GeV. It is possible that this enhanced residual emission is due to an imperfect FB template for this particular energy range. We observe that the authors of Ref.\ \cite{TheFermi-LAT:2017vmf} derived their FB template using gamma-ray data with energies between $[1,10]$ GeV. Application of their technique may be capable of better accommodating the small-scale features of the data in the $[11.8,158.1]$ GeV energy range. Similarly to what is shown in the second column of Fig.~\ref{fig:Residuals}, we also show in Fig.~\ref{fig:fluxprofiles} a comparison of the data and the best fit model components in a way that the relative importance of each of the components included in the fit can immediately be seen. Both these figures show that our model provides a good fit to the data.

A fit to the nuclear bulge template preferred a two component spectral model over a single one. Initially, we attempted to fit a PL to only the \textit{Fermi}-LAT data taken from the nuclear bulge. We found a best-fit slope of $\alpha=2.40\pm 0.03$ and normalization $N_0=(9.8\pm 2.4)\times 10^{-11}$ MeV$^{-1}$ cm$^{-2}$ s$^{-1}$ with $\mbox{p-value}=0.045$ ($\chi^2=22.7$ for $15-2$ degrees of freedom). Although this p-value passes the usual acceptance condition $\mbox{p-value}>0.001$, when the resulting PL formula is extrapolated to higher energies, it clearly overshoots the H.E.S.S.~measurements from the same region (left-hand side panel of Fig.~\ref{Fig:NuclearbulgeSpectrum}). We also tested a PLE spectrum and obtained $\mbox{p-value}=0.0003$ ($\chi^2_{\rm PLE} = 36.7$ for $15-3$ degrees of freedom ), which is less than the usual acceptance value.

Next, we included the H.E.S.S.~data points~\cite{Aharonian:2006au} to our $\chi^2$ test so as to make the model consistent with higher energy data from the same region. We therefore opted to add a PLE spectral model to account for data at around $\sim 1$ GeV and a PL model for $E\geq11$ GeV. This composite model consisting of a  PLE+PL was preferred to a single PL with $3.4\sigma$ statistical significance ($\Delta \chi^2=\chi^2_{\rm PL}-\chi^2_{\rm PLE}$=17.7 for four extra parameters). Best fitting spectral parameters corresponding to the PLE component were found to be $\Gamma=1.6\pm0.2$, $E_{\rm cut}=5.3\pm 1.4$ GeV and $L=(3.9\pm 0.5)\times 10^{36}$ erg s$^{-1}$ calculated in the energy range $E=[100,500000]$ MeV. As for the PL component we found $N_0=(5.6\pm 1.5)\times 10^{-11}$ MeV$^{-1}$ cm$^{-2}$ s$^{-1}$, $\alpha=2.4\pm 0.1$ for $E\geq 11$ GeV~\footnote{This value is motivated by the known typical $E_{\rm cut}$ of the resolved MSPs. }. While the H.E.S.S ROI ($|l|<0.8^\circ$ and $|b|<0.3^\circ$) is indeed close to that of the nuclear bulge template, the H.E.S.S ROI could be slightly smaller. A dedicated observation of exactly the nuclear bulge region with H.E.S.S or the forthcoming Cherenkov Telescope Array (CTA) could make the consistency across the full gamma-ray energy range much better. Similarly to the boxy bulge component, the best-fit PLE parameters are consistent with spectrum of resolved MSPs. These results are summarized in the right-hand side panel of Fig.~\ref{Fig:NuclearbulgeSpectrum}. 

A simple astrophysical model that can match the characteristics of the preferred PLE+PL model is an unresolved population of MSPs in the nuclear bulge in addition to a population of energetic electrons from star formation activity and the same MSPs. Prompt gamma-rays from MSPs could account for the PLE spectrum seen at \textit{Fermi}-LAT energies and a population of electrons injected from the same region could explain both the high-energy tail observed at energies $>10$ GeV \textit{and} the H.E.S.S.~Galactic ridge spectrum (see for example Ref.~\cite{Guepin:2018jkb} for a variation of this picture). Hadronic gamma-rays are also expected to contribute to the Galactic ridge at some level. A combination of hadronic and leptonic scenarios for this region have been explored in Ref.~\cite{Macias:2014sta}. However, our new results for the nuclear bulge region motivates a much more detailed modeling that self-consistently accounts for hadronic gamma-rays from the CMZ region as well as a new population of electrons from MSPs and star formation activity. This particular topic will be addressed in future work.

It is useful to compare the mass-to-light ratios of our hypothesized boxy bulge and nuclear bulge MSP populations with others residing in different systems\footnote{We caution that this comparison assumes that the ratio of MSP mass to stellar mass is constant across the different regions considered, which can only be approximately correct.}. The stellar mass of the Galactic bulge is estimated to be around $1.5 \times 10^{10}$ solar masses ($M_{\odot}$) and for the nuclear bulge $1.4 \times 10^9$ $M_{\odot}$. Bringing together the bulge mass estimates with our best-fit luminosities of the boxy bulge and nuclear bulge components, the luminosity per stellar mass for the boxy bulge is found to be $(1.5 \pm 0.3)\times10^{27}$ erg s$^{-1}$ $M_{\odot}^{-1}$ while for the nuclear bulge it is found to be $(2.8 \pm 0.4)\times10^{27}$ erg s$^{-1}$ $M_{\odot}^{-1}$. Thus, we find a luminosity per stellar mass for the nuclear bulge that is $\sim 2$ times higher than for the boxy bulge. This is qualitatively consistent with a picture in which the MSPs of the boxy bulge are systematically older than those in the nuclear bulge, and hence less luminous per stellar mass. The higher luminosity per stellar mass of the nuclear bulge might also be due to additional energy being injected from star formation activity on top of an old MSP population.

\section{Discussion and conclusions}\label{sec:discussion}

In this work we have performed a detailed reevaluation of the systematic uncertainties affecting the properties of the Galactic Center Excess (GCE). We have utilized hydrodynamical interstellar gas and dust maps~\cite{Macias:2016nev} in order to account for uncertainties associated with gas-correlated gamma-ray emission. In addition, we have employed, for the first time, the new 3D inverse-Compton (IC) maps\footnote{Some of
the new 3D IC maps dramatically improve the total log-likelihood of baseline model. This demonstrates the great constraining potential of the \textit{Fermi}-LAT data from the inner Galaxy for constructing improved CRs and ISRF models.} that abandon the assumption of Galactocentric cylindrical symmetry adopted in most previous analyses of the same region. Improved low-latitude \textit{Fermi} bubbles maps have also been included. By considering this wide range of maps and uncertainties, we have confirmed recent studies~\cite{Macias:2016nev,Bartels2017} that reported a positive correlation between the stellar distribution of the Galactic bulge and the GCE.

A recent study by the \textit{Fermi} collaboration~\cite{TheFermi-LAT:2015kwa} has raised some concerns with regard to whether the GCE is the result of incomplete modeling of the IC component in the inner Galaxy. However, our results reveal that such concerns are unwarranted. Even though the new 3D IC maps~\cite{Porter:2017vaa} include the bulge/bar for the CR source and ISRF densities, an additional stellar bulge component in the form of the nuclear bulge + boxy bulge is still required by the \textit{Fermi}-LAT data. This is consistent with the hypothesis of an unresolved population of MSPs in the Galactic Center region. Since CRs interacting with interstellar gas and the ISRF produce IC morphologies at $\sim 1$ GeV that are different to  both those of the CR source distribution and the target material, statistical analyses of the GCE would still be required to add template model maps to account for the prompt gamma-ray emission from MSPs that closely match their source distribution.

\begin{figure}[t!]
\centering
\includegraphics[scale=0.5]{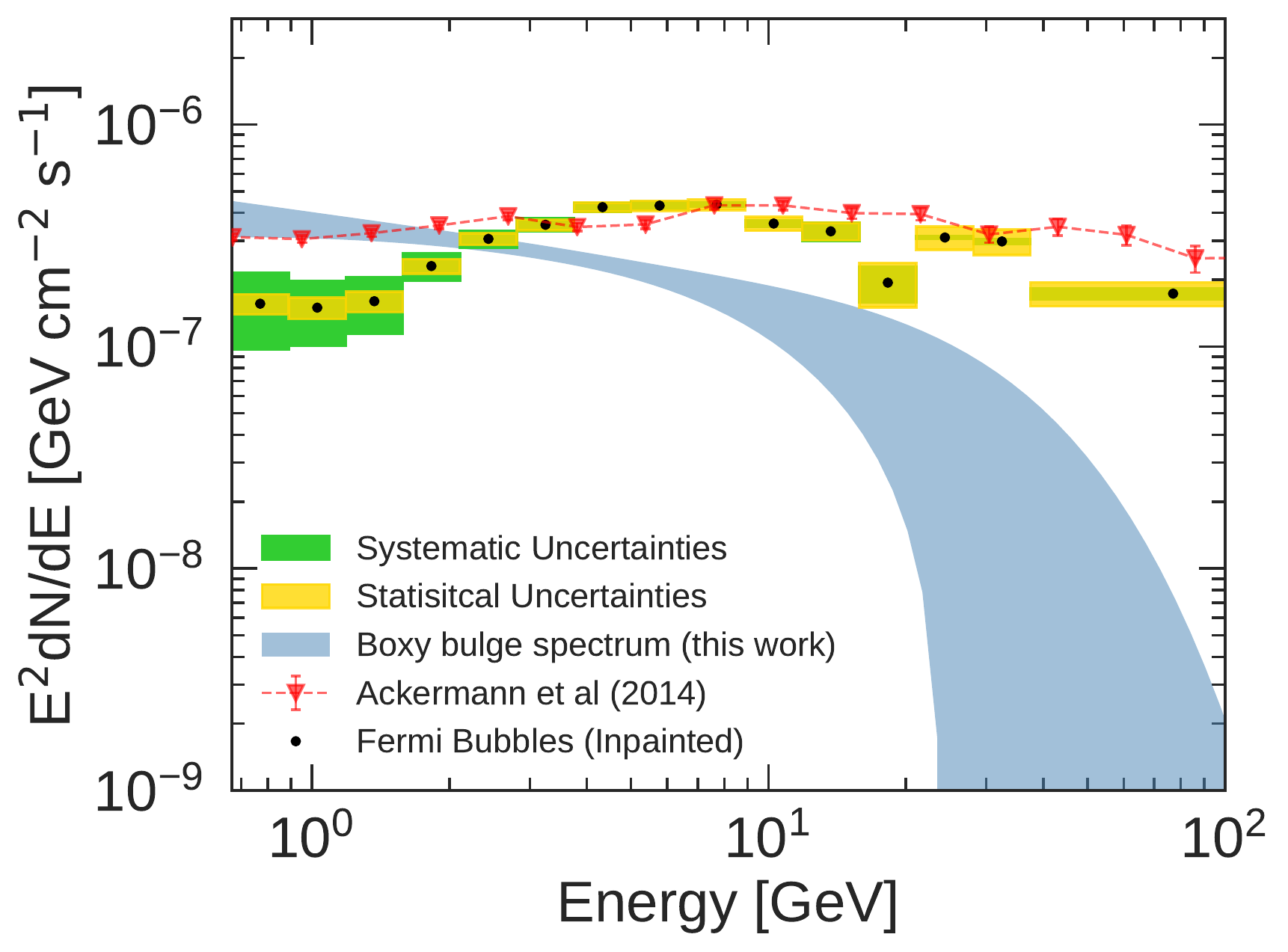} \caption{\label{Fig:FermiBubblesSpectrum} \textbf{Fermi Bubbles spectrum:} Inferred bin-by-bin fluxes (black points) and corresponding statistical (yellow) and systematic (green) error bars. The systematic errors show the flux dispersion when the maximum-likelihood analyses are repeated using the alternative 3D and 2D IC maps presented in Ref.~\cite{Porter:2017vaa}. For comparison purposes, we show the butterfly plot of the boxy bulge spectrum of our Fig.~\ref{Fig:BoxybulgeSpectrum}. The red points are  the overall high-latitude ($|b|>10^\circ$) flux extracted from Fig.~18 of Ref.~\cite{FermilatBubblespaper}.}
\end{figure}

Although the X-shape~\cite{Ness:2016aaa} structure of the Galactic bulge is estimated to contain some $\sim 40$\% of the mass of the bulge stars, these stars spend only a small fraction of their orbital time within the X arms~\cite{Bland-Hawthorn2016,Nataf:2017}. Therefore, at any given time, about 5--10\% of bulge stars are observed to be within the X arms~\cite{Cao:2013dwa,Nataf:2017,Simionetal:2017}. Under the assumption that gamma-ray sources are kinematically similar to stars, we would therefore expect that the X arms would contribute $\sim 10\%$ of the Galactic bulge gamma-ray luminosity. In the maximum-likelihood analyses presented in Table~\ref{Tab:GCElikelihoods}, we found that there is statistical evidence for the X-shape bulge in gamma-rays, but that this was weaker than that of the whole boxy bulge structure. In our step-wise statistical procedure, only new sources that increased the log-likelihood the most were allowed to be included in the subsequent baseline model.

In Ref.~\cite{Macias:2016nev}, using a RoI of size $15^\circ \times 15^\circ$ and only the standard 2D IC maps, some of us showed that the GCE was better fit by either the nuclear bulge + X-bulge or nuclear bulge + boxy bulge. In that work both stellar templates were found to have similar statistical significances. Now with our larger RoI and improved IC maps used in this analysis, we are finding that the significance of the nuclear bulge + X-bulge is smaller than that for the nuclear bulge + boxy bulge template. However, we note that it is not at present possible to nest the boxy bulge template with the X-bulge template in the same maximum-likelihood fit, which is necessary for a robust comparison. This is because the boxy bulge model already contains a substantial fraction---if not all---of the X-bulge stars. Our new results strongly encourage a careful revaluation of the luminosities associated to the X-bulge and boxy bulge stellar components using stellar maps that are physically self-consistent.
\begin{figure}[t!]
\centering
\begin{tabular}{cc}
\includegraphics[scale=0.43]{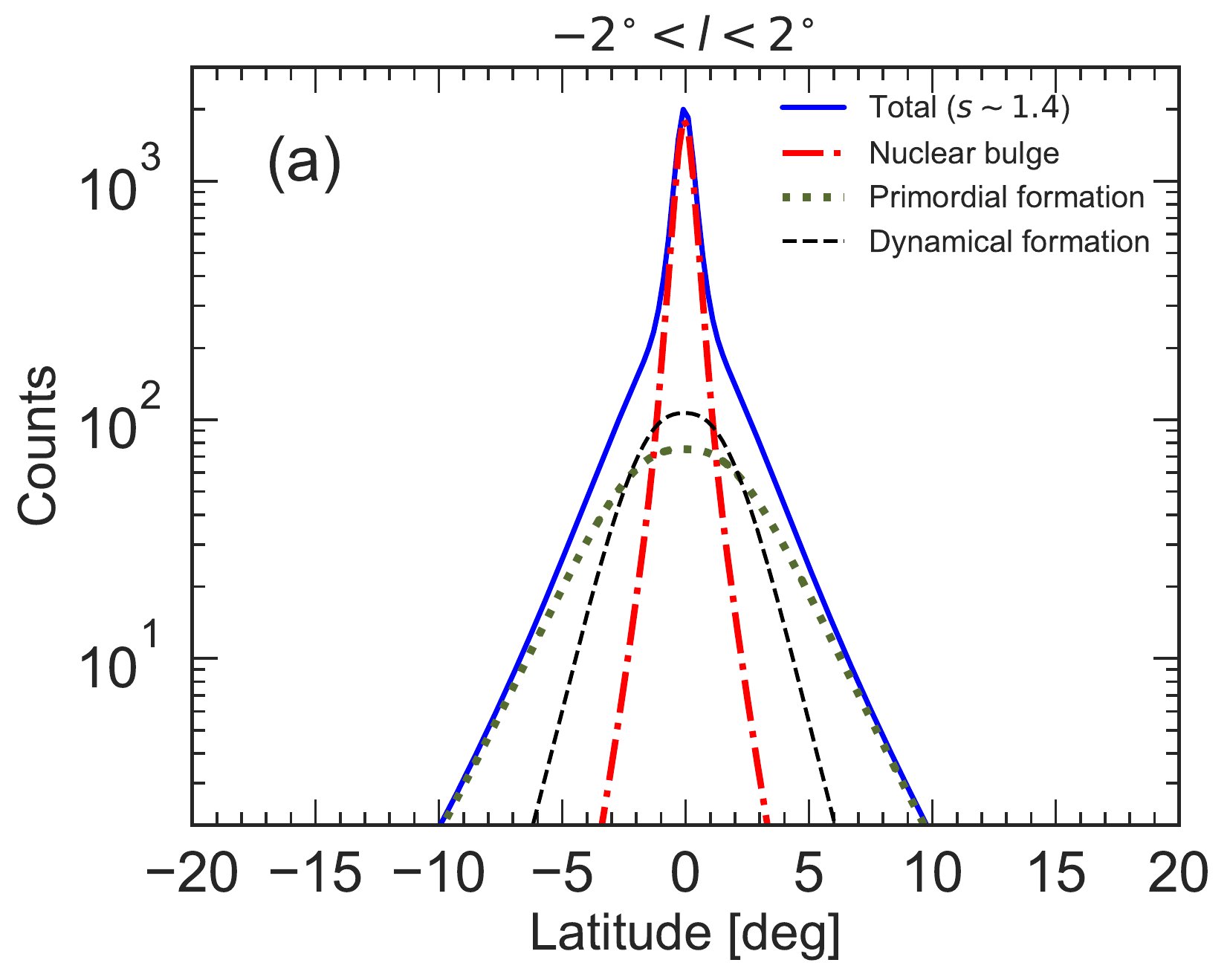} & \includegraphics[scale=0.43]{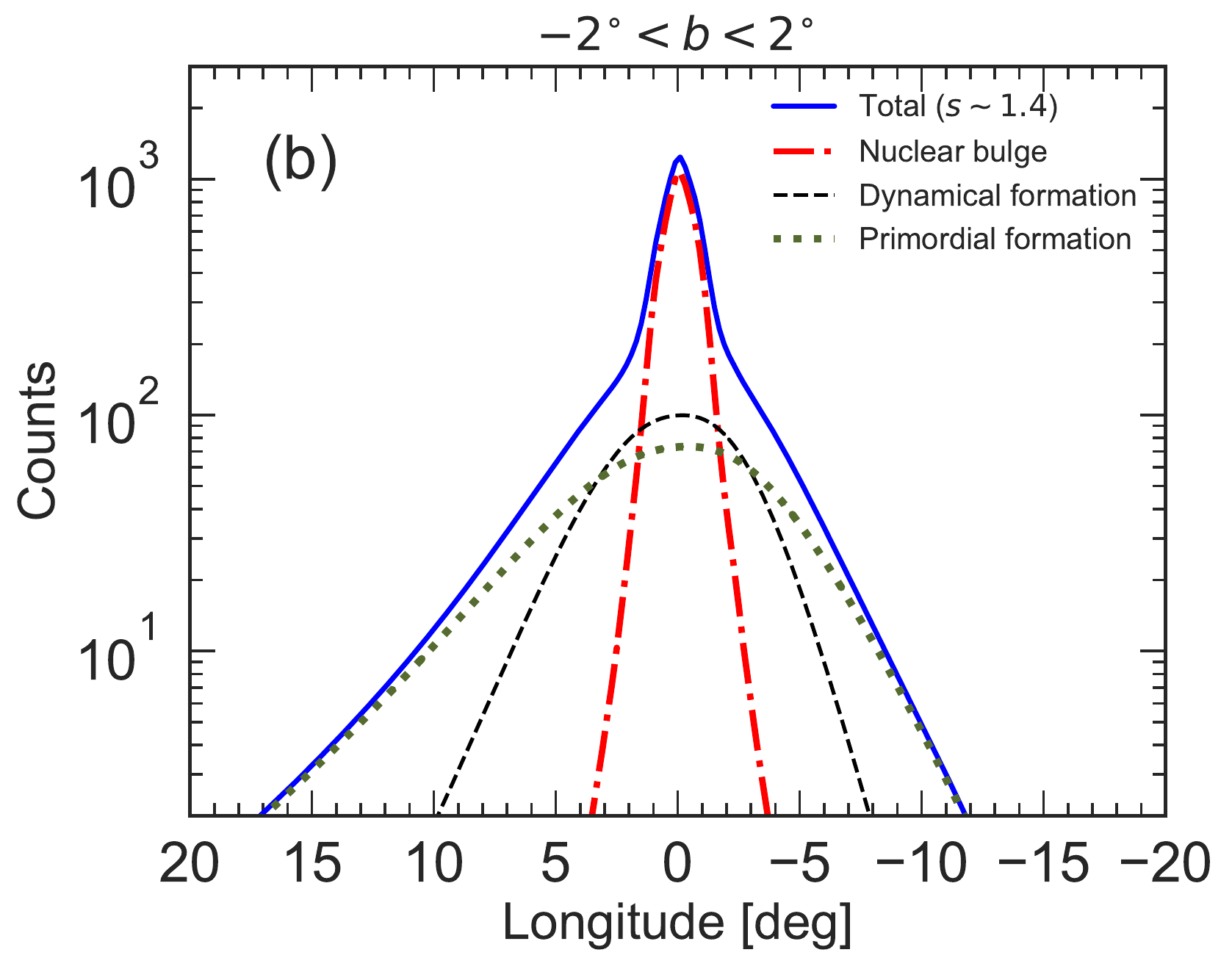} 
\end{tabular}
\caption{\label{Fig:primordialfraction} \textbf{Best-fit emission generated by MSPs formed through the dynamical and primordial formation channels:} See also Fig.~\ref{Fig:BPslopeprofile} for descriptions.  Panels (a) and (b) show the relative contribution of the primordial and dynamical MSPs formation channel. This was obtained by simultaneously including in the fit a $\rho^1_{\rm bar} (R,\phi,z)$ and $\rho^2_{\rm bar} (R,\phi,z)$ map and then estimating their corresponding best-fit gamma-ray luminosities. When the flux fractions calculated this way are superimposed, the total flux profile is approximately described by $\rho^{1.4}_{\rm bar} (R,\phi,z)$, in agreement with results shown in Fig.~\ref{Fig:BPslopeprofile}. The red solid, gren dotted and black dashed lines show the best-fit nuclear bulge, primordial formation and  dynamical formation fractions, respectively. We find that primordially formed MSPs contribute $52\% \pm 21\%$ of the Galactic bulge luminosity. }
\end{figure}

In order to test whether the stellar models can be improved by allowing more flexibility to the templates, we rotated the boxy bulge template around the central position in increments of $9^\circ$ and computed the log-likelihood at each orientation. The results from this test indicate that the boxy bulge in its original orientation is consistent and that the data is sufficiently constraining to differentiate the East-West and North-South asymmetries present in the stellar bulge model (see Fig.~\ref{Fig:Rotation}).

Similarly, with the aim of testing the GCE for any possible translational shifts with respect to the location of the stellar bulge, we have moved the central position of the boxy bulge along latitude and longitude in steps of $0.4^\circ$. Statistically significant shifts were found toward positive latitudes at $b=0.72^{\circ}\pm 0.07^\circ$ and negative longitudes at $l=-1.69^{\circ} \pm 0.12^{\circ}$ (see Figs.~\ref{Fig:translationLon} and \ref{Fig:translationLat}). The former is relatively small and is likely due to a positive excess resulting from the suppression of a cluster of point sources in the 2FIG~\cite{Fermi-LAT:2FIG} catalog around the region $l=0^\circ$ and $b\sim[0^\circ,2^\circ]$. However, the later prefers a shifted center at larger angular distance and increased statistical confidence. This result is in line with a similar such analysis performed with the red clump giant stellar data as seen in Table 1 of Ref.~\cite{Cao:2013dwa}, where a Galactic westward translation of around 0.3$^\circ$ was found. Although the preferred shifted positions in the two studies differ by $\Delta l\sim 1.4^\circ$, it is worth noting that in the region around $b\sim 0^\circ$ and $l\sim [-0^\circ,-2^\circ]$ there is a population of young massive stellar objects that might be emitting significant gamma-ray flux which, by construction, the analysis of Ref.~\cite{Cao:2013dwa} does not account for. This is because the red clump giants are systematically older stellar objects. In our view the results of these empirical tests emphasize the validity of the proposal for a stellar origin for the GCE, within the systematic limitations of the current maps. In summary, we find that the gamma-ray data highly prefers a Galactic bulge model that is consistently positioned and aligned within one or two degrees of the known Galactic bulge~\cite{Freudenreich:1997bx}. All other translational and rotational variations result in significantly poorer fits to the data.   

Perhaps the most striking result to emerge from the GCE data is that any model predicting a spherically symmetric morphology is robustly ruled out (see Fig.~\ref{Fig:NFWsignificance}). This, of course, includes the annihilation of weakly-interacting dark matter models but also current versions of the disrupted globular cluster scenario~\cite{Gnedin:2013cda,Brandt:2015ula,Fragione:2017rsp}. Any successful disrupted globular cluster model must explain both the spherical distribution of surviving globular clusters in the Galaxy as well as the boxy/peanut distribution of MSPs in the bulge. Taken together, these two observations seem to disfavor this scenario (at least in its current form~\cite{Gnedin:2013cda,Brandt:2015ula,Fragione:2017rsp}) as the main formation mechanism for MSPs in the Galactic bulge.

We have also tested the primordial and dynamical formation scenarios for the Galactic bulge MSPs. In our detailed morphological analysis of the GCE, we have found a significant statistical preference ($7.6\sigma$) for an \textit{admixture formation} scenario in which a fraction of the MSPs are formed through the primordial channel ($52\% \pm 23\%$) and  the remaining one through the dynamical channel. Due to systematic uncertainties in the Galactic diffuse emission model, our maximum-likelihood method assigns almost identical log-likelihood values to a pure primordial formation as well as to a pure dynamical formation scenario (see Fig.~\ref{Fig:BPslopeprofile}). However, models whose spatial morphology start to depart from those predicted by these two scenarios are statistically disfavored by the data. This is an important check that lends further support to the MSP theory for the GCE because we expect both components to exist in the MSP population~\cite{Voss:2006az}.

Reference~\cite{Eckner:2017oul} used the results of a correlation between globular cluster gamma-ray luminosity and the stellar encounter rate~\cite{Huietal:2010} to infer the energetics and morphology of the dynamical formation scenario for MSPs in the Galactic bulge. That study predicts a contribution to the bulge luminosity relative to the primordial gamma-ray luminosity at the percent level ($\lesssim 5\%$). In contrast, using our maximum-likelihood procedure we find that the best-fit contribution for this channel is $\sim 48\%\pm 21\%$ (see Fig.~\ref{Fig:primordialfraction}).  This was obtained by simultaneously including in the fit a $\rho^1_{\rm bar} (R,\phi,z)$ and $\rho^2_{\rm bar} (R,\phi,z)$ map and then estimating their corresponding best-fit gamma-ray luminosities\footnote{The fractional luminosity error was obtained by adding in quadrature the inferred energy fluxes at each energy bin. We also note that in this case there are 15 new degrees of freedom (one for each band of the $\rho_{\rm bar}^2$ template) while in estimating $s$ there was only one new degree of freedom.}. 
 Figure~\ref{Fig:primordialfraction} shows how the flux fractions computed this way adds up to approximately our best-fit stellar density slope $s\sim 1.4$ obtained with the slope $s$ scan of Fig.~\ref{Fig:BPslopeprofile}. Although our detailed morphological analysis does not seem to corroborate the luminosity fraction predicted by Ref.~\cite{Eckner:2017oul}, we caution that a more direct comparison with that study is not at present possible given that the authors have assumed a spherically symmetric model (see Sec.~3 in~\cite{Eckner:2017oul}) for the distribution of the bulge stars. Such stellar distributions are thought to be not realistic~\cite{Bland-Hawthorn2016,Nataf:2017} and are robustly ruled out by the gamma-ray data~\cite{Macias:2016nev,Bartels2017}.

In agreement with most previous analyses of the GCE~\cite{Abazajian:2012pn,Gordon:2013vta,Macias:2013vya,Huang:2013pda,Abazajian:2014fta,Abazajian:2014hsa,Calore:2014xka,Daylan:2014rsa,TheFermi-LAT:2015kwa}, we have found that the spectrum of the boxy bulge is consistent with that expected from an unresolved  population of MSPs in the Galactic bulge (see Fig.~\ref{Fig:BoxybulgeSpectrum}). In particular, we confirm earlier results~\cite{Linden:2016rcf} that claimed the detection for a high energy component above $\sim 10$ GeV associated to the GCE. As can be seen in Fig.~\ref{Fig:FermiBubblesSpectrum}, this high-energy tail has a different spectrum from that of the \textit{Fermi} Bubbles spectrum. Such a component could be the result of IC emission of electrons and positrons injected by the same Galactic bulge MSP population~\cite{Linden:2016rcf}.

Furthermore, we identified a corresponding high-energy tail also in the nuclear bulge template (Fig.~\ref{Fig:NuclearbulgeSpectrum}). As far as we are aware, this is the first time that a high-energy tail has been detected in this smaller angular region. A fit to the \textit{Fermi} and H.E.S.S.~observations from this site preferred a two component spectral model (although with a low confidence of $3.5\sigma$) comprising a power-law with an exponential cutoff model plus a simple power-law for energies $\gtrsim 10$ GeV. An important implication of this result is that energetic electrons
emitted by MSPs, and possibly star forming activity, could potentially contribute a significant fraction of the TeV gamma-rays detected by H.E.S.S.

Our results show that the Galactic Center Excess in gamma rays is correlated with the bulge of the Milky Way. The most likely explanation for this correlation is that the unresolved bulge millisecond pulsars are responsible for the Galactic Center Excess. 

\vspace{2mm}

\begin{acknowledgments}
We are grateful to Kevork Abazajian, Shin'ichiro Ando, Anthony M.~Brown, Henry Freudenreich, Ryan Keeley, Dmitry Malyshev, Harrison Ploeg, Troy Porter, Nicholas Rodd, Tracy Slatyer, and Deheng Song for useful comments and discussions during the project. This work was supported by World Premier International Research Center Initiative (WPI Initiative), MEXT, Japan. OM was partially supported by JSPS KAKENHI Grant Numbers JP17H04836, JP18H04340 and JP18H04578. SH thanks support by the U.S.~Department of Energy under award number de-sc0018327. MK was supported by NSF Grant No. PHY1620638. DMN was supported by the Allan C. and Dorothy H. Davis Fellowship. OM acknowledges the high performance computer center at Kavli IPMU for providing computational resources and specially to Tatsuhiro Yano for technical support that have contributed to the results reported in this paper. 
\end{acknowledgments}

\bibliography{references}

\providecommand{\href}[2]{#2}\begingroup\raggedright\begin{thebibliography}{10}

\bibitem{Genzel:2010zy}
R.~Genzel, F.~Eisenhauer, and S.~Gillessen, {\it {The Galactic Center Massive
  Black Hole and Nuclear Star Cluster}},  {\em Rev. Mod. Phys.} {\bf 82} (2010)
  3121--3195, [\href{http://arxiv.org/abs/1006.0064}{{\tt arXiv:1006.0064}}].

\bibitem{Krumholz:2015aa}
M.~Krumholz and D.~Kruijssen, {\it {A dynamical model for the formation of gas
  rings and episodic starbursts near galactic centres}},  {\em Mon. Not. Roy.
  Astron. Soc.} {\bf 453} (2015) 739,
  [\href{http://arxiv.org/abs/1505.07111}{{\tt arXiv:1505.07111}}].

\bibitem{Ponti:2015tya}
G.~Ponti et~al., {\it {The XMM–Newton view of the central degrees of the
  Milky Way}},  {\em Mon. Not. Roy. Astron. Soc.} {\bf 453} (2015), no.~1
  172--213, [\href{http://arxiv.org/abs/1508.04445}{{\tt arXiv:1508.04445}}].

\bibitem{Muno:2003ne}
M.~P. Muno, F.~K. Baganoff, M.~W. Bautz, W.~N. Brandt, P.~S. Broos, E.~D.
  Feigelson, G.~P. Garmire, M.~R. Morris, G.~R. Ricker, and L.~K. Townsley,
  {\it {A deep Chandra catalog of x-ray point sources toward the Galactic
  Center}},  {\em Astrophys. J.} {\bf 589} (2003) 225--241,
  [\href{http://arxiv.org/abs/astro-ph/0301371}{{\tt astro-ph/0301371}}].

\bibitem{Bertone:2004pz}
G.~Bertone, D.~Hooper, and J.~Silk, {\it {Particle dark matter: Evidence,
  candidates and constraints}},  {\em Phys. Rept.} {\bf 405} (2005) 279--390,
  [\href{http://arxiv.org/abs/hep-ph/0404175}{{\tt hep-ph/0404175}}].

\bibitem{Goodenough:2009gk}
L.~Goodenough and D.~Hooper, {\it {Possible Evidence For Dark Matter
  Annihilation In The Inner Milky Way From The Fermi Gamma Ray Space
  Telescope}},  \href{http://arxiv.org/abs/0910.2998}{{\tt arXiv:0910.2998}}.

\bibitem{Hooper:2010mq}
D.~Hooper and L.~Goodenough, {\it {Dark Matter Annihilation in The Galactic
  Center As Seen by the Fermi Gamma Ray Space Telescope}},  {\em Phys. Lett.}
  {\bf B697} (2011) 412--428, [\href{http://arxiv.org/abs/1010.2752}{{\tt
  arXiv:1010.2752}}].

\bibitem{Boyarsky:2010dr}
A.~Boyarsky, D.~Malyshev, and O.~Ruchayskiy, {\it {A comment on the emission
  from the Galactic Center as seen by the Fermi telescope}},  {\em Phys. Lett.}
  {\bf B705} (2011) 165--169, [\href{http://arxiv.org/abs/1012.5839}{{\tt
  arXiv:1012.5839}}].

\bibitem{Abazajian:2012pn}
K.~N. Abazajian and M.~Kaplinghat, {\it {Detection of a Gamma-Ray Source in the
  Galactic Center Consistent with Extended Emission from Dark Matter
  Annihilation and Concentrated Astrophysical Emission}},  {\em Phys. Rev.}
  {\bf D86} (2012) 083511, [\href{http://arxiv.org/abs/1207.6047}{{\tt
  arXiv:1207.6047}}]. [Erratum: Phys. Rev.D87,129902(2013)].

\bibitem{Gordon:2013vta}
C.~Gordon and O.~Macias, {\it {Dark Matter and Pulsar Model Constraints from
  Galactic Center Fermi-LAT Gamma Ray Observations}},  {\em Phys. Rev.} {\bf
  D88} (2013), no.~8 083521, [\href{http://arxiv.org/abs/1306.5725}{{\tt
  arXiv:1306.5725}}]. [Erratum: Phys. Rev.D89,no.4,049901(2014)].

\bibitem{Macias:2013vya}
O.~Macias and C.~Gordon, {\it {Contribution of cosmic rays interacting with
  molecular clouds to the Galactic Center gamma-ray excess}},  {\em Phys. Rev.}
  {\bf D89} (2014), no.~6 063515, [\href{http://arxiv.org/abs/1312.6671}{{\tt
  arXiv:1312.6671}}].

\bibitem{Huang:2013pda}
W.-C. Huang, A.~Urbano, and W.~Xue, {\it {Fermi Bubbles under Dark Matter
  Scrutiny. Part I: Astrophysical Analysis}},
  \href{http://arxiv.org/abs/1307.6862}{{\tt arXiv:1307.6862}}.

\bibitem{Abazajian:2014fta}
K.~N. Abazajian, N.~Canac, S.~Horiuchi, and M.~Kaplinghat, {\it {Astrophysical
  and Dark Matter Interpretations of Extended Gamma-Ray Emission from the
  Galactic Center}},  {\em Phys. Rev.} {\bf D90} (2014), no.~2 023526,
  [\href{http://arxiv.org/abs/1402.4090}{{\tt arXiv:1402.4090}}].

\bibitem{Abazajian:2014hsa}
K.~N. Abazajian, N.~Canac, S.~Horiuchi, M.~Kaplinghat, and A.~Kwa, {\it
  {Discovery of a New Galactic Center Excess Consistent with Upscattered
  Starlight}},  {\em JCAP} {\bf 1507} (2015), no.~07 013,
  [\href{http://arxiv.org/abs/1410.6168}{{\tt arXiv:1410.6168}}].

\bibitem{Calore:2014xka}
F.~Calore, I.~Cholis, and C.~Weniger, {\it {Background Model Systematics for
  the Fermi GeV Excess}},  {\em JCAP} {\bf 1503} (2015) 038,
  [\href{http://arxiv.org/abs/1409.0042}{{\tt arXiv:1409.0042}}].

\bibitem{Daylan:2014rsa}
T.~Daylan, D.~P. Finkbeiner, D.~Hooper, T.~Linden, S.~K.~N. Portillo, N.~L.
  Rodd, and T.~R. Slatyer, {\it {The characterization of the gamma-ray signal
  from the central Milky Way: A case for annihilating dark matter}},  {\em
  Phys. Dark Univ.} {\bf 12} (2016) 1--23,
  [\href{http://arxiv.org/abs/1402.6703}{{\tt arXiv:1402.6703}}].

\bibitem{TheFermi-LAT:2015kwa}
{\bf Fermi-LAT} Collaboration, M.~Ajello et~al., {\it {Fermi-LAT Observations
  of High-Energy $\gamma$-Ray Emission Toward the Galactic Center}},  {\em
  Astrophys. J.} {\bf 819} (2016), no.~1 44,
  [\href{http://arxiv.org/abs/1511.02938}{{\tt arXiv:1511.02938}}].

\bibitem{Abazajian:2010zy}
K.~N. Abazajian, {\it {The Consistency of Fermi-LAT Observations of the
  Galactic Center with a Millisecond Pulsar Population in the Central Stellar
  Cluster}},  {\em JCAP} {\bf 1103} (2011) 010,
  [\href{http://arxiv.org/abs/1011.4275}{{\tt arXiv:1011.4275}}].

\bibitem{OLeary:2015qpx}
R.~M. O'Leary, M.~D. Kistler, M.~Kerr, and J.~Dexter, {\it {Young Pulsars and
  the Galactic Center GeV Gamma-ray Excess}},
  \href{http://arxiv.org/abs/1504.02477}{{\tt arXiv:1504.02477}}.

\bibitem{OLeary:2016cwz}
R.~M. O'Leary, M.~D. Kistler, M.~Kerr, and J.~Dexter, {\it {Young and
  Millisecond Pulsar GeV Gamma-ray Fluxes from the Galactic Center and
  Beyond}},  \href{http://arxiv.org/abs/1601.05797}{{\tt arXiv:1601.05797}}.

\bibitem{Gnedin:2013cda}
O.~Y. Gnedin, J.~P. Ostriker, and S.~Tremaine, {\it {Co-Evolution of Galactic
  Nuclei and Globular Cluster Systems}},  {\em Astrophys. J.} {\bf 785} (2014)
  71, [\href{http://arxiv.org/abs/1308.0021}{{\tt arXiv:1308.0021}}].

\bibitem{Brandt:2015ula}
T.~D. Brandt and B.~Kocsis, {\it {Disrupted Globular Clusters Can Explain the
  Galactic Center Gamma Ray Excess}},  {\em Astrophys. J.} {\bf 812} (2015),
  no.~1 15, [\href{http://arxiv.org/abs/1507.05616}{{\tt arXiv:1507.05616}}].

\bibitem{Cholis:2015dea}
I.~Cholis, C.~Evoli, F.~Calore, T.~Linden, C.~Weniger, and D.~Hooper, {\it {The
  Galactic Center GeV Excess from a Series of Leptonic Cosmic-Ray Outbursts}},
  {\em JCAP} {\bf 1512} (2015), no.~12 005,
  [\href{http://arxiv.org/abs/1506.05119}{{\tt arXiv:1506.05119}}].

\bibitem{Cholis:2014lta}
I.~Cholis, D.~Hooper, and T.~Linden, {\it {Challenges in Explaining the
  Galactic Center Gamma-Ray Excess with Millisecond Pulsars}},  {\em JCAP} {\bf
  1506} (2015), no.~06 043, [\href{http://arxiv.org/abs/1407.5625}{{\tt
  arXiv:1407.5625}}].

\bibitem{Hooper:2015jlu}
D.~Hooper and G.~Mohlabeng, {\it {The Gamma-Ray Luminosity Function of
  Millisecond Pulsars and Implications for the GeV Excess}},  {\em JCAP} {\bf
  1603} (2016), no.~03 049, [\href{http://arxiv.org/abs/1512.04966}{{\tt
  arXiv:1512.04966}}].

\bibitem{Ploeg:2017vai}
H.~Ploeg, C.~Gordon, R.~Crocker, and O.~Macias, {\it {Consistency Between the
  Luminosity Function of Resolved Millisecond Pulsars and the Galactic Center
  Excess}},  {\em JCAP} {\bf 1708} (2017), no.~08 015,
  [\href{http://arxiv.org/abs/1705.00806}{{\tt arXiv:1705.00806}}].

\bibitem{Bartels:2018xom}
R.~T. Bartels, T.~D.~P. Edwards, and C.~Weniger, {\it {Bayesian Model
  Comparison and Analysis of the Galactic Disk Population of Gamma-Ray
  Millisecond Pulsars}},  \href{http://arxiv.org/abs/1805.11097}{{\tt
  arXiv:1805.11097}}.

\bibitem{Lee:2015fea}
S.~K. Lee, M.~Lisanti, B.~R. Safdi, T.~R. Slatyer, and W.~Xue, {\it {Evidence
  for Unresolved $\gamma$-Ray Point Sources in the Inner Galaxy}},  {\em Phys.
  Rev. Lett.} {\bf 116} (2016), no.~5 051103,
  [\href{http://arxiv.org/abs/1506.05124}{{\tt arXiv:1506.05124}}].

\bibitem{Bartels:2015aea}
R.~Bartels, S.~Krishnamurthy, and C.~Weniger, {\it {Strong support for the
  millisecond pulsar origin of the Galactic center GeV excess}},  {\em Phys.
  Rev. Lett.} {\bf 116} (2016), no.~5 051102,
  [\href{http://arxiv.org/abs/1506.05104}{{\tt arXiv:1506.05104}}].

\bibitem{Horiuchi:2016zwu}
S.~Horiuchi, M.~Kaplinghat, and A.~Kwa, {\it {Investigating the Uniformity of
  the Excess Gamma rays towards the Galactic Center Region}},  {\em JCAP} {\bf
  1611} (2016), no.~11 053, [\href{http://arxiv.org/abs/1604.01402}{{\tt
  arXiv:1604.01402}}].

\bibitem{Linden:2016rcf}
T.~Linden, N.~L. Rodd, B.~R. Safdi, and T.~R. Slatyer, {\it {High-energy tail
  of the Galactic Center gamma-ray excess}},  {\em Phys. Rev.} {\bf D94}
  (2016), no.~10 103013, [\href{http://arxiv.org/abs/1604.01026}{{\tt
  arXiv:1604.01026}}].

\bibitem{Macias:2016nev}
O.~Macias, C.~Gordon, R.~M. Crocker, B.~Coleman, D.~Paterson, S.~Horiuchi, and
  M.~Pohl, {\it {Galactic bulge preferred over dark matter for the Galactic
  centre gamma-ray excess}},  {\em Nat. Astron.} {\bf 2} (2018), no.~5
  387--392, [\href{http://arxiv.org/abs/1611.06644}{{\tt arXiv:1611.06644}}].

\bibitem{Bartels2017}
R.~Bartels, E.~Storm, C.~Weniger, and F.~Calore, {\it {The Fermi-LAT GeV excess
  as a tracer of stellar mass in the Galactic bulge}},  {\em Nat. Astron.} {\bf
  2} (2018), no.~10 819--828, [\href{http://arxiv.org/abs/1711.04778}{{\tt
  arXiv:1711.04778}}].

\bibitem{Dwek:1995xu}
E.~Dwek, R.~G. Arendt, M.~G. Hauser, T.~Kelsall, C.~M. Lisse, S.~H. Moseley,
  R.~F. Silverberg, T.~J. Sodroski, and J.~L. Weiland, {\it {Morphology, near
  infrared luminosity, and mass of the galactic bulge from Cobe dirbe
  observations}},  {\em Astrophys. J.} {\bf 445} (1995) 716.

\bibitem{Freudenreich:1997bx}
H.~T. Freudenreich, {\it {Cobe's galactic bar and disk}},  {\em Astrophys. J.}
  {\bf 492} (1998) 495--510, [\href{http://arxiv.org/abs/astro-ph/9707340}{{\tt
  astro-ph/9707340}}].

\bibitem{LopezCorredoira:1999dg}
M.~Lopez-Corredoira, P.~L. Hammersley, F.~Garzon, E.~Simonneau, and T.~J.
  Mahoney, {\it {Inversion of stellar statistics equation for the galactic
  bulge}},  {\em Mon. Not. Roy. Astron. Soc.} {\bf 313} (2000) 392,
  [\href{http://arxiv.org/abs/astro-ph/9911182}{{\tt astro-ph/9911182}}].

\bibitem{Babusiaux:2005zi}
C.~Babusiaux and G.~Gilmore, {\it {The Structure of the Galactic bar}},  {\em
  Mon. Not. Roy. Astron. Soc.} {\bf 358} (2005) 1309--1321,
  [\href{http://arxiv.org/abs/astro-ph/0501383}{{\tt astro-ph/0501383}}].

\bibitem{Ness:2016aaa}
M.~Ness and D.~Lang, {\it {The X-shaped Bulge of the Milky Way revealed by
  WISE}},  {\em Astrophys. J.} {\bf 152} (2016), no.~1 14,
  [\href{http://arxiv.org/abs/1603.00026}{{\tt arXiv:1603.00026}}].

\bibitem{Garzon:1997cs}
F.~Garzon, M.~Lopez-Corredoira, P.~Hammersley, T.~J. Mahoney, X.~Calbet, and
  J.~E. Beckman, {\it {A major star formation region in the receding tip of the
  stellar galactic bar}},  {\em Astrophys. J.} {\bf 491} (1997) L31,
  [\href{http://arxiv.org/abs/astro-ph/9710081}{{\tt astro-ph/9710081}}].

\bibitem{Hammersley:2000mx}
P.~L. Hammersley, F.~Garzon, T.~Mahoney, M.~Lopez-Corredoira, and M.~A.~P.
  Torres, {\it {Detection of the old stellar component of the major galactic
  bar}},  {\em Mon. Not. Roy. Astron. Soc.} {\bf 317} (2000) L45,
  [\href{http://arxiv.org/abs/astro-ph/0007232}{{\tt astro-ph/0007232}}].

\bibitem{Porter:2017vaa}
T.~A. Porter, G.~Johannesson, and I.~V. Moskalenko, {\it {High-Energy Gamma
  Rays from the Milky Way: Three-Dimensional Spatial Models for the Cosmic-Ray
  and Radiation Field Densities in the Interstellar Medium}},  {\em Astrophys.
  J.} {\bf 846} (2017), no.~1 67, [\href{http://arxiv.org/abs/1708.00816}{{\tt
  arXiv:1708.00816}}].

\bibitem{Casandjian:andFermiLat2016}
F.~{Acero}, M.~{Ackermann}, M.~{Ajello}, A.~{Albert}, L.~{Baldini},
  J.~{Ballet}, G.~{Barbiellini}, D.~{Bastieri}, R.~{Bellazzini}, E.~{Bissaldi},
  E.~D. {Bloom}, R.~{Bonino}, E.~{Bottacini}, T.~J. {Brandt}, J.~{Bregeon},
  P.~{Bruel}, and et~al., {\it {Development of the Model of Galactic
  Interstellar Emission for Standard Point-source Analysis of Fermi Large Area
  Telescope Data}},  {\em Astrophys. J. Supp.} {\bf 223} (Apr., 2016) 26,
  [\href{http://arxiv.org/abs/1602.07246}{{\tt arXiv:1602.07246}}].

\bibitem{TheFermi-LAT:2017vmf}
{\bf Fermi-LAT} Collaboration, M.~Ackermann et~al., {\it {The Fermi Galactic
  Center GeV Excess and Implications for Dark Matter}},  {\em Astrophys. J.}
  {\bf 840} (2017), no.~1 43, [\href{http://arxiv.org/abs/1704.03910}{{\tt
  arXiv:1704.03910}}].

\bibitem{Wolleben:2007}
M.~{Wolleben}, {\it {A New Model for the Loop I (North Polar Spur) Region}},
  {\em Astrophys.~J.} {\bf 664} (July, 2007) 349--356,
  [\href{http://arxiv.org/abs/0704.0276}{{\tt arXiv:0704.0276}}].

\bibitem{Fermi-LAT:2FIG}
{\bf Fermi-LAT} Collaboration, M.~Ajello et~al., {\it {Characterizing the
  population of pulsars in the inner Galaxy with the Fermi Large Area
  Telescope}},  {\em Submitted to: Astrophys. J.} (2017)
  [\href{http://arxiv.org/abs/1705.00009}{{\tt arXiv:1705.00009}}].

\bibitem{Nishiyama2015}
S.~Nishiyama, K.~Yasui, T.~Nagata, T.~Yoshikawa, H.~Uchiyama, R.~Schodel,
  H.~Hatano, S.~Sato, K.~Sugitani, T.~Suenaga, J.~Kwon, and M.~Tamura, {\it
  Magnetically confined interstellar hot plasma in the nuclear bulge of our
  galaxy},  {\em Astrophys. J. Lett.} {\bf 769} (2013), no.~2 L28.

\bibitem{R12}
T.~P. {Robitaille}, E.~{Churchwell}, R.~A. {Benjamin}, B.~A. {Whitney},
  K.~{Wood}, B.~L. {Babler}, and M.~R. {Meade}, {\it {A self-consistent model
  of Galactic stellar and dust infrared emission and the abundance of
  polycyclic aromatic hydrocarbons}},  {\em A\&A} {\bf 545} (Sept., 2012) A39,
  [\href{http://arxiv.org/abs/1208.4606}{{\tt arXiv:1208.4606}}].

\bibitem{Wilandetal:1994}
J.~L. {Weiland}, R.~G. {Arendt}, G.~B. {Berriman}, E.~{Dwek}, H.~T.
  {Freudenreich}, M.~G. {Hauser}, T.~{Kelsall}, C.~M. {Lisse}, M.~{Mitra},
  S.~H. {Moseley}, N.~P. {Odegard}, R.~F. {Silverberg}, T.~J. {Sodroski}, W.~J.
  {Spiesman}, and S.~W. {Stemwedel}, {\it {COBE Diffuse Background Experiment
  Observations of the Galactic Bulge}},  {\em Astrophys. J.} {\bf 425} (Apr.,
  1994) L81.

\bibitem{Binneyetal:1997}
J.~{Binney}, O.~{Gerhard}, and D.~{Spergel}, {\it {The photometric structure of
  the inner Galaxy}},  {\em Mon. Not. Roy. Astron. Soc.} {\bf 288} (June, 1997)
  365--374, [\href{http://arxiv.org/abs/astro-ph/9609066}{{\tt
  astro-ph/9609066}}].

\bibitem{Cao:2013dwa}
L.~Cao, S.~Mao, D.~Nataf, N.~J. Rattenbury, and A.~Gould, {\it {A New
  Photometric Model of the Galactic Bar using Red Clump Giants}},  {\em Mon.
  Not. Roy. Astron. Soc.} {\bf 434} (2013), no.~1 595--605,
  [\href{http://arxiv.org/abs/1303.6430}{{\tt arXiv:1303.6430}}].

\bibitem{WeggandGerhard:2013}
C.~{Wegg} and O.~{Gerhard}, {\it {Mapping the three-dimensional density of the
  Galactic bulge with VVV red clump stars}},  {\em Mon. Not. Roy. Astron. Soc.}
  {\bf 435} (Nov., 2013) 1874--1887,
  [\href{http://arxiv.org/abs/1308.0593}{{\tt arXiv:1308.0593}}].

\bibitem{Bland-Hawthorn2016}
J.~{Bland-Hawthorn} and O.~{Gerhard}, {\it {The Galaxy in Context: Structural,
  Kinematic, and Integrated Properties}},  {\em Ann. Rev. of A \& A} {\bf 54}
  (Sept., 2016) 529--596, [\href{http://arxiv.org/abs/1602.07702}{{\tt
  arXiv:1602.07702}}].

\bibitem{Samland:2003}
M.~{Samland} and O.~E. {Gerhard}, {\it {The formation of a disk galaxy within a
  growing dark halo}},  {\em Astron. and Astrophys.} {\bf 399} (Mar., 2003)
  961--982, [\href{http://arxiv.org/abs/astro-ph/0301499}{{\tt
  astro-ph/0301499}}].

\bibitem{Shenetal:2010}
J.~{Shen}, R.~M. {Rich}, J.~{Kormendy}, C.~D. {Howard}, R.~{De Propris}, and
  A.~{Kunder}, {\it {Our Milky Way as a Pure-disk Galaxy-A Challenge for Galaxy
  Formation}},  {\em Astrophys. J. Lett.} {\bf 720} (Sept., 2010) L72--L76,
  [\href{http://arxiv.org/abs/1005.0385}{{\tt arXiv:1005.0385}}].

\bibitem{Kunderetal:2016}
A.~{Kunder}, R.~M. {Rich}, A.~{Koch}, J.~{Storm}, D.~M. {Nataf}, R.~{De
  Propris}, A.~R. {Walker}, G.~{Bono}, C.~I. {Johnson}, J.~{Shen}, and Z.-Y.
  {Li}, {\it {Before the Bar: Kinematic Detection of a Spheroidal Metal-poor
  Bulge Component}},  {\em Astrophys. J. Lett.} {\bf 821} (Apr., 2016) L25,
  [\href{http://arxiv.org/abs/1603.06578}{{\tt arXiv:1603.06578}}].

\bibitem{Skrutskie:2006}
M.~F. {Skrutskie}, R.~M. {Cutri}, R.~{Stiening}, M.~D. {Weinberg},
  S.~{Schneider}, J.~M. {Carpenter}, C.~{Beichman}, R.~{Capps}, T.~{Chester},
  J.~{Elias}, J.~{Huchra}, J.~{Liebert}, C.~{Lonsdale}, D.~G. {Monet},
  S.~{Price}, P.~{Seitzer}, T.~{Jarrett}, J.~D. {Kirkpatrick}, J.~E. {Gizis},
  E.~{Howard}, T.~{Evans}, J.~{Fowler}, L.~{Fullmer}, R.~{Hurt}, R.~{Light},
  E.~L. {Kopan}, K.~A. {Marsh}, H.~L. {McCallon}, R.~{Tam}, S.~{Van Dyk}, and
  S.~{Wheelock}, {\it {The Two Micron All Sky Survey (2MASS)}},  {\em
  Astrophys. J.} {\bf 131} (Feb., 2006) 1163--1183.

\bibitem{McWilliam:2010}
A.~{McWilliam} and M.~{Zoccali}, {\it {Two Red Clumps and the X-shaped Milky
  Way Bulge}},  {\em Astrophys. J.} {\bf 724} (Dec., 2010) 1491--1502,
  [\href{http://arxiv.org/abs/1008.0519}{{\tt arXiv:1008.0519}}].

\bibitem{Nataf:2010}
D.~M. {Nataf}, A.~{Udalski}, A.~{Gould}, P.~{Fouqu{\'e}}, and K.~Z. {Stanek},
  {\it {The Split Red Clump of the Galactic Bulge from OGLE-III}},  {\em
  Astrophys. J. lett.} {\bf 721} (Sept., 2010) L28--L32,
  [\href{http://arxiv.org/abs/1007.5065}{{\tt arXiv:1007.5065}}].

\bibitem{Laplace}
``Repair damaged image pixel using {Laplace} interpolation.''
  \url{https://community.wolfram.com/groups/-/m/t/873396}.
\newblock Accessed: 2018-12-18.

\bibitem{Malyshev:2012mb}
D.~Malyshev, {\it {Spectral components analysis of diffuse emission
  processes}},  \href{http://arxiv.org/abs/1202.1034}{{\tt arXiv:1202.1034}}.

\bibitem{Galprop}
``Galprop.'' \url{http://galprop.stanford.edu}.
\newblock Accessed: 2018-10-15.

\bibitem{ackermannajelloatwood2012}
M.~Ackermann et~al., {\it Fermi-lat observations of the diffuse gamma-ray
  emission: Implications for cosmic rays and the interstellar medium},  {\em
  Astrophys. J.} {\bf 750} (2012), no.~1 3.

\bibitem{3FGL}
{\bf Fermi-LAT} Collaboration, F.~{Acero}, M.~{Ackermann}, M.~{Ajello},
  A.~{Albert}, et~al., {\it {Fermi Large Area Telescope Third Source Catalog}},
   {\em Astrophys. J. Supp.} {\bf 218} (June, 2015) 23,
  [\href{http://arxiv.org/abs/1501.02003}{{\tt arXiv:1501.02003}}].

\bibitem{SelfLiang87}
S.~G. {Self} and K.~{Liang}, {\it Asymptotic properties of maximum likelihood
  estimators and likelihood ratio tests under nonstandard conditions},  {\em
  Journal of the American Statistical Association} {\bf 82} (1987), no.~398
  605--610.

\bibitem{Cholis:2018}
B.~Balaji, I.~Cholis, P.~J. Fox, and S.~D. McDermott, {\it {Analyzing the
  Gamma-Ray Sky with Wavelets}},  {\em Phys. Rev.} {\bf D98} (2018), no.~4
  043009, [\href{http://arxiv.org/abs/1803.01952}{{\tt arXiv:1803.01952}}].

\bibitem{Yusef-Zadeh:2009}
F.~{Yusef-Zadeh}, J.~W. {Hewitt}, R.~G. {Arendt}, B.~{Whitney}, G.~{Rieke},
  M.~{Wardle}, J.~L. {Hinz}, S.~{Stolovy}, C.~C. {Lang}, M.~G. {Burton}, and
  S.~{Ramirez}, {\it {Star Formation in the Central 400 pc of the Milky Way:
  Evidence for a Population of Massive Young Stellar Objects}},  {\em
  Astrophys. J.} {\bf 702} (Sept., 2009) 178--225,
  [\href{http://arxiv.org/abs/0905.2161}{{\tt arXiv:0905.2161}}].

\bibitem{Immeretal:2012}
K.~{Immer}, F.~{Schuller}, A.~{Omont}, and K.~M. {Menten}, {\it {Recent star
  formation in the inner Galactic bulge seen by ISOGAL. II. The central
  molecular zone}},  {\em A \& A} {\bf 537} (Jan., 2012) A121,
  [\href{http://arxiv.org/abs/1111.3295}{{\tt arXiv:1111.3295}}].

\bibitem{Koepferletal:2015}
C.~M. {Koepferl}, T.~P. {Robitaille}, E.~F.~E. {Morales}, and K.~G. {Johnston},
  {\it {Main-sequence Stars Masquerading as Young Stellar Objects in the
  Central Molecular Zone}},  {\em Astrophys. J.} {\bf 799} (Jan., 2015) 53,
  [\href{http://arxiv.org/abs/1411.4646}{{\tt arXiv:1411.4646}}].

\bibitem{Longmore:2018}
S.~{Longmore} and J.~M.~D. {Kruijssen}, {\it {Constraints on the Distribution
  of Gas and Young Stars in the Galactic Centre in the Context of Interpreting
  Gamma Ray Emission Features}},  {\em Galaxies} {\bf 6} (May, 2018) 55,
  [\href{http://arxiv.org/abs/1805.06287}{{\tt arXiv:1805.06287}}].

\bibitem{TheFermi-LAT:2PC}
{\bf Fermi-LAT} Collaboration, A.~A. Abdo et~al., {\it {The Second Fermi Large
  Area Telescope Catalog of Gamma-ray Pulsars}},  {\em Astrophys. J. Suppl.}
  {\bf 208} (2013) 17, [\href{http://arxiv.org/abs/1305.4385}{{\tt
  arXiv:1305.4385}}].

\bibitem{Gnedin:2002un}
O.~Y. Gnedin, H.~Zhao, J.~E. Pringle, S.~M. Fall, M.~Livio, and G.~Meylan, {\it
  {The unique history of the globular cluster omega centauri}},  {\em
  Astrophys. J.} {\bf 568} (2002) L23--L26,
  [\href{http://arxiv.org/abs/astro-ph/0202045}{{\tt astro-ph/0202045}}].

\bibitem{Hobbs:2005yx}
G.~Hobbs, D.~R. Lorimer, A.~G. Lyne, and M.~Kramer, {\it {A Statistical study
  of 233 pulsar proper motions}},  {\em Mon. Not. Roy. Astron. Soc.} {\bf 360}
  (2005) 974--992, [\href{http://arxiv.org/abs/astro-ph/0504584}{{\tt
  astro-ph/0504584}}].

\bibitem{Pfahl:2001df}
E.~Pfahl, S.~Rappaport, and P.~Podsiadlowski, {\it {A comprehensive study of
  neutron star retention in globular clusters}},  {\em Astrophys. J.} {\bf 573}
  (2002) 283, [\href{http://arxiv.org/abs/astro-ph/0106141}{{\tt
  astro-ph/0106141}}].

\bibitem{Brandt:1994rr}
W.~N. Brandt and P.~Podsiadlowski, {\it {The Effects of High-Velocity Supernova
  Kicks on the Orbital Properties and Sky Distributions of Neutron Star
  Binaries}},  \href{http://arxiv.org/abs/astro-ph/9412023}{{\tt
  astro-ph/9412023}}.

\bibitem{Huietal:2010}
C.~Y. {Hui}, K.~S. {Cheng}, and R.~E. {Taam}, {\it {Dynamical Formation of
  Millisecond Pulsars in Globular Clusters}},  {\em Astrophys. J.} {\bf 714}
  (May, 2010) 1149--1154, [\href{http://arxiv.org/abs/1003.4332}{{\tt
  arXiv:1003.4332}}].

\bibitem{Gonthier:2018ymi}
P.~L. Gonthier, A.~K. Harding, E.~C. Ferrara, S.~E. Frederick, V.~E. Mohr, and
  Y.-M. Koh, {\it {Population syntheses of millisecond pulsars from the
  Galactic Disk and Bulge}},  {\em Astrophys. J.} {\bf 863} (2018), no.~2 199,
  [\href{http://arxiv.org/abs/1806.11215}{{\tt arXiv:1806.11215}}].

\bibitem{Voss:2006az}
R.~Voss and M.~Gilfanov, {\it {A study of the population of LMXBs in the bulge
  of M31}},  {\em Astron. Astrophys.} (2006)
  [\href{http://arxiv.org/abs/astro-ph/0610649}{{\tt astro-ph/0610649}}].
  [Astron. Astrophys.468,49(2007)].

\bibitem{Fragione:2017rsp}
G.~Fragione, F.~Antonini, and O.~Y. Gnedin, {\it {Disrupted Globular Clusters
  and the Gamma-Ray Excess in the Galactic Centre}},  {\em Mon. Not. Roy.
  Astron. Soc.} {\bf 475} (2018), no.~4 5313--5321,
  [\href{http://arxiv.org/abs/1709.03534}{{\tt arXiv:1709.03534}}].

\bibitem{Nataf:2017}
D.~M. {Nataf}, {\it {Was the Milky Way Bulge Formed from the Buckling Disk
  Instability, Hierarchical Collapse, Accretion of Clumps, or All of the
  Above?}},  {\em Publications of the Astronomical Society of Australia} {\bf
  34} (Sept., 2017) e041, [\href{http://arxiv.org/abs/1708.01262}{{\tt
  arXiv:1708.01262}}].

\bibitem{Aharonian:2006au}
{\bf H.E.S.S.} Collaboration, F.~Aharonian et~al., {\it {Discovery of
  very-high-energy gamma-rays from the galactic centre ridge}},  {\em Nature}
  {\bf 439} (2006) 695--698, [\href{http://arxiv.org/abs/astro-ph/0603021}{{\tt
  astro-ph/0603021}}].

\bibitem{Guepin:2018jkb}
C.~Guépin, L.~Rinchiuso, K.~Kotera, E.~Moulin, T.~Pierog, and J.~Silk, {\it
  {Pevatron at the Galactic Center: Multi-Wavelength Signatures from
  Millisecond Pulsars}},  {\em JCAP} {\bf 1807} (2018), no.~07 042,
  [\href{http://arxiv.org/abs/1806.03307}{{\tt arXiv:1806.03307}}].

\bibitem{Macias:2014sta}
O.~Macias, R.~Crocker, C.~Gordon, and S.~Profumo, {\it {Cosmic ray models of
  the ridge-like excess of gamma rays in the Galactic Centre}},  {\em Mon. Not.
  Roy. Astron. Soc.} {\bf 451} (2015), no.~2 1833--1847,
  [\href{http://arxiv.org/abs/1410.1678}{{\tt arXiv:1410.1678}}].

\bibitem{FermilatBubblespaper}
{\bf Fermi-LAT} Collaboration, M.~Ackermann et~al., {\it {The Spectrum and
  Morphology of the $Fermi$ Bubbles}},  {\em Astrophys. J.} {\bf 793} (2014),
  no.~1 64, [\href{http://arxiv.org/abs/1407.7905}{{\tt arXiv:1407.7905}}].

\bibitem{Simionetal:2017}
I.~T. {Simion}, V.~{Belokurov}, M.~{Irwin}, S.~E. {Koposov},
  C.~{Gonzalez-Fernandez}, A.~C. {Robin}, J.~{Shen}, and Z.~Y. {Li}, {\it {A
  parametric description of the 3D structure of the Galactic bar/bulge using
  the VVV survey}},  {\em Mon. Not. Roy. Astron. Soc.} {\bf 471} (Nov., 2017)
  4323--4344, [\href{http://arxiv.org/abs/1707.06660}{{\tt arXiv:1707.06660}}].

\bibitem{Eckner:2017oul}
C.~Eckner et~al., {\it {Millisecond pulsar origin of the Galactic center excess
  and extended gamma-ray emission from Andromeda - a closer look}},  {\em
  Astrophys. J.} {\bf 862} (2018), no.~1 79,
  [\href{http://arxiv.org/abs/1711.05127}{{\tt arXiv:1711.05127}}].

\end{thebibliography}\endgroup
\bibliographystyle{JHEP}

\end{document}